\documentclass[twocolumn,aps,10pt,prx,notitlepage,preprintnumbers,superscriptaddress,amsmath,amssymb]{revtex4-2}
\usepackage{graphicx} 
\usepackage{xcolor}
\usepackage{circuitikz}
\usepackage{soul}
\usepackage{enumitem}

\newcommand{\rhot}{\tilde \rho}
\newcommand{\wmax}{w_{0}}
\newcommand{\rhop}{{\rho}_{\rm rf}}
\newcommand{\Df}{D_{ \varphi}}
\newcommand{\Dq}{D_{q}}
\newcommand{\norm}[1]{\lVert#1\rVert}

\newcommand{\cP}{\mathcal{P}}
\newcommand{\erf}{{\rm erf}}
\newcommand{\erfc}{{\rm erfc}}
\newcommand{\nint}{{\rm round}}

\newcommand{\ket}[1]{\lvert #1 \rangle }
\newcommand{\bra}[1]{\langle #1 \rvert }

\newcommand{\Tb}{T}

\newcommand{\fr}{f_{\rm r}}

\newcommand{\addFN}[1]{#1} 

\newcommand{\be}{\begin{equation}}
\newcommand{\ee}{\end{equation}}
\newcommand{\Tr}{{\rm Tr}}
\newcommand{\sgn}{{\rm sgn}}
 
\newcommand{\cU}{\mathcal{U}}
\newcommand{\trnorm}[1]{\norm{#1}_{\rm tr}}
\newcommand{\Ts}{t_{\rm s}}
\newcommand{\zs}{z_{\rm rev}}
\newcommand{\trev}{t_{\rm rev}}
\newcommand{{\ULC}}{U_{\text{LC}}}

\renewcommand{\mod}{\,\,{\rm mod}\,\, }

\newcommand{\ws}{w_{\rm s}} 
\newcommand{\bs}{\boldsymbol}
\graphicspath{{./}}
\newcommand{\flc}{f_{\rm LC}}
\newcommand{\TLC}{\tau_{\rm LC}}

\newcommand{\lamp}{{\lambda_0}}

\newcommand{\kth}{\kappa} 

\newcommand{\xiq}{\xi_q}
\newcommand{\lth}{\lambda}

\begin{document} 

\title{Self-correcting GKP qubit and gates  in a driven-dissipative circuit}
\author{
Frederik Nathan}
\affiliation{Center for Quantum Devices and NNF Quantum Computing Programme, Niels Bohr Institute,  University of Copenhagen, 2100 Copenhagen, Denmark}
\affiliation{
Department of Physics and Institute for Quantum Information and Matter, California Institute of Technology, Pasadena, CA 91125, USA}
\author{Liam O'Brien}
\affiliation{
Department of Physics and Institute for Quantum Information and Matter, California Institute of Technology, Pasadena, CA 91125, USA}
\author{Kyungjoo Noh}
\affiliation{AWS Center for Quantum Computing, Pasadena, CA, 91125, USA}
\author{Matthew H. Matheny}
\affiliation{AWS Center for Quantum Computing, Pasadena, CA, 91125, USA}
\author{Arne L. Grimsmo}
\affiliation{AWS Center for Quantum Computing, Pasadena, CA, 91125, USA}
\author{Liang Jiang}
\affiliation{AWS Center for Quantum Computing, Pasadena, CA, 91125, USA}
\affiliation{Pritzker School of     Molecular Engineering, The University of Chicago, Chicago, Illinois 60637, USA}
\author{Gil Refael}
\affiliation{
Department of Physics and Institute for Quantum Information and Matter, California Institute of Technology, Pasadena, CA 91125, USA}
\affiliation{AWS Center for Quantum Computing, Pasadena, CA, 91125, USA}

\begin{abstract}
\addFN{We show that a self-correcting GKP qubit can be realized with a high-impedance LC circuit coupled to a resistor and a Josephson junction   via a controllable switch. When activating the switch in a particular  
stepwise pattern,  the resonator  relaxes into a subspace of GKP states that encode a protected qubit. Under continued operation, the resistor dissipatively error-corrects the qubit against bit flips  {and} decoherence  by absorbing noise-induced entropy. 
We show that  this leads  to an exponential enhancement of   coherence time ($T_1$ and $T_2$), even in the presence of extrinsic noise, imperfect control, and  device parameter variations.
We show the qubit supports  exponentially robust single-qubit Clifford gates,  implemented via appropriate control of the switch, and  readout/initialization via supercurrent measurement. 
The qubit's self-correcting properties allows it to  operate at $\sim 1\, {\rm K}$ temperatures and  resonator Q factors down to $\sim 1000$  for realistic parameters, and make it amenable to parallel control through global control signals.
{We discuss how the effects of quasiparticle poisoning---potentially, though not necessarily, a  limiting factor---might be mitigated.} 
We finally demonstrate that a related device supports a self-correcting magic $T$ gate.} 
\end{abstract}

\maketitle

Quantum error correction is a crucial element     in quantum computing, due to the inevitability of noise from,  e.g.,  uncontrolled degrees of freedom, imperfect control, or fluctuations of device parameters~\cite{Shor_1995,Calderbank_1996,Steane_1996,Knill_1998,Raussendorf_2007,Fowler_2012,acharya_2023}.
Many approaches---such as surface codes---rely on {\it active  correction}, which eliminate noise-induced entropy via readout/feedback~\cite{Fowler_2012,Terhal_2015,acharya_2023}. 
Requirements for  rapid readout, extensive  control, and  complex device architectures, make the scalability of these approaches a significant  challenge~\cite{Preskill_2018,pauka_2019,acharya_2023}. 
On the other hand, classical bits are often intrinsically stable due to {\it dissipation}~\cite{Bravyi_2009,Bombin_2013}:  in a magnetic 
hard-disk, e.g., noise-induced magnetic fluctuations  are  damped dissipatively before  they accumulate  to generate bit flips, leading to extreme robustness.
Similarly harnessing  dissipation for quantum error correction  is a  challenging, but desirable, goal~\cite{Barnes_2000,Kraus_2008,Verstraete_2009,Leghtas_2013,reiter_2017,Touzard_2018,de_neeve_error_2020,fluhmann_2019,perez_improved_2020,harrington_engineered_2022,grimm_2020,gyenis_2021,sellem_2023_gkp}.

\begin{figure}[h!]
    \includegraphics[width=0.99\columnwidth]{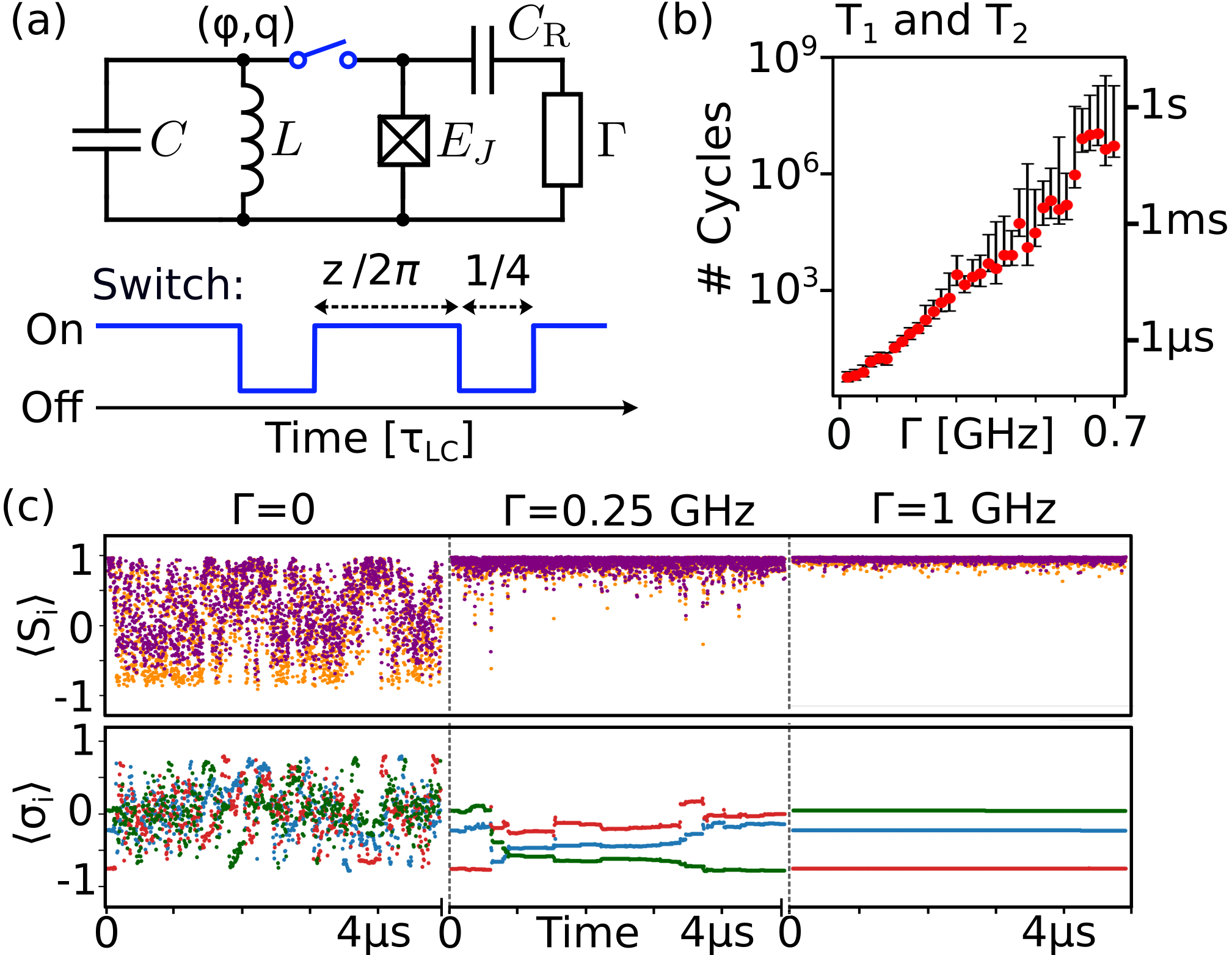}
    \caption{\addFN{\textbf{Self-correcting GKP qubit}. 
    (a) Circuit diagram and protocol, with $z $  an integer and $\TLC$  the  oscillation period of the LC resonator. 
    (b) Simulated qubit lifetime with  extrinsic charge noise  (strength $10^{-12}\,e^2/\rm{Hz}$; see Sec. \ref{sec:charge_noise}), versus resistor-induced loss rate, $\Gamma$ [see Eq.~\eqref{eq:gamma}], for parameters  $L=10\,\mu{\rm H}$, $C=60\,{\rm fF}$, $z=2$, $E_J/h = 200\,{\rm GHz} $, and resistor temperature  $T=40\,{\rm mK}$. 
    Data points are obtained via averaging over  50-100  Universal Lindblad equation (ULE) trajectories, with error bars 
    indicating $95\%$ confidence interval from bootstrap resampling. Note that bit flip and dephasing  appear symmetrically in the protocol, implying relaxation and coherence times are {\it identical}: $T_1=T_2$.
    (c) Expectations of {GKP stabilizers $S_1$ and $S_2$ (purple,  orange) and logical 
       operators $\sigma_x , \sigma_y,\sigma_z$  (red, blue,  green)  [see Eqs.~(\ref{Pauli_ops},\ref{eq:Stabilizers})] for individual,  randomly selected     ULE trajectories at  $3$ values of $\Gamma$.
    }}}
    \label{fig_1}
\end{figure}

Here we propose an architecture for a  dissipatively error-corrected  qubit. 
The device   consists of an LC resonator with impedance close to $h/2e^2 \approx 12.91\,{\rm k}\Omega$,  connected to a Josephson junction and a dissipative element through a controllable switch, which is activated via a stepwise protocol [Fig.~\ref{fig_1}(a)]. 
Quantum information is encoded in a thermal {ensemble} 
of generalized  Gottesman-Kitaev-Preskill (GKP) states~\cite{Gottesman_2000} (see Sec.~\ref{sec:Quantum_Info} and  Fig.~\ref{fig:crenellation} for details), and can be accessed and initialized via the Josephson junction supercurrent.
When the  Josephson energy is larger than   temperature and LC frequency,
  the qubit enters  a regime of {\it dissipative error correction} (DEC), where noise-induced fluctuations are damped dissipatively without affecting the   encoded information.
Thereby, the resistor error corrects  the qubit  against bit flip {\it and} dephasing.

\addFN{
Our simulations show that DEC can leads to exponential  increase of the coherence time that can extend   beyond  $10^6$ oscillator cycles, 
even with extrinsic charge or flux noise present [Fig.~\ref{fig_1}(bc)]. 
We confirm this  exponential scaling analytically [Eq.~\eqref{eq:error_rate_result}].
Its self-correcting properties give the  qubit finite  tolerance for manufactoring {and control} imperfections, making the qubit amenable to parallel control through global control signals, and, potentially, calibration-free operation. Additionally, the self-corrrecting properties allow the qubit to operate at relatively large temperatures ($\sim 1\,{\rm K}$) with imperfect resonators ($Q\sim 1000$) for realistic circuit parameters (see also  Table~\ref{tab:parameters} below).}

\addFN{DEC protects the qubit against phase-space local noise, generated by finite-order polynomials of mode quadratures.
This generic class of noise includes flux and charge noise, along with uncontrolled deviations of device parameters and  control signals.  Notably, quasiparticle poisoning events may fall outside this category, and hence potentially---though not necessarily---provide a limiting factor for the qubit's stability. 
We discuss possible consequences of quasiparticle poisoning and mitigation strategies in Sec.~\ref{sec:experimental_considerations} below.}


Interestingly, 
the  qubit  supports a native set of 
Clifford gates, implemented via control of the switch. The gates are topologically robust~\cite{conrad2024latticesgatescurvesgkp}, and  dissipatively corrected by the resistor, making them exponentially insensitive to control noise (see Fig.~\ref{fig:gates}). 
We also find that a different encoding  results in a native  self-correcting T gate by a similar mechanism~\cite{magic_states}.

\addFN{Achieving an efficient  switched Josephson junction 
is a key technological challenge for  the qubit.
We estimate the required switch   time 
proportional to the resonator inductance  $L$ as $\delta t_{\rm max} \sim 8 \, {\rm ps} \times L\,[\mu{\rm H}]$.  We thus expect the  {device to be feasible} for  {$L$} in the  $  1\text{--}5\, \mu$H range and switch rise  times in the $5\,\text{--}30\,{\rm ps}$ range.
For reference, high-impedance resonators have been realized with  $L=2.5\,\mu{\rm H}$~\cite{Pechenezhskiy_2020}, and pulse-train generators  with  $\lesssim 10\,{\rm ps}$ rise times have been available for  decades~\cite{Afshari_2005}.
In Sec.~\ref{sec:experimental_considerations} we discuss possible routes to integrating such pulses with a switchable Josephson junctions. Importantly, even if insufficient for dissipative error correction,  an imperfect switch may still efficiently prepare and stabilize GKP states.
Realizing a rapidly  switchable Josephson junction may thus carry a significant reward, by enabling  a self-correcting qubit with exponentially-scaling lifetime, a protected set of gates, high-operating temperature, and an intrinsic tolerance for manufactiuring and control inconsistencies.}

\addFN{In recent years, multiple works have succesfully  realized GKP states in trapped-ion, or circuit-QED devices through readout/feedback 
\cite{fluhmann_2019,campagne-ibarcq_quantum_2020,Eickbusch_2022,Sivak_2023}, including protocols  emulating dissipation~\cite{LachanceQuirion_2023,deNeeve_2022}.
There have also been proposals  for deterministic protocols  generating GKP states through coherent driving protocols~\cite{kolesnikow_2023_GKP,conrad_2021_GKP} or circuits featuring gyrators~\cite{Rymarz_2021}, and more recently,  related proposals have been proposed for dissipatively stabilizing GKP states through   bath engineering via frequency combs \cite{sellem_2023_gkp} and qubit resets \cite{LachanceQuirion_2023}. 
A key advantage of our proposal is that it realizes dissipative error correction  with {\it generic} thermodynamic baths and a  stepwise switch activation protocol, offering a complementary approach with potentially simpler realizations. 
The native  self-correcting single-qubit Clifford gates may also simplify the qubit's  integration in a quantum processor.}

The rest of the paper is organized as follows: 
in Sec.~\ref{sec:Quantum_Info} we describe how we encode a GKP  qubit  in an LC resonator; in Sec.~\ref{sec:Setup}, we introduce the  self-correcting  GKP qubit, discuss its basic operating  principles and demonstrate its support of topologically-robust single-qubit Clifford gates. 
In Sec.~\ref{sec:Stabilization}, we  \addFN{analytically demonstrate the exponential protection of the encoded information}. 
In Sec.~\ref{sec:T_gate}, we discuss a scheme for generating dissipatively-corrected $T$ gates in a related architecture.
In Sec. \ref{sec:Readout}, we discuss implementations of readout/initialization. 
In Sec.~\ref{sec:params} we estimate the device criteria, operation timescales, and noise tolerances, summarized in   Table~\ref{tab:parameters}. 
In Sec.~\ref{sec:Numerics} we  provide data from numerical simulations of the device. 
We conclude with a discussion in Sec.~\ref{sec:Discussion}.

\section{Encoding of quantum information}
\label{sec:Quantum_Info}
\begin{figure}
    \centering
    \includegraphics[width=0.99\columnwidth]{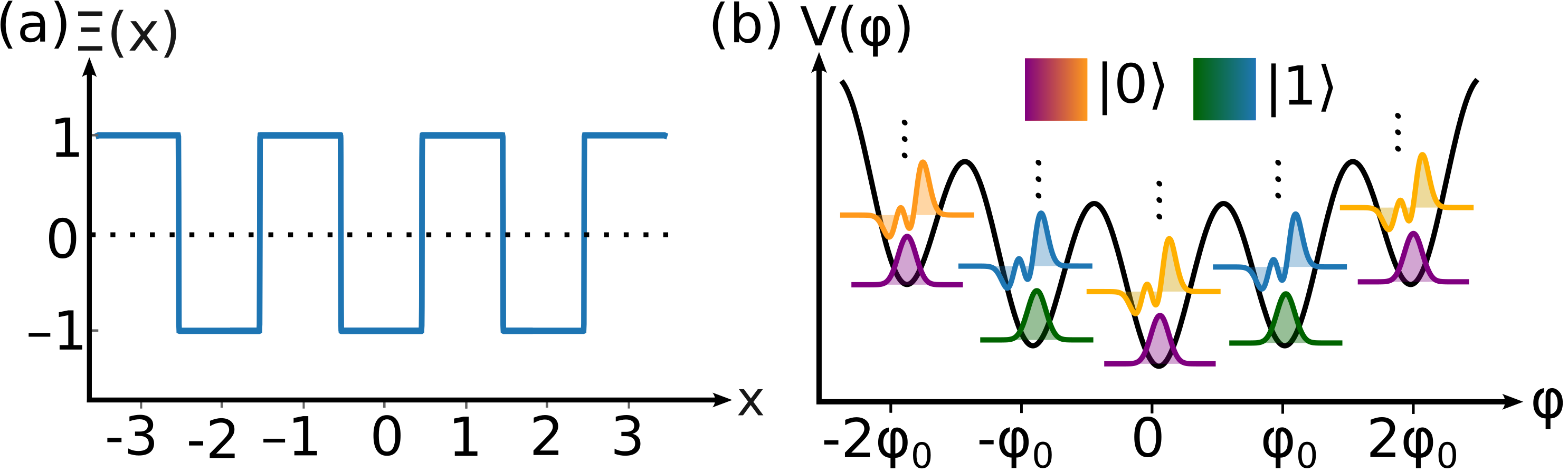}
    \caption{\textbf{GKP encoding} (a) Crenellation function $\Xi$ used to encode the quantum information  via  Eqs.~\eqref{Pauli_ops}.
    \addFN{(b) Examples of logical $|0\rangle$ (purple, orange) and  $|1\rangle$ (green, blue) states. Black curve indicates the flux potential from the inductor and Josephson junction (when active).}
}
    \label{fig:crenellation}
\end{figure}
The qubit is encoded in thermal \addFN{ensembles} of GKP states in an LC resonator~\cite{Gottesman_2000}.  In terms of the resonator flux $\varphi$ and charge $q$, the GKP states  have their Wigner function  support confined  near integer multiples of $\varphi/\varphi_0$ and $q/e$, where $e$ denotes the electron charge, and $\varphi_0 =h/2e$ the flux quantum. 
{The parities of $\varphi/\varphi_0$ and  $q/e$} define the $\sigma_z$ and $\sigma_x$ logical operators, respectively, via
\be 
\sigma_z = {\Xi(\varphi/\varphi_0)},\quad 
\sigma_x = {\Xi(q/e)}, \quad
\sigma_y = -i \sigma_z \sigma_x,
\label{Pauli_ops}
\ee 
where 
$ 
\Xi(x)\equiv \sgn\cos(\pi x)
$ 
denotes  the {\it  crenellation function} (see Fig.~\ref{fig:crenellation}),   
which takes value $1$ when the closest integer to $x$ is even and value $-1$ if the closest integer to $x$ is odd~\cite{Fourier_expansion,Pantaleoni_2020}.
Since $\Xi(x)=-\Xi(x+1)$, the 3 operators above satisfy the Pauli anticommutation relations 
\be 
\{\sigma_i,\sigma_j\}=2\delta_{ij},
\ee 
and hence form a valid qubit observable.
We can encode a $\nu$-dimensional qudit in an analogous fashion; see Sec.~\ref{sec:T_gate} for an example. 
The modular encoding in Eq.~\eqref{Pauli_ops} allows {\it thermally mixed}  physical states to encode {\it pure} logical states [see Fig.~\ref{fig:crenellation}(b)]. 
This key feature   underlies the exponential stability of our  qubit.

GKP-encoded information is protected against 
noise induced by finite-order polynomials of $\varphi$ and $q$, such as charge/flux noise and photon loss---here termed {\it local noise}. 
The protection  emerges because local noise generates a continuous flow of the system's  Wigner function. 
The logical operators $\{\sigma_i\}$ are  unaffected by this flow  as long as the system's Wigner function support does not leak across the domain boundaries located at $\varphi=(n_1+1/2)\varphi_0$ and $q=(n_2+1/2)e$ for integers $n_1$ and $n_2$. 
{Put another way}, the encoded information is protected as long as  the phase-space support of the system remains confined in the span of high-eigenvalue eigenstates of  the two {\it GKP stabilizers} 
\be 
S_1 = \cos\left({2\pi \varphi}/{\varphi_0}\right), \quad S_2 = \cos(2\pi  q/e)\,.
\label{eq:Stabilizers}
\ee 
Henceforth we refer to the mutual high-eigenvalue subspace  of $S_1$ and $S_2$ as the {\it code subspace}, and to states within the code subspace as (generalized) GKP states.

In circuit-QED, GKP states can be realized as phase-coherent superpositions of states confined deep within the wells of a Josephson potential. 
To ensure $\langle S_2\rangle \approx 1$, a GKP state's restriction to a single well must  be  approximately  identical for nearby wells, up to  a well-parity dependent   relative amplitude, which encodes the quantum information. 
The logical states of the qubit, $|0\rangle$ and $|1\rangle$,   correspond to  GKP  states with full support in even and odd wells, respectively.


\section{Self-correcting  qubit}
\label{sec:Setup}
\label{Model_circuit}
We now show how GKP states can be dissipatively  stabilized and error corrected in the circuit device in  Fig.~\ref{fig_1}(a). We also  demonstrate that the device  supports native, protected single-qubit Clifford gates.

\subsection{Device}
The device  consists of  an LC resonator connected   via a switch to a Josephson junction and, capacitively, to a dissipative element.  
Here the dissipative element can, e.g., be a resistor or a transmission line connected to an external reservoir; for simplicity we refer to it from here on as a resistor.
The resulting  circuit is  described by 
\begin{equation}
    H(t) = \frac{\varphi^2}{2L}
    + \frac{q^2}{2C} -\ws(t)\left[E_J \cos\left(\frac{2\pi \varphi}{\varphi_0}\right)+\frac{q Q_{\rm R}}{C_{\rm R}}\right] +H_{\rm R},
    \label{H_LCJ_classical}
\end{equation} 
where $L$ and $C$ denote the inductance and capacitance of the LC circuit,  $E_J$ the Josephson energy of the junction, while $\ws(t)$  defines the time-dependence of the switch. 
Additionally,  $H_{\rm R}$ denotes the resistor Hamiltonian,  $C_{\rm R}$ the coupler capacitance, and  $Q_{\rm R}$ denotes the fluctuating  charge on the resistor-side of the coupler. 
\addFN{Due to its self-correcting properties the qubit  may also operate with the resistor connected permanently to the LC resonator, see Sec.~\ref{sec:params} for details.}
\subsection{Phase revival trick}
\label{sec:stabilization_overview}
To see  how   GKP states emerge in the device, first note that  activating the switch (setting $\ws=1$) causes the system to dissipatively relax in the cosine  wells from the Josephson potential, confining it in the high-eigenvalue subspace of $S_1$.
We  can   stabilize $S_2$ by subsequently {\it deactivating} the switch for a quarter of the LC oscillation cycle, $\TLC\equiv 2\pi \sqrt{LC}$. 
In this deactivated interval, the Hamiltonian  generates a $\pi/2$ rotation of phase space that interchanges $\varphi$ and $q$ up to  a rescaling set by the resonator impedance, $ \sqrt{L/C}$. 
Setting  
\be  
    \sqrt{{L}/{C}} \approx  {h}/{2e^2} 
    \label{impedance_cond}
\ee 
ensures that  $\varphi/\varphi_0$ is mapped to  $q/e$ and vice versa  (up to a sign), leading to an effective interchange of  $S_1$ and $S_2$ [see Eq.~\eqref{eq:Stabilizers}]. 
Hence, at the end of the deactivated segment, the distirbution of  charge $q$ is confined  in peaks near multiples of $e$, resulting in $\langle S_2\rangle\approx 1$. 
Reactivating the switch   will again confine  the system in the high-eigenvalue subspace of $S_1$, due to relaxation into the wells of the Josephson potential. The system will  be confined in the code subspace  at this point if $\langle S_2\rangle$ retains its near-unit value over the second switch-activated interval. 
\addFN{
As a key discovery,  we uncover a revival mechansim that ensures that this is the case if 
the reactivated-switch segment  has duration given by an integer multiple of the {\it revival time}, 
\be 
\trev = \sqrt{LC}. 
\ee }

\addFN{The revival mechanism arises because dephasing of wavefunction components in distinct Josephson potential wells  correspond to diffusion of $q$ away from multiples of $e$. 
Due to its capacitative coupling, the resistor can only generate such charge diffusion continuously and indirectly, through the inductor of the circuit.  
In particular, the  diffusion can be slow enough to allow the peaks of the charge distribution to  remain well-defined and non-overlapping on the timescale it takes for flux to relax  (see Sec.~\ref{sec:Stabilization} for details).}

While wavefunction components in distinct wells remain phase-coherent in the presence of the \textit{resistor}, they do acquire  deterministic relative phase factors, due to their different \textit{inductance energies}, corresponding to the vertical offset of  wells  in Figs.~\ref{fig:crenellation}(b)~and~\ref{fig:wavefunctions}(a). 
In the switch-reactivated segment, these phase  factors  initially cause the expectation of $ S_2$, which translates the wavefunction by $\pm$ 2 wells, to  decay to zero. 
However,   the inductance energy  for well $n$ is given by $n^2 \varepsilon_L$, with $\varepsilon_L = \varphi_0^2/2L$. 
Since \be 
(n^2 \mod 4) = (n \mod 2)\quad {\rm for}\quad n \in \mathbb Z ,
\label{eq:moduloresult}\ee 
wells with the same parity have inductance energies  congruent modulo $4\varepsilon_L $.  
As a result, the phase factors from wells of the same parity  {\it realign} at integer multiples of a {\it revival time} $2\pi \hbar /4\varepsilon_ L =\trev$. 
At these instances,  $S_2$ {revives}. This revival mechanism is clearly    demonstrated in our   numerical simulations---see {Fig.~\ref{fig:measurement}(e) and shaded regions in} {Fig.~\ref{fig:stab_fig}(c).}

The above considerations lead  us to conclude that {the system is stabilized in the GKP code subspace} by two cycles of the  switch protocol
\begin{equation}
\ws(t) \approx  \left\{
    \begin{array}{lrcl}
            1, &    0       &\leq t<        & \Ts\\ 
            0, &    \Ts     &\leq  t<       &  \Ts + \frac{1}{4}\TLC
    \end{array} \right.
    \quad 
            \Ts = \zs\,\trev ,
    \label{eq:f_time_dependence}
\end{equation}
with $\zs\in\mathbb Z$. From here on, we refer to the   $w_{\rm s}=1$ and $w_{\rm s}=0$ segments  as the {\it stabilizer} and {\it free} segments, respectively.

\addFN{We can also realize a $\nu$-dimensional GKP qudit with the scheme above by setting $\sqrt{{L}/{C}}=\nu h/4e^2$, and fixing the stabilizer segment duration to an integer multiple of $2\pi \hbar /\nu^2\varepsilon_L$. This    results in the free segment generating an exchange of the stabilizers of a $\nu$-dimensional square-lattice GKP qudit $S_1$ and $S_2^{(\nu)}\equiv \cos(\nu\pi q/e)$~\cite{Gottesman_2000}, and the stabilizer segment confining the system in the high-eigenvalue subspace of $S_1$ while  preserving the expectation value of $S_2^{(\nu)}$ via the phase revival mechanism described above.} 
\label{subsec:exp_lifetimes}

\begin{figure}
    \centering
    \includegraphics[width=0.99\columnwidth]{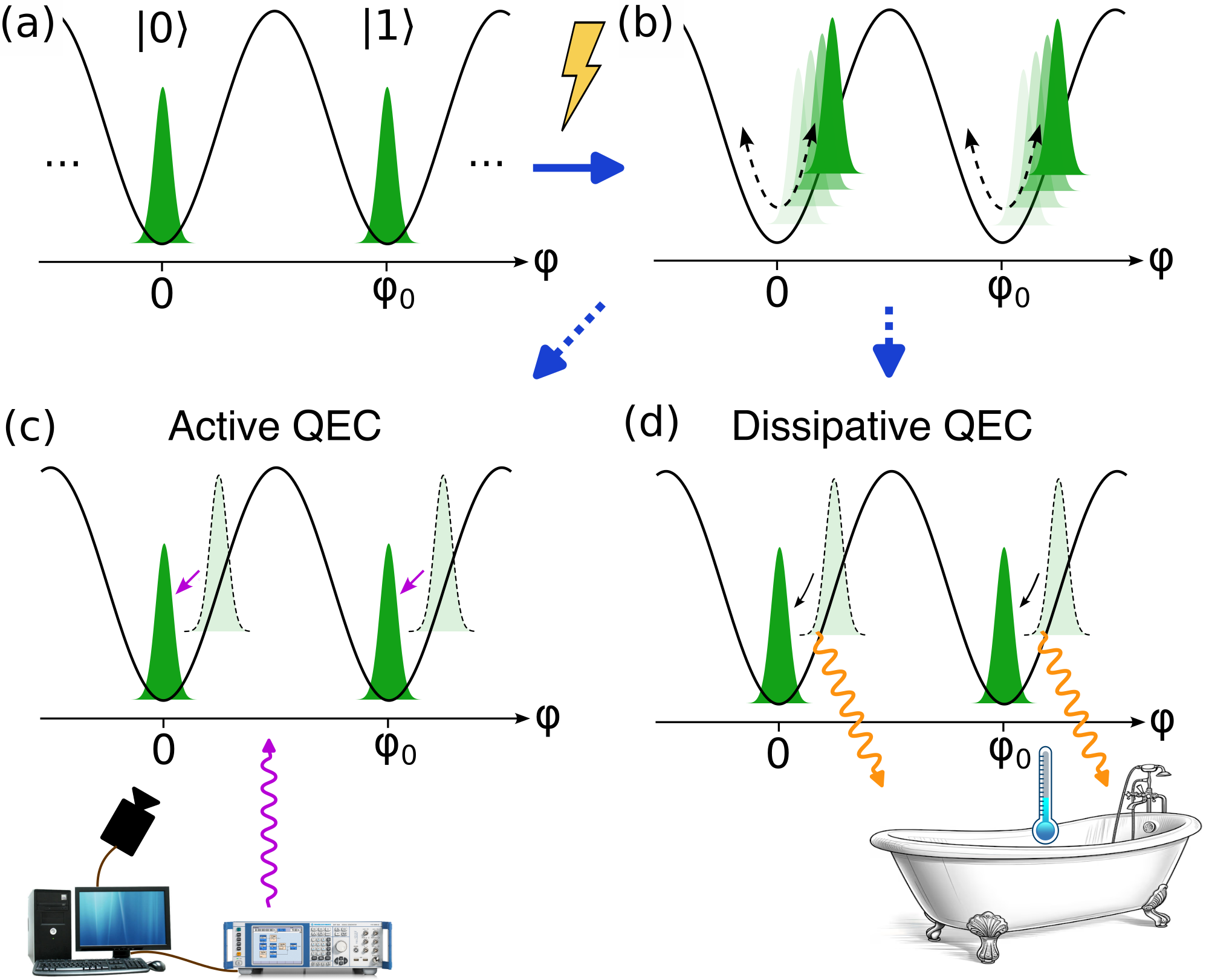}
    \caption{\addFN{\textbf{Illustration of dissipative quantum error correction}  
    (a) Quantum information  is encoded in the   well parity of the Josephson potential. (b) Noise excites fluctuations in the potential that  will  cause a bit flip via spill-over between  wells, if accumulating. (c) With {\it active quantum error correction},  a detector monitors for fluctuations and counter-steers against any with  appropriate control signals, thereby absorbing noise-induced entropy  analogously to Maxwell's demon. 
    (d) With {\it dissipative quantum error correction},    noise-induced entropy is absorbed by thermodynamic reservoirs, via dissipative damping of  fluctuations}. } 
    \label{fig:dec}
\end{figure}
\subsection{Dissipative error correction}
The switch cycle in Eq.~\eqref{eq:f_time_dependence} is a {\it dissipative quantum error correction (DEC) protocol}, in which noise-induced entropy is absorbed  by the resistor, leading to  protection the  encoded information (see Fig.~\ref{fig:dec}). 
Specifically, when repeated, the protocol  in Eq.~\eqref{eq:f_time_dependence}  resets  any state into the code subspace every two cycles.
Recall from Sec.~\ref{sec:Quantum_Info} that local noise only generates  logical errors if  causing the system's phase-space support to leak through the domain boundaries at $\varphi = (n_1+1/2)\varphi_0$ and $q=(n_2+1/2)e$  for $n_1,n_2\in \mathbb Z$ [see Eq.~\eqref{Pauli_ops}].
Logical errors can  thus only occur if such leakage occurs within two cycles when starting from a stabilized state. 
For noise  weaker than this threshold, \addFN{logical error rates are exponentially suppressed. We support this claim with analytic and numerical discussions below  [see Sec.~\ref{sec:Stabilization} and Sec.~\ref{sec:dec_numerics}].} 

The DEC-based operating principle of  the qubit  contrasts to conventional quantum error  correction, which utilizes a readout/feedback apparatus for removal of noise induced entorpy.
This offers several advantages: in particular, the qubit can be stabilized without any need for readout or feedback control. Additionally, the   dissipative error correction makes the qubit resilient to {imperfections of the protocol and device}, such as parameter mistargeting and   control noise, since these imperfections can be viewed on par with other extrinsic noise sources, and be   dissipatively corrected.
The qubit thus has finite tolerances for inconsitencies of device and driving parameters, which we  estimate in Sec.~\ref{sec:params} (see Tab.~\ref{tab:parameters}).

We finally remark that the DEC protocol  protects the qubit against local noise, but may  fail to protect against  nonlocal noise  (arising from infinite-order polynomials of $\varphi$ and $q$, which {act  nonlocally} in phase space). 
Particularly  relevant  are {quasiparticle poisoning, phase slips, and} uncontrolled cooper pair tunneling, which  translate $q$ or $\varphi$ by integer multiples of $e$ or $\varphi_0$, and hence act nonlocally in phase space. 
As a result, the timescales for uncontrolled cooper pair tunneling, phase slips, and quasiparticle poisoning may  provide  upper limits on the qubit lifetime.
Mitigating these noise sources is thus crucial to achieve significant lifetime enhancement.

\subsection{Self-correcting gates}
\label{sec:scgates}
\addFN{
Here we demonstrate that the qubit natively supports protected single-qubit Clifford gates. 
In partiuclar, we show how an $S$ and Hadamard ($H$) gate are naturally generated by the stabilizer and free segments. 
}

\addFN{
The $S$ gate is generated through the    phase  revival mechanism discussed  in Sec. \ref{sec:stabilization_overview}.
Specifically, at the $z$th phase revival,   states  in even wells have acquired a phase factor $(-i)^z$ relative to states in odd wells, since thir inductance energies  are congruent to $0$ and $\varepsilon_L$ modulo $4\varepsilon
_L$, respectively [see Eq.~\eqref{eq:moduloresult}]. 
This assignment of phase factors is equivalent to   $z$ applications of the $S$ gates, $S=e^{-i\frac{\pi}{4} \sigma_z}$.
A $H$ gate is generated by  the free segment, which interchanges $\varphi/\varphi_0$ and $q/e$ (up to a sign), which swaps $\sigma_x$ and $\sigma_z$, equivalent to the action of  a $H$ gate.
Arbitrary single-qubit Clifford control can be achieved via appropriate interspersing of stabilizer and free segment, e.g., by 
 varying $\zs$, and not necessarily ordering stabilizer and free segments in  an alternating pattern. 
 }

\addFN{
The $S$ and $H$ gates are self-correcting and topologically robust~\cite{conrad2024latticesgatescurvesgkp} in the sense that the output logical state is invariant under any smooth and phase-space local perturbation of the protocol that keeps the  final state within the GKP code subspace.
{As a result, the infidelity from control noise is {\it exponentially suppressed} when the noise strength  below a characteristic  threshold scale determined by the circuit parameters. Specifically, the error rate from control noise  decreases exponentially with the ratio of the  threshold scale relative to the noise strength [see Sec.~\ref{sec:charge_noise}, and, in particular, Eq.~\eqref{eq:error_vs_mistiming}].}
In Sec.~\ref{sec:gate_numerics}, we verify the exponential protection of the gates in simulations. 
}

\addFN{
A consequence of the results above is that odd  $\zs$ causes the  switch protocol to cyclically permute $\sigma_x$, $\sigma_y$, and $\sigma_z$---possibly with an alternating sign.
In this sense, our device can be viewed as a dissipative phase-locked oscillator, or Floquet time crystal~\cite{Holthaus_1994,Sacha_2015,Khemani_2016,Else_2016b,Zhang_2017,Gong_2018,Loerch_2019,Nathan_2020a}, whose  emergent periodicity is controlled by $\zs \mod 4$. 
Choosing  $\zs$ odd  causes  the $3$ logical operators of the qubit to appear on equal footing, implying that  phase and amplitude errors are treated symmetrically, and $T_1 = T_2$.
{The symmetry above moreover means that any noise channel on the  qubit is depolarizing.}}

\section{Exponential scaling of lifetime}
\label{sec:Stabilization}
\label{Stabilization}

\begin{figure}
    \centering
    \includegraphics[width=0.99\columnwidth]{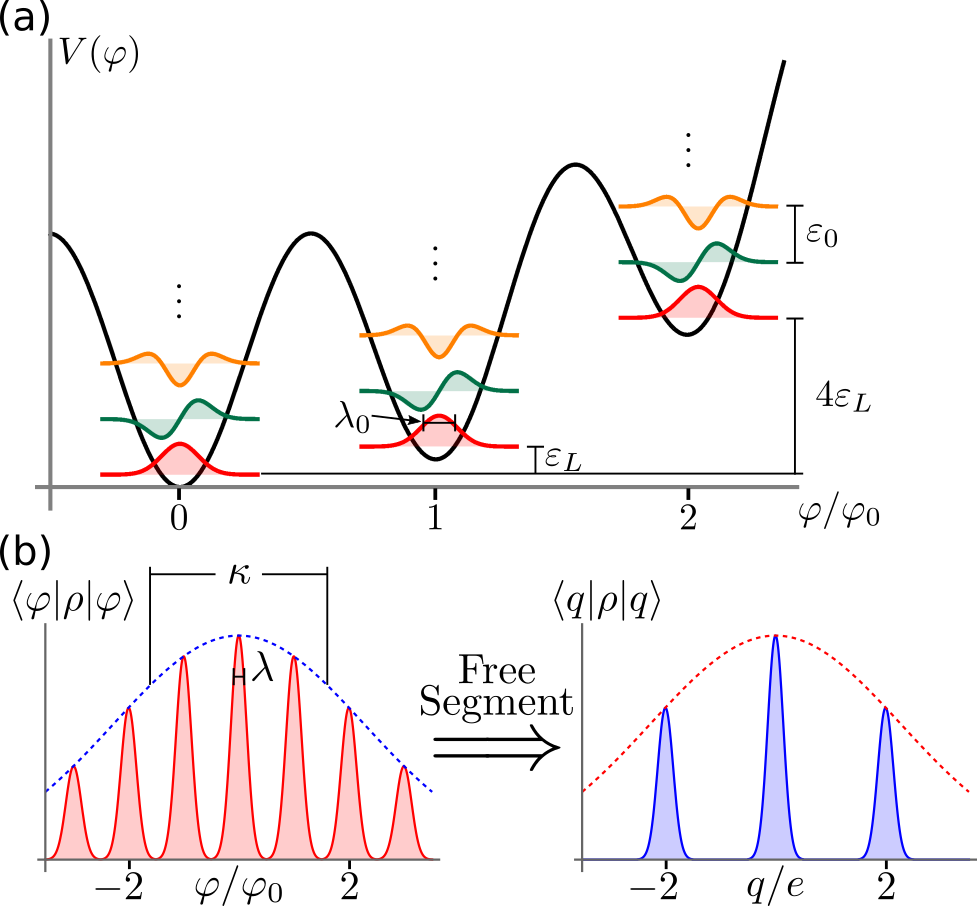}
\caption{\textbf{Characteristic scales of the qubit} 
(a) Approximate low-energy eigenstates of the circuit Hamiltonian during the stabilizer segment [i.e.,  $w_s(t) = 1$]. 
To  leading order in $h\flc/E_J$, the eigenstates are  harmonic oscillator eigenstates with vacuum fluctuation width $\lambda_0 \varphi_0$, with $\lambda_0 = (h\flc/4\pi^3E_J)^{1/4}$, centered near $m\varphi_0$ for each $m\in \mathbb Z$. 
The corresponding   excitation energy is givne by  $\varepsilon_0 = \sqrt{4e^2E_J/C}$ 
with an overall energy shift $m^2\varepsilon
_L$ in well $m$, where $\varepsilon_L = \varphi_0^2/2L$. 
(b) Left: probability density of $\varphi$ for a typical GKP state generated by our protocol (red), corresponding to a logical state $\frac{1}{\sqrt{2}}(|0\rangle +|1\rangle)$. The state is a thermal ensemble of the low-lying eigenstates of each well in panel (a), with thermal fluctuation width $\lambda = \lambda_0\sqrt{\coth(2\epsilon_0/k_{\rm B}T)}$, weighted by an envelope function (dashed line). Right: Result of evolving the state under the free segment.
This mapping fixes the width of the envelope to be $\kappa = \sqrt{\coth(2\varepsilon_0/k_{\rm B}T)}/\pi \lambda_0$ (see Appendix~\ref{app:envelope}).} 
\label{fig:wavefunctions}
\end{figure}

\addFN{We now demonstrate the expontial scaling of qubit lifetime in regimes where $E_J \gg k_{\rm B}T, h\flc$, where $T$ denotes the temperature of the resistor and $\flc = 1/2\pi\sqrt{LC}$ the bare LC frequency.
Below, we identify sufficient conditions for the emergence of a stable fixed point in the GKP code subspace, where the  relaxation during the stabilizer segment only causes exponentially weak decoherence between the wells of the cosine  potential from the Josephson junction.  
In  this regime, the logical error rate per protocol cycle, $p_{\rm error}$ is  exponentially suppressed, and satisfies
\be 
p_{\rm error} \lesssim \exp\left[\frac{-1}{k_{\rm B}T/E_J+\pi^2 \lambda_0^2}\right],
\label{eq:error_rate_result}
\quad \lambda_0 \equiv   \left(\frac{ h\flc  }{4\pi^3 E_J}\right)^{1/4}. 
\ee  
Here $\lambda_0$ denotes the vacuum fluctuation width in the cosine potential in units of $\varphi_0$ [see Fig. \ref{fig:wavefunctions}(a)], and defines a bare GKP squeezing parameter of the protocol.
Here and below $a \lesssim b$ indicates that $a$ is bounded by $b$ up to a subleading prefactor; in particular, $a \lesssim b$  allows $a$ to be much smaller than $b$. 
We thus emphasize that the bound above is not necessarily tight, and that the actual error rate may be exponentially smaller than the right-hand side above. }
\addFN{
When $ E_J/k_{\rm B}T \ll \lambda_0$, the first term in the numerator above dominates, and the error rate is of order $e^{- E_J/k_{\rm B}T}$; i.e.,  the error rate bound scales as an Arrhenius law. The $ E_J/k_{\rm B}T$ and $\lambda_0^2$ terms can thus be viewed as the contribution of the error rate from thermal and quantum fluctuations, respectively. 
}
\addFN{Importantly, we expect the exponentially scaling logical error rate to persist in the presence of device/protocol imperfections and external noise; in Sec.~\ref{sec:params} we estimate noise tolerances based on these considerations.}

\addFN{The remainder of this section is devoted to establishing Eq.~\eqref{eq:error_rate_result}, and identifying sufficient (not necessarily required) conditions under which it holds. Technical details are provided in Appendices~\ref{seca:heff_derivation}-\ref{app:error_rate_bound}. The discussion  is of technical nature, and readers can move directly to the more physically-oriented sections \ref{sec:T_gate}-\ref{sec:Discussion} without loss of continuity. }

\addFN{\subsection{Fixed point in code subspace } 
\label{sec:fixed_point}
We first demonstrate the protocol has a stable fixed point in the GKP code subspace.
To this end, we consider the case where the system is initialized in the subspace of mutual eigenstates  of $S_1$  and $S_2$ with  eigenvalues greater than some  $s_0>-1$, and $\varphi$ support confined in the range $|\varphi|\ll \varphi_0 E_J/\varepsilon_L$, where the wells of the Josephson  potential remain well-defined.
Below, we identify {\it sufficient} criteria under which the protocol takes any such initial state to a fixed point where the $S_1$ and $S_2$ support of the system remains exponentially confined near $1$ [Eq.~\eqref{eq:fp_conditions}]. 
}

\addFN{
\subsubsection{Relaxation in stabilizer segment}
We first consider how the state of the system  $\rho(t)$ evolves in the first  stabilizer segment.
During this step, the   circuit including the resistor is described by 
\be 
H_{\rm s}\equiv \frac{\varphi^2}{2L}+\frac{q^2}{2C}-E_J \cos\left(\frac{2\pi\varphi}{\varphi_0}\right) + \frac{qQ_{\rm R}}{C_{\rm R}}  + H_{\rm R}. \label{eq:hs_def}
\ee 
The wells of the cosine potential from the Josephson junction  are approximately described as harmonic LC oscillators with 
with capacitance $C $ and effective inductance $L_{\rm eff} = \varphi_0^2 /4\pi^2E_J$---see Fig. \ref{fig:wavefunctions}(a). 
The characteristic excitation energy in the  wells is given by $\varepsilon_0 \equiv  \hbar/\sqrt{CL_{\rm eff}}=\sqrt{4\pi h\flc  E_J}$, while the vacuum fluctuation width of $\varphi$ in the wells, $\delta \varphi_0 \equiv  [\hbar^2 L_{\rm eff}/4C ]^{1/4}$ is given by $\lambda_0 \varphi_0$, with $\lambda_0$ defined in Eq.~\eqref{eq:error_rate_result}.
}

\addFN{
During the stabilizer segment, the dissipation from the resistor causes the  $\varphi$ support of $\rho(t)$ to relax into the wells of the cosine potential from the Josephson junction.
The characteristic rate for this process, $ \Gamma$,  can be found from the low-temperature power spectral density of the resistor $J_0(\omega)\equiv \frac{1}{2\pi}\int dt e^{i\omega t}\langle Q_{\rm R}(t)Q_{\rm R}(0)\rangle$ via Fermi's golden rule~\cite{Breuer,GardinerZoller,Nathan_2020b}: 
$\Gamma = (\hbar C_{\rm R})^{-2}2\pi  J_0([E_{n'}-E_n]/\hbar)|\langle \psi_{n'}|q|\psi_n\rangle|^2$ in the low temperature limit; here $|\psi_n\rangle$ a given ground state of the nondissipative part of $H_{\rm s}$ confined inside the wells of the Josephson potential, and $|\psi_n'\rangle$ the eigenstate corresponding to its first excited state [see Fig.~\ref{fig:wavefunctions}(a)], while $E_n$ and $E_{n'}$ denote the corresponding energies. Noting  that $E_n-E_n'\approx \varepsilon_0$, and that $|\psi_n\rangle$ and $|\psi_n'\rangle$  are adjacent Hermite functions with characteristic width $\lamp \varphi_0$,  such that $q|\psi_n\rangle = \frac{\hbar}{\sqrt{2}\lamp\varphi_0}|\psi_{n'}\rangle$, and  using $C=e/2\pi \flc\varphi_0 $, the relaxation rate during the stabilizer segment is thus given by 
\be 
\Gamma =  \frac{ e^2}{C_{\rm R}^2 \pi \hbar^2 }\frac{J(\varepsilon_0/\hbar)} {\lambda_0^2}.
\label{eq:gamma}
\ee}

\addFN{
Since the system is already confined inside the wells of the cosine potential ($S_1>s_0$) at the onset of the stabilizer segment, leakage of probability support between the wells is exponentially supressed during the relaxation~\cite{Ankerhold_1995}.  
As a result, the flux probability distribution in  well $m$---defined as the  interval  $(m-1/2)\varphi_0<\varphi < (m+1/2)\varphi_0$---relaxes to a state approximately identical to the thermal steady-state of an LC resonator  with excitation energy $\varepsilon_0$ and vacuum fluctuation width $\delta \varphi_0$, up to a preactor accounting for the near-conserved total  probability support within each well. The thermal flux distribution of an LC resonator is given by  a Gaussian with characteristic width $\delta \varphi = \lth\varphi_0/\sqrt{2}$, where 
\be
\lth = \lambda _0\sqrt{\coth(2 \varepsilon_0/k_{\rm B}T)},
\ee 
This dimensionless number 
can be viewed as a {\it thermally renormalized} GKP squeezing parameter serves
as the small parameter in our analysis below. 
}

\addFN{The above results imply that, after the characteristic relaxation time $1/\Gamma$, the  flux distribution of the system, $p_{\varphi}(x,t)\equiv \Tr[\rho(t)\delta(\varphi-x)]$, is given by
\be 
p_\varphi(x,t) \approx \sum_m \frac{p_m(t)}{\sqrt{2\pi}\lth\varphi_0} e^{-\frac{(x-m\varphi_0)^2}{\lth^2\varphi_0^2}}, \label{eq:flux_distribution_stabseg}
\ee 
where $p_m(t) = \int_{[m-1/2]\varphi_0}^{[m+1/2]\varphi_0}dx p_{\varphi}(x,t) $ denotes the near-conserved steady-state  probability support  in well $m$---see left panel of Fig. \ref{fig:wavefunctions}(b) for an illustration.
Specifically, the leakage of probability support between the wells of the cosine potential is of order $(\varepsilon_0/h) e^{-2 E_J/k_{\rm B}T}$~\cite{Ankerhold_1995}, implying that  $p_m(t)$ are constants of motion in the steady-state, up to $\mathcal O(e^{-2 E_J/k_{\rm B}T})$ corrections. }



\subsubsection{Preservation of inter-well coherence}
\addFN{
We now  demonstrate that  phase-coherence is maintained between different wells of the cosine potential during the stabilizer segment.
To  account for the deterministic phase revival mechanism described in Sec.~\ref{sec:stabilization_overview}, we first transform to a comoving frame generated  by the unitary transformation
\be 
V(t) \equiv e^{-i\, \nint\left({\varphi}/{\varphi_0}\right)^2 \varepsilon_L t/\hbar},
\label{eq:cmftransf}\ee 
where $\nint(x)$ gives the integer closest to $x$. 
Evidently, 
$V^\dagger(t)$ assigns a phase factor $e^{i\, m^2 {\varepsilon_L t}/{\hbar}}$ to states in well $m$, thus canceling the phase factor from the offset of inductance energies described in Sec.~\ref{sec:stabilization_overview}.
Reflecting the revival mechanism described there, the  system's density matrix in the rotating frame, $ V^\dagger(t) \rho(t)V(t)$, coincides with that in the lab frame, $\rho(t)$,  at multiples of the revival time $t_{\rm rev}\equiv 2\pi \hbar /4\varepsilon_L$, up to applications  of a number of  $S$ gates. 
To see this, note that, for $n\in \mathbb Z$, $V(nt_{\rm rev}) = e^{-i\frac{\pi}{2}n  \nint(\varphi/\varphi_0)^2 }$. 
Using Eq.~\eqref{eq:moduloresult} and $\nint^2(x) \mod 4 = \frac{1}{2}(1-\Xi(x))$, we find  $V(nt_{\rm rev}) = e^{-i \frac{\pi}{4}n [1- \Xi(\varphi/\varphi_0)]}$. Recognizing $\Xi(\varphi/\varphi_0)$ as $\sigma_z$ [Eq.~\eqref{Pauli_ops}], we thus obtain 
\be 
V(nt_{\rm rev})=e^{\frac{-i\pi}{4}n(1-\sigma_z)} \quad {\rm for}\quad n \in \mathbb Z .
\label{eq:sgate}\ee 
Since $e^{-i\frac{\pi}{4}\sigma_z}$ defines the  logical $S$ gate operator, we see that  $V(t_{\rm rev})$ generates an $S^\dagger$ gate and a  global phase. 
Since  $[S_2,\sigma_z]=0$, the  probability distributions for $S_2$ in the lab  [$\rho(t)$] and comoving [$V^\dagger (t)\rho(t)V(t)$] frames coincide at integer multiples of the revival time $\trev$.}

\newcommand{\Hp}{{ H}_{\rm rf}}
\newcommand{\Hpz}{{\tilde H}}

\addFN{
We  now show that $ V^\dagger(t)\rho(t)V(t)$ remains in the high-eigenvalue subspace of $S_2$ during the entire stabilizer segment. 
To this end, we first consider the Hamiltonian in the comoving frame 
$\Hp(t) = V^\dagger(t)[H_{\rm s}+i\hbar \partial_t ]V(t)$~\footnote{Specifically, $\Hp$ is defined such that 
$\partial_t V^\dagger \rho V = -i[  \Hp,V^\dagger \rho V]$.}. 
We now introduce the {\it quasiflux} operator, 
\be \bar \varphi\equiv  \varphi  \ \mod \varphi_0.
\ee 
with    branch cut chosen at $ \pm \varphi_0/2$ such that $\varphi =\bar \varphi+ \varphi_0\nint(\varphi/\varphi_0)$. 
Using $\varphi =\bar  \varphi +\varphi_0 \nint(\varphi/\varphi_0)$, a straightforward calculation yields
$\Hp(t) = \Hpz +\Delta \Hp(t)$
with 
\begin{align}
\Hpz &=\Hpz_0 + \nint\Big(\frac{\varphi}{\varphi_0}\Big) \frac{\varphi_0\bar \varphi }{L},
\label{eq:heff}
\end{align}
where  $ 
\Delta \Hp =-i V^\dagger \left[ \frac{q^2}{2C} -\frac{qQ_{\rm B}}{C_{\rm R}},V\right]$ and 
\begin{align}
\Hpz_0&=\frac{\bar \varphi^2}{2L} - \frac{q^2}{2C}-E_J \cos\left(\frac{2\pi \bar \varphi}{\varphi_0}\right) +\frac{q Q_{\rm R}}{C_{\rm R}}  + H_{\rm R},
\label{eq:hpz0def}\end{align}
Importantly, 
$[\bar \varphi,S_2]=0$~\cite{S2comm}, implying $[\Hpz_0,S_2]=0$.}

\addFN{
We now show that $\Delta \Hpz$ can be neglected: recall $V(t)$ assigns a piecewise-constant phase factor to eigenstates of $\varphi$. 
Since $q=-i\hbar \partial_\varphi$,  the $\varphi$-support of $\Delta H$ is  thus confined to  infinitesimally small neighbourhoods surrounding  $ \varphi =(z+1/2)\varphi_0$ for each $z\in \mathbb Z$.
Since the support of the system in this region is of order $e^{-2 E_J/k_{\rm B} T}$~\cite{Ankerhold_1995}, $\Delta \Hp$ can thus be neglected at the cost of an exponentially suppressed correction. 
Indeed, in Appendix~\ref{seca:heff_derivation}, we show  that
the time-evolution generated by $\Hpz$, $\rhot(t)\equiv e^{-i\Hpz t}\rho(0)e^{i\Hpz t}$ remains exponentially close to $V^\dagger\rho V$ at all times, $\trnorm{V^\dagger(t)\rho(t)V(t)-\rhot(t)}\lesssim \mathcal O(e^{- E_J/k_{\rm B }T}),$
with $\trnorm{\cdot}$ denoting the trace norm. 
}

\addFN{
Having neglected $\Delta H(t)$, we consider the  evolution of  $S_2$ in the comoving frame, 
$
\langle {\tilde S}_2\rangle \equiv  \Tr[\tilde \rho S_2]$. 
Recalling that  $[S_2,\bar H]= 0 $ and $\partial_t \tilde \rho = -\frac{i}{\hbar}[\tilde H,\tilde \rho]$, explicit calculation shows 
$
\partial_t\langle {\tilde S}_2\rangle  = - \Tr\left[ \frac{2 \bar \varphi\varphi_0}{   \hbar L} \sin(2\pi q/e)\tilde \rho\right] .
$
Using that $|\Tr[\rho A]|\leq \sqrt{\Tr[A^\dagger A \rho]}$ for any density matrix $\rho$, and  that $\sin^2(x)\leq 1$, this implies 
\be 
|\partial_t \langle\tilde S_2\rangle| \leq \frac{2\pi }{\varphi_0\trev}\sqrt{\langle \bar \varphi^2\rangle}. 
\ee 
where we used   $\trev = \pi \hbar L/\varphi_0^2$.
From this result, we conclude two things: first, since $\langle \bar \varphi^2\rangle\leq \varphi_0^2/4$,  $|\partial_t \langle\tilde S_2\rangle| \leq \pi/\trev$ at all times. 
Secondly,  after the relaxation the the quasi-thermal steady-state in Eq.~\eqref{eq:flux_distribution_stabseg}, where $\langle \bar \varphi^2\rangle= \lambda\varphi_0$,  $|\partial_t \langle\tilde S_2\rangle| \leq 2\pi \lambda /\trev$. 
Thus $\langle \tilde S_2\rangle$ should remain near-constant on the revival timescale $\trev$, provided $\lambda \ll 1$.}

\addFN{
In Appendix~\ref{app:speed_limit}, we extend the result above to demonstrate that not only the expectation value, but the   entire {\it probability support} of $S_2$ can remain near-stationary  in the comoving frame over a revival time when $\lambda \ll 1$. 
Specifically,  for any $k>0$, the  cumulative probability support for $S_2$, 
${\cP}_2(s,t) \equiv \Tr[\theta(s-S_2)\rhot(t)]$ satisfies 
\be 
{\cP}_2(s,t) \lesssim e^{-k[s_0-s-\Delta s_k(t)]}, 
\label{eq:quasicarge_confinement}
\ee 
where   $\Delta s_k(t)\equiv  - \int_0^t  d\tau  v_k(\tau )$, and 
\be 
v_k(t)\equiv \frac{\Tr[\rhot_k(t)\bar \varphi \sin\!\frac{2\pi q}{e}]}{\varphi_0 t_{\rm rev}/2\pi} ,\quad\!\!  \rhot_k\equiv  \frac{e^{\frac{-kS_2}{2}} \rhot(t)e^{\frac{-kS_2}{2}}}{\Tr\big[e^{{-kS_2}} \rhot(t)\big] }
\label{eq:velocity_def}
\ee
For each $k$,  $v_k(t)$ defines a maximal  velocity for the spread of $S_2$ support, beyond which it must decrease exponentially at rate $k$.
Hence, it defines a speed limit for inter-well dephasing, akin to the Lieb-Robinson velocity.
In the regime $k \gg 1$,  the $S_2$ support of $ \rhot(t)$ must be  confined above $s_2 - \Delta s_k (t)$. In this sence inter-well coherence is maintained as long as 
$ 
\Delta s_k(t) < 1 -s_0
$
for large $k$. 
}

\addFN{
We now  show that $\Delta s_k(t)$ remains much smaller than $1$ during the entire stabilizer segment when  
\be 
\lambda \ll 1\quad{\rm  and}\quad  \Gamma \trev \ll 1.
\label{eq:fp_conditions}
\ee 
To this end, we consider the $\bar \varphi$-support range  of $\rho(t)$,  $\Delta \varphi(t)$~\footnote{This may, e.g., be defined such that  $ \rho(t)$ has its $\bar \varphi$ support confined in the interval $|\bar \varphi|\leq \Delta \varphi(t)$, up to a $[-\Delta \varphi(t),\Delta \varphi(t)]$ up to  a 
residual weight of $\erf(c)$, for some $c$ we fix at a $\mathcal O(1)$ value sufficiently large that the residual can be neglected. E.g. setting $c = 6$ leads to an error of order $\erfc(6) \sim 10^{-17}$. }.
The evolution is unaffected  if,  at time $t$, we restrict the  system to the subspace with $|\bar \varphi|\leq \Delta \varphi(t)$~\cite{truncationscheme}.
Next, we use that $|\Tr[X \rho]|\leq \norm{X}$ for any density matrix $\rho$, with $\norm{X}$ the singular value norm of $X$. 
Using that the truncation above leads to $\norm{\bar \varphi \sin(2\pi q/e)}\leq \Delta \varphi(t)$ at time $t$, and noting that $\rhot_k(t)$ is   a density matrix,
 we find from Eq.~\eqref{eq:velocity_def} that 
\be 
v_k(t) \leq  \frac{2 \pi}{t_{\rm rev}}\frac{\Delta \varphi(t)}{\varphi_0}.
\ee 
We now  note that   $\Delta  \varphi(t)$ is trivially bounded by $ \varphi_0/2$. Moreover, during the stabilizer segment, $\Delta \varphi(t)$ should relax to   $\sim  2\pi \lth \varphi_0$ on the characteristic relaxation timescale $1/\Gamma$~\footnote{For instance, with the trunctation scheme in footnote~\cite{truncationscheme}, it would relax to $c\lambda \varphi_0$,  with $c$ the $\mathcal O(1)$ cutoff scale introduced there.}.
Thus,  $v_k(t)$ is bounded by a number that  relaxes from at most $\pi/\trev$ to $2\pi \lambda/\trev$ on the timescale $1/\Gamma$, implying 
$
\Delta s_k(t) \lesssim \frac{\pi}{\Gamma \trev}+2\pi  \lth  \frac{ t}{\trev} .
\label{eq:deltasresult}
$ 
Evidently, $\Delta s_k(t)$ remains much smaller than $1$ on the timescale $\trev$  if the conditions in Eq.~\eqref{eq:fp_conditions} are satifsied. 
We again emphasize that these are sufficient, but not necessarily required conditions for preservation of phase-coherence during the stabilizer segment.
In particular, phase-coherence may remain preserved even when $\Gamma \trev \lesssim 1 $.
}


\addFN{
\subsubsection{Fixed point}
We now demonstrate that a stable fixed point emerges in the code subspace under the conditions in Eq.~\eqref{eq:fp_conditions}.}

\addFN{
We first consider the evolution of the cumulative probability distribution for $S_1$, $ {\cP}_1(s,t)\equiv \Tr[\theta(s-S_1)\tilde \rho(t)]$; note that $V(t)$ commutes with $S_1$, implying the distributions for $S_1$ are identical in the lab ($\rho$) and comoving ($\tilde \rho$) frames. 
Due to the Gaussian confinement of $\varphi$ near integer multiples of $\varphi_0$ [Eq.~\eqref{eq:flux_distribution_stabseg}], 
the cumulative probability distribution for $S_1$, $ {\cP}_1(s,t)\equiv \Tr[\theta(s-S_1)\tilde \rho(t)]$ is given by $\approx \erfc(\arccos(s)/2\pi \lambda)$ at the end of each stabilizer segment.
Using $\erfc(x)\sim e^{-x^2}$ and $\arccos(s) \sim \sqrt{2-2s}$, we can rewrite this to  
\be 
{\cP}_1(s,t)\lesssim  \exp\left[\frac{s-1}{2\pi^2 \lambda^2 }\right].
\label{eq:p1}
\ee 
This result holds  for $ t\gg 1/\Gamma$. }

\addFN{
We next  note that   the free segment interchanges $S_2$ and $S_1$. 
Hence, at the onset of the second and all subsequent stabilizer segments, $\cP_2(s,t)$ equals the right-hand side of Eq.~\eqref{eq:p1}, 
implying that the $S_2$ support of the system is initially exponentially confined near $1$ in the regime $\lambda \ll 1$.
Moreover, under the conditions in Eq.~\eqref{eq:fp_conditions},  the probability support for $S_2$ is effectively constant during  stabilizer segment.
As a result, we find 
\be 
 {\cP}_2(s,t)\lesssim \exp\left[\frac{s-1}{2\pi^2 \lambda^2 }\right]
\label{eq:s2fixedpoint}
\ee 
in  the second, and all subsequent, stabilizer segments; this result is obtained  systmatically in Appendix~\ref{app:speed_limit}. }

\addFN{
We finally recall again that the free segment interchanges the probability distributions for $S_1$ and $S_2$,  we conclude that at the onset of the  the third, and all subsequent stabilizer segments, $ {\cP}_1(s,t)\lesssim \exp\left[\frac{s-1}{2\pi^2 \lambda^2 }\right]$. 
Noting that this is consistent with the steady-state flux distribution in the stabilizer segment [Eq.~\eqref{eq:flux_distribution_stabseg}], we anticipate little change of this distribution during the stabilizer segment, and hence  we expect that Eq.~\eqref{eq:p1} remains satisfied throughout the this, and all subsequent stabilizer segments. 
As a result, we conclude that two cycles of the the driving protocol takes the system to a fixed  point satisfying Eqs.~\eqref{eq:p1}-\eqref{eq:s2fixedpoint}. }

\addFN{
\subsection{Exponential scaling of lifetime}
We finally  demonstrate the exponential suppression  of the logical error rate quoted in Eq.~\eqref{eq:error_rate_result}. }

\addFN{
First note use  that the  density matrix  in the comoving frame, $\tilde \rho$,  remains exponentially confined in the  code subspace during 
the stabilizer segment [Eqs.~\eqref{eq:p1},~\eqref{eq:s2fixedpoint}]. 
From the considerations in Sec.~\ref{sec:Quantum_Info}, this should imply that the expectation values of the logical operators in the state $ \rhot(t)$ remain exponentially invariant during the stabilizer segment. 
In Appendix~\ref{app:error_rate_bound},  we confirm this intuition via explicit computation, showing that, for $i=1,2,3$, and throughout the stabilizer segment
\be 
|\Tr[\sigma_i \rhot(t)]-\Tr[\sigma_i \rhot(0)]| \lesssim \exp\left[
\frac{-1}{k_{\rm B}T/E_J+\pi^2\lambda_0^2}\right].
\label{eq:logical_er_result}\ee 
Here we suppressed power-law prefactors on the right hand side; see Appendix~\ref{app:error_rate_bound} for expressions involving these. 
We now recall  and that $V^\dagger(t)\rho(t)V(t)$ differs from $\rhot(t)$ by  a $\mathcal O(e^{- E_J/k_{\rm B}T})$ correction, and that $V(\Ts)$ is identicial to the $\zs$th power of the logical $S$ gate operator. 
Hence, the expectation values of the logical operators at the end of the stabilizer segment ($t=\Ts$) relate to those at the onset ($t=0$) through  
\be 
\langle{\boldsymbol \sigma}(\Ts)\rangle =  {{\mathcal S}}^{\zs}\langle {\boldsymbol \sigma}(0)\rangle+ \delta{\boldsymbol\sigma}_{\rm error}, 
\ee 
where ${{\mathcal S}}$ denotes the $3\times 3$ Bloch sphere rotation matrix corresponding to an $S$ gate~\footnote{Specifically,  ${{\mathcal S}}_{ij}= \delta_{iz}\delta_{jz}+\epsilon_{ijz}$, with  $\epsilon_{ijk}$ the Levi-Civita symbol}, and 
\be 
|\delta{\bs  \sigma}_{\rm error}| \lesssim \exp\left[-\frac{1}{k_{\rm B} T/E_J+4\pi \lambda_0^2}\right].
\ee 
Here, we suppressed prefactors on the right hand side  which scale as a power law with $\lambda_0^{-1}$ and $\frac{E_J}{k_{\rm B}T}$, and thus are subdominant in the limit $\lambda_0^{-1},\frac{E_J}{k_{\rm B}T} \gg 1$.
Identifying $p_{\rm error} =|\delta{\bs  \sigma}|$, leads to Eq.~\eqref{eq:error_rate_result}.}


\section{Protected $T$ gate with alternative encoding}
 \label{sec:T_gate}

Here we show how a protected $T$ gate can be realized 
using an alternative configuration of the  qubit, based on an alternative encoding scheme, and a resonator with impedance $\sqrt{L/C}=h/e^2$~\cite{magic_states}.
The  $T$ gate can be realized  using the quasi-modular  logical operators $\bar{\bs \sigma} = (\bar\sigma_x,\bar\sigma_y,\bar\sigma_z)$, where 
\be 
\bar\sigma_z = \Xi(\varphi/\varphi_0),\quad\bar \sigma_x =e^{-i\frac{q}{2e}}  I I_{\rm m}, \quad \bar\sigma_y = -i\bar\sigma_z\bar\sigma_x.
\label{eq:nu4_paulis}
\ee 
Here $\Xi(x)$ denotes the crenellation function,  while 
$I$ and 
$I_{\rm m}$ denote the phase space and {\it modular} inversion operators, respectively: letting $|\phi\rangle$ denote the $\varphi$-eigenstate with eigenvalue $\phi$, these  are defined by $I|\varphi\rangle =|-\varphi\rangle$ and $I_{\rm m}|z\varphi_0 +\delta \varphi\rangle=I_{\rm m}|z\varphi_0-\delta \varphi\rangle$ for $|\delta \varphi|\leq \varphi_0/2$ and $z\in \mathbb Z$. 
The logical operators above    satisfy  $\{\bar \sigma_i,\bar \sigma_j\} = \delta_{ij}$, and hence form a valid qubit observable
\footnote{To see this,  note that, for $z\in \mathbb Z$ and $|\delta \varphi|\leq \varphi_0/2$, $\bar\sigma_x |z\varphi_0+\delta \varphi\rangle = |-(z-1)\varphi_0+\delta \varphi\rangle$ and $\bar \sigma_z|z\varphi_0+\delta \varphi \rangle =(-1)^z|z\varphi_0+\delta \varphi \rangle$. 
Noting that the states $|z\varphi_0+\delta \varphi \rangle$ form a complete orthormal basis, the relations above imply $\bar \sigma_x^2 =\bar \sigma_z^2 = 1$ and $\bar \sigma_x \bar \sigma_z = - \bar \sigma_z\bar \sigma_x$, implying that the anticommutator relations hold.}.

With the above encoding, the code subspace is spanned by  $4$ families of  states, with support near $\varphi \mod 4\varphi_0 = \zeta\varphi_0 $ for $\zeta=0,1,2$, or $3$, respectively. 
The computational spaces  split up into two sectors:  logical operators do not couple states with $\zeta \in \{0,1\}$ to states with $\zeta\in \{2,3\}$. 
We use the first sector as  the computational space for the system, with $\zeta =0,1$ resulting in  eigenvalues $1$ and $-1$ of $\bar \sigma_z$. 
States with $\zeta=2,3$ can be considered  non-computational.
The logical operators $\{\bar \sigma_i\}$ have stabilizers  $S_1$ and $\bar S_2=\cos(4\pi q/e)$. 
The quasi-modular encoding above can  be thus dissipatively stabilized  by the device and protocol in Sec.~\ref{sec:stabilization_overview}  by setting $\sqrt{L/C}=h/e^2$.
The revival of $\bar S_2$ in the stabilizer  segment is again ensured by picking the stabilizer segment duration to be an integer multiple of $\TLC/2\pi$: to see this, note that the  inductance energy   is given by $
\varepsilon_L = \pi h\flc/4$, implying that  states in  well  $m$ acquire a relative phase factor    $e^{-i \zs \pi m^2/4}$ after the stabilizer segment. 
Since $(n+4)^2 - n^2 \in 8\mathbb Z$, revival of $\bar S_2$ is ensured for each integer  choice of $\zs$.

The $T$ gate emerges because   the $|\zeta \rangle$ logical state (for $\zeta \in\{0,1\}$) has support in wells where $m\mod 4 =\zeta $.
Since   $m \mod 4 =\zeta  $ implies $m^2 \mod 8 = \zeta$ for $\zeta\in\{0,1\}$, the $|\zeta \rangle$ logical state of the qubit   acquires a  phase factor $e^{-i\pi \zeta /4}$ during the stabilizer segment. 
A stabilizer segment with $\zs \in 8\mathbb Z+1$  hence generates a $T$ gate \footnote{This $T$ gate is the hermitian conjugate of the usual definition of the $T$ gate, which would have $-1$-eigenstates of $\sigma_z$ acquire a phase of $\sqrt{i}$.
The gate is exponentially protected by the same mechanism that the $S$ gate is for the configuration discussed in the previous sections}.

%
Unlike the modular encoding in Eq.~\eqref{Pauli_ops},  the quasi-modular encoding above does not appear to support a native, protected Hadamard gate, and hence also not universal gates.  
The $T$ protocol could be used as a high-fidelity magic factory 
that generates  magic states on the physical qubit level  with exponentially-suppressed infidelity. 


\section{readout and initialization} 
\label{sec:Readout}

\begin{figure}
\includegraphics[width=0.99\columnwidth]{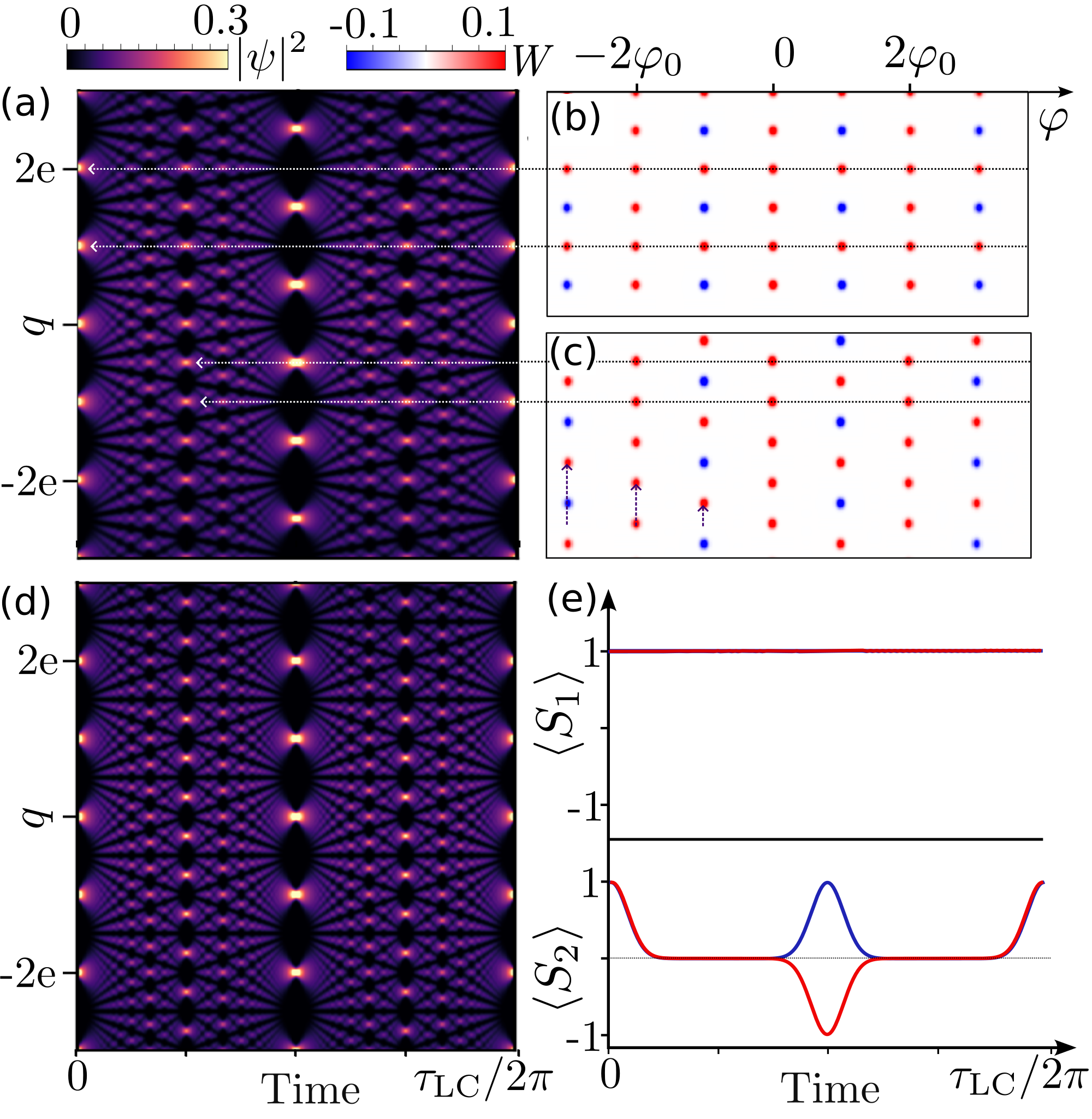}
    \caption{\textbf{Evolution of logical states during stabilizer segment}. Data are obtained with parameters $E_J/h=200\,{\rm GHz}$ and $\flc = 0.82\,{\rm GHz}$, in the absence of  the resistor and charge noise. 
    (a) Evolution of charge support $p(q)$  during the stabilizer segment for a logical $|1\rangle$-state.
    (b) Wigner function of the system at $t=0$  and (c) at $t = \TLC/(8\pi)$. 
    Horizontal arrows indicate correspondence to peaks in panel (a), and purple arrows indicate shear-drift of peaks in Wigner function.
    (d) Evolution of charge probability distribution for  a logical $|0\rangle$-state.
    (e) Evolution of stabilizers during the stabilizer segment for the $|0\rangle$  (red) and $|1\rangle$ state (blue).}
    \label{fig:measurement}
\end{figure}

We now describe a possible way to measure
 $\sigma_z$ via the supercurrent in the Josephson junction.
The protocol can also be used for initialization in a \addFN{logical $|X\rangle$ state (i.e., a state with $\langle \sigma_x\rangle =1$)}. 
The protocol consists of  the following sequence: 
\begin{enumerate}
    \item Activate the switch ($\ws=1$) 
     for a duration $\TLC/4\pi$ 
     \item  Deactivate the switch ($\ws =0$) for a duration $\TLC/4$ 
    \item 
    Reactivate the switch and measure  the squared supercurrent  in the Josephson junction, $I^2 = \left({2e} E_J/\hbar\right)^2\sin^2(2\pi \varphi/\varphi_0)$,  e.g., from the frequency shift of an adjacent transmon  due to the induced magnetic field~\cite{danilin_2018,budoyo_2020}. 
    In this interval,  the relaxation time $\Gamma^{-1}$ (controlled through $C_{\rm R}$) must be longer than the detection time of the  device.
    \item If the average squared supercurrent is larger than $I_c^2/4$, the system was in a $|1\rangle$ logical  state ($\langle \sigma_x\rangle = -1$) at the onset of the readout protocol. 
    If not, the system was in a $|0\rangle$ logical state at the onset.
    \item 
    ({\it For initialization}):
    Deactivate the switch again after a duration $z_{\rm readout}/(2\pi \flc)$ where $z_{\rm readout}$ is an integer large enough to ensure successful measurement of the supercurrent. 
    If no supercurrent is detected, the system is  initialized in a logical state with $\langle \sigma_x\rangle=1$ at the end of the protocol. If not, run the ordinary stabilization protocol for a few cycles, and repeat the steps above. 
\end{enumerate}

The readout protocol exploits a characteristic peak structure  that emerges  in the   charge support 
of GKP states during the stabilizer segment, $p_q(x)\equiv \Tr[\delta(q-x)\rho]$.
At times  $t =a\TLC /(2\pi b )$  for  $a,b\in \mathbb Z$, the charge support of $\rho $ is confined near multiples of $q= e/b$. 
If  $b$ is furthermore even, the parity of the multiple  reveals the $\sigma_z$:  a logical $|s\rangle$ state, with $\sigma_z$-eigenvalue  $(-1)^s$, will have charge support confined near  $(2n+s)e/b$ for $n\in \mathbb Z$.
The peak structure is evident in     Fig.~\ref{fig:measurement}(ad), where we plot  the evolution of $p(q)$ during the stabilizer segment starting from two different logical states with $\sigma_z$ eigenvalue $-1$  (a) and  $1$ (d)\footnote{ 
    The initial state consisted of well ground states with a Gaussian envelope, and had wavefunction  $\Psi_s(\varphi)$ for $s=0,1$, where $\Psi_s(\varphi)=  \sum_k e^{-8\pi^2 \lamp^2 k^2}e^{-(\varphi-[2k+s]\varphi_0)/2\lamp^2\varphi_0^2}$ (up to a normalization).}.
The    structure  emerges due to  a  shear-drift  of the peaks of the system's  Wigner function, $W(\varphi,q)$ during the stabilizer segment. 
At time $t$,  all peaks  located at flux  $\varphi = n\varphi_0 $   have shifted in the $q$-direction by an amount $2\pi n e \flc t/2$. 
This causes distinct Wigner function peaks to align in the $\varphi$ direction for rational $2\pi \flc t$, leading to emergence of peaks in the charge probability distribution, $p(q)= \int d\varphi W(\varphi,q)$, as illustrated in  Fig.~\ref{fig:measurement}(a-c).
The mechanism is discussed in further detail in   Appendix~\ref{Appendix_Measurement}.

The readout protocol exploits the peak structure 
as follows:    
Step $1$ of the protocol evolves the system  until a time  $t=\TLC/4\pi$ ($a=1,b=2)$, where a logical  $|s\rangle$ state will have its $q$ support confined near $(n+s/2)e$ for integer $n$. 
Step $2$   maps $ q/e$ to $-\varphi/\varphi_0$, implying that the $\varphi$support of the resulting state is confined near $(n+s/2)\varphi_0$ for $n\in \mathbb Z$. 
If the system is in a $|1\rangle$  logical state, the physical  state will  have its $\varphi$-support confined near the maxima of the Josephson potential. 
During step $3$, the system will thus decay to the ground state of the Josephson potential, leading to a detectable supercurrent signal. 
On the other hand, for a $|0\rangle$ logical state, the system will be deep in the wells of the Josephson potential at the onset of step $3$, and no supercurrent will be detected. 

The protocol above can be used for initialization: by starting from a random initial state, a few cycles of the stabilization protocol  first drives the system into the code subspace. 
Subsequently, the readout protocol is applied.
If no supercurrent was detected, the system is known to have even support in all wells after the readout protocol, i.e.,  to be in a logical $|X\rangle = \frac{1}{\sqrt{2}}(|0\rangle +|1\rangle)$ state.  
If a nonzero supercurrent is detected, the system is in an undetermined state (since the relaxation into the wells will cause the measured $|1\rangle$ state to  be destroyed, and replaced by a mixed logical state). 
In that case, the stabilization/readout protocol is repeated, until no supercurrent is detected. 

Importantly, noise from the readout apparatus does not affect the stability of the qubit during its normal operation. Specifically, the squared supercurrent $I^2=(2eE_J/\hbar)^2\sin^2(2\pi \varphi/\varphi_0)$  \addFN{commutes with the logical operators $\{\sigma_i\}$}; noise coupled to it hence cannot decohere the encoded information. 
Indeed, the supercurrent readout  can work as a syndrome detector, if switched on during the normal operation of the qubit. 

\newcommand{\pmax}{p_{\rm error}} 

\section{Noise tolerance, device requirements, and timescales}
\label{sec:params}
\begin{table}[]
        \centering
        \addFN{
        \begin{tabular}{|c||c|c|c|}
    \hline 
        \multicolumn{4}{|c|}{{\bf Parameters}}  \\ \hline 
        $L$                                   &$2.5\mu H$&      $4.5 \mu{\rm H}$& $10 \mu{\rm H}$   \\     
        $C$                                 & $ 15 \,{\rm fF}$  &      $27 \,{\rm fF}$      & $60 \,{\rm fF} $         \\ 
        $E_J/h$                     &      $200\,{\rm GHz}$        & $9 0\,{\rm GHz}$      & $200\,{\rm GHz}$          \\ 
        $\Gamma$     &      $2\,{\rm GHz} $   &$1\,{\rm GHz}$      & $0.5\,{\rm GHz}$          \\ 
        $\zs$                       &      $8 $          &$9$           & $8 $       \\
        \hline
        $\flc$                      &         $.82 \, {\rm GHz}$   &  $.46\, {\rm GHz}$      &  $.21\, {\rm GHz}$          \\ 
        $\lamp$ & $0.08$ & $0.08$ & $0.05$\\ 
         \hline\hline 
        \multicolumn{4}{|c|}{{\bf Device requriements}}  \\ 
     
        \hline 
        Switch rise time  & $16 $ ps  & $31 $ ps  & $45$ ps           \\ 
        LC Quality factor  & $520$ & $920$ & $2\, 000$  \\
        Max. Temperature & $0.58\,{\rm K}$ & $0.23\,{\rm K}$ & $0.9\,{\rm K}$\\ 
        Accuracy of $L/ C$       & $8\% $  & $9\% $ & $6\% $                    \\
                     \hline
\multicolumn{4}{c}{}\\[-1em]
\hline
        \multicolumn{4}{|c|}{{\bf Noise thresholds}}  \\\hline 
        Charge noise &  $4\times 10^{-12}\frac{e^2}{\rm Hz}$ & $7\times 10^{-12} \frac{e^2}{\rm Hz} $ &$2\times 10^{-11}\frac{e^2}{\rm Hz}$        \\ 
        Flux noise   & $3\times 10^{-12}\frac{\varphi_0^2}{\rm Hz}  $ &  $2\times 10^{-12}\frac{\varphi_0^2}{\rm Hz}$ & $5\times 10^{-12}\frac{\varphi_0^2}{\rm Hz}$            \\ 

        \hline

\multicolumn{4}{c}{}\\[-1em]\hline 
        \multicolumn{4}{|c|}{{\bf Operation timescales}}  \\ 
        \hline
        Protocol cycle          & $2\,{\rm ns}$ & $4\,{\rm ns}$ & $7\,{\rm ns}$                   \\ 
        $H$ Gate 
& $0.3\,{\rm ns}$ & $0.55\,{\rm ns}$ & $1.2\,{\rm ns}$                \\ 
        $S$ Gate            & $0.2\,{\rm ns}$ & $0.35\,{\rm ns}$ & $0.8\,{\rm ns}$             \\ 
        Readout             & $75\,{\rm ns}$ & $75\,{\rm ns}$ & $75\,{\rm ns}$                 \\ 
        Initialization      & $310\,{\rm ns}$ & $310\,{\rm ns}$ & $320\,{\rm ns}$  
         \\ 
            \hline 
    \end{tabular}
    \caption{
    \textbf{Operating regime  and operation timescales for $3$ parameter scenarios}. Estimate are 
    obtained in Sec.~\ref{sec:params}, for an error tolerance of $\pmax=0.0003$ per cycle (4 standard deviations). 
    Parameters $L,C$,   $E_J$, $\Gamma$, and $\zs$ denote the inductance, capacitance, and Josephson energy, resistor-induced loss rate, and stabilizer segment duration in units of the revival time $\trev\equiv \sqrt{LC}$, respectively.  $\flc$ and $\lamp$  denote the derived LC frequency and GKP squeezing parameters. 
    The  charge and flux noise thresholds denote  the estimate maximal tolerated  power-spectral density for a white-noise charge or flux signal. 
     For noise tolerances and device requirements below the  listed thresholds, we expect   the  qubit lifetime to remain exponentially long.
     \addFN{The parameters in column 1 are closest to experimental access, with $L = 2.5\,\mu{\rm H}$, $C\approx 2.7\,{\rm fF}$ resonators achieved in Ref.~\cite{Pechenezhskiy_2020}, and   pulse train generators available with rise times below $10\,{\rm ps}$~\cite{Afshari_2005}.}}
          \label{tab:parameters}
}
\end{table}

\addFN{
Here  characterize the qubit's perforfmance in the presence of device/control imperfectins or noise, and estimate  the parameter requirements for the qubit. 
We also estimate relevant operational  timescales.}
\addFN{We list our estimated device  requirements, noise tolerances, and operational timescales for 
$3$ different parameter scenarios in  Table~\ref{tab:parameters};  each requirement/threshold is estimated  for a tolerated logical error probability per cycle $\pmax\approx 0.0003$  corresponding to  $4$ standard deviations (see below for details).} 

\subsection{Device requirements}
\addFN{We first estimate  the device requirements, focusing on Josephson energy, impedance, temperature, control resolution, and quality factor of the LC resonator.}


\addFN{
\textbf{Josephson energy and temperature.} We estimate the minimal required Josephson energy using the error bound estimate in Eq.~\eqref{eq:error_rate_result}, $ p_{\rm error} \lesssim \exp(-\frac{-1}{k_{\rm B}T/E_J+\pi^2 \lambda_0^2 })$, where $  \lamp = \left(\frac{h\flc}{4\pi^3 E_J}\right)^{1/4}$.
The  error rate  $\pmax \sim 0.0003$ hence requires $(k_{\rm B}T/E_J)+\pi^2 \lambda_0^2 \lesssim 1/8$, leading to the following  estimate for the minimal required Josephson energy and maximal temperature: 
\be 
E_{J}\gtrsim 60 h\flc, \quad k_{\rm B}T\lesssim {E_J/8-\sqrt{E_J h\flc }}
\ee 
We emphasize that the thresholds above are obtained from the error estimate in Eq.~\eqref{eq:error_rate_result}, which may be a relatively loose upper bound. The qubit may thus operate at Josephson energies smaller than, and temperatures larger than, the thresholds we identify above.}

\addFN{\textbf{Switch rise time.} We next estimate the condition on the switch rise time. 
A finite rise time $\Delta t$ causes the free segment to effectively mistarget the $\pi/2$ rotation of phase space by an angle  $\delta \theta$ where $\delta \theta\sim  2\pi \flc \Delta t/2$.
We estimate the induced error rate to be given by the total phase support displaced by more than $1/2$ in phase-space symmetric units where $e=\varphi_0=1$. 
In Appendix~\ref{app:envelope}}\addFN{, we show the  system's Wigner function envelope  is  a Gaussian  with  standard deviation $\kappa/\sqrt{2}$ in these units, where $\kappa = \sqrt{\coth(2\varepsilon_0/k_{\rm B} T)}/\pi\lamp$, and $\varepsilon_0 =   \sqrt{4\pi  E_Jh\flc } $ (see  also Fig.~\ref{fig:wavefunctions}).
This leads us to estimate $p_{\rm error}$ as the probability weight of this Gaussian envelope beyond distance $1/2\delta \theta$. 
Using $1/2\delta \theta \sim 1/2\pi \flc \Delta t$, and working in the limit of low temperatures where $\kappa \approx 1/\pi \lamp$,  leading to 
\be 
p_{\rm error}(\Delta t) \sim e^{-\frac{\lamp^2}{4 \flc ^2 \Delta t^2}}. 
\label{eq:error_vs_mistiming}
\ee 
For comparison, we have plotted this estimate together with numerically obtained error rates in Fig.~\ref{fig:gates}.
For a logical error rate corresponding to $p_{\rm error}=  0.0003$, which requires $E_J \gtrsim 60$, such that $\lamp \sim 0.1$, we hence estimate a minimal rise time of order $\Delta t_{\rm max} \sim  \frac{0.02}{\flc}$.
We emphasize that the switch must  be completely turned off in the bulk of  the free segment to ensure preservation of logical information. If this is not the case, spurious tunneling of Cooper pairs may  induce irrecoverable logical errors. We thus expect the qubit to only be resilient to switch mistiming, with imperfect switch deactivation  causing logical errors (hough  still allowing generation of GKP states---see Sec.~\ref{sec:experimental_considerations}).}

\addFN{
\textbf{Impedance}. 
Next, we consider the tolerance for mistargeting the impedance, $Z=\sqrt{L/C}$.
A  finite deviation of impedance $\delta Z \equiv Z-h/2e^2$  leads to a squeezing of the Wigner function over a free segment by the factor $2 \delta Z e^2/h$. 
Analogously to our condition for control resolution, we estimate the error rate to be given by the phase space support of the system in the region where the squeezing induced displacement exceeds $1/2$ in units with $\varphi_0=e=1$. 
Recalling that the system's Wigner function has a Gaussian envelope of width $\kappa \sim 1/\pi \lamp$, this leads to 
\be 
p_{\rm error}(\delta Z) \sim e^{-\frac{\pi^2 \lamp^2}{4[2e^2\delta Z /h]^2}} 
\ee 
With $\pmax=0.0003$ and $\lamp \sim 0.1$ (as required to reach this error rate), this leads to a window of $\sim 700 \Omega$ (i.e., $\sim 5\%$ tolerance for relative deviations). 
Because of the square-root,  the tolerance for relative deviations of  $L$ and $C$ is twice that of $Z$, up to $\sim 10\%$.
}

\addFN{
\textbf{Quality factor}. We next consider the consequences of a finite $Q$ factor of the LC resonator caused by uncontrolled capacitative coupling to its surrounding environment---i.e.,  photon loss. 
During the stabilizer segment, a capacitative coupling to an external environment  is {beneficial}; indeed such a coupling is leveraged by our dissipative stabilization protocol.
During the free segment, the capacitive coupling on the other hand results in a uniform loss rate of photons from the LC resonator, at the rate $\gamma = 2\pi \flc /Q$, where $Q$ is the corresponding quality factor
~\cite{GardinerZoller}. %
Working in
units where $\varphi_0=e=1$, photon loss generates  simultaneous uniform diffusion and   shrinkage of phase space, with diffusion constant $\gamma \coth(h\flc/2k_{\rm B}T)$ and shrinkage rate $\gamma$~\cite{Isar_1991}. 
At the end of the free segment, after a duration $1/4\flc $, photon loss has thus shrunk phase space by a factor $\approx (1-\gamma /4\flc)$, and diffused it with a diffusion kernel of variance  $\Delta= \sqrt{\gamma \coth(h\flc/2k_{\rm B}T)/4\flc}$. 
We estimate the logical error rate as the phase space weight  displaced by more than $1/2$ by the diffusion kernel or the total phase space weight displaced by more than $1/2$ by the phase space shrinkage, whichever is largest. 
Recalling that the envelope of the Wigner function is a Gaussian with standard deviation $\kappa/\sqrt{2}$, this leads to the estimate $p_{\rm error} \sim \max\{e^{\frac{-1}{8\Delta ^2}},e^{-\frac{4\flc^2}{\kappa^2\gamma^2}}\}$. Using the expression for $\Delta$ along with $Q=2\pi/\gamma\TLC$
\be 
p_{\rm error}(Q) \sim \max \left\{\ e^{-\frac{Q}{4\pi  }\tanh\frac{h\flc}{2k_{\rm B }T}}\ , \  e^{-\lambda_0^2 Q^2}\ \right\}.
\ee 
Working in the regime $\flc\sim 1\,{\rm GHz}$, $T \sim 100\,{\rm mK}$ and for $p_{\rm error}\sim 0.0003$ (requiring $\lamp\gtrsim 0.1$), this leads to the condition $Q_{\rm min}\sim 450$.}
\addFN{This relatively mild condition raises the possiblity that  the dissipative element can permanently connected to the resonator, possibly in combination with appropriate filtering of bath modes.}

\addFN{
\subsection{Tolerance for noise}
\label{sec:charge_noise}
Here analyze the effects of flux and charge noise. 

\textbf{Charge noise.}
We  model charge noise  as  a fluctuating charge $\xi(t)$  capacitively coupled to the circuit through  $H_{q}(t)=\xi(t)q/C$.
For simplicity, we assume  $\xi(t)$   a white-noise signal, with 
uniform power spectral density $\gamma_q/2\pi $, such that $\langle \xi(t)\xi(t')\rangle = \gamma_q \delta(t-t')$. 
On its own,  $H_q(t)$ generates Brownian motion  of the flux, $\varphi$,  with  diffusion constant $\Dq  
=\gamma_q/C^2 $~\cite{GardinerZoller}. 
Within the stabilizer segment,  dissipation from the resistor counteracts the diffusion,  driving $\varphi$ towards the minima of the flux potential, $\varphi=n\varphi_0$. 
We expect the qubit to remain stable in the stabilizer segment if the effective diffusion length within the resistor-induced relaxation  time $\Gamma^{-1}$ is much smaller than $\varphi_0/2$, i.e., if $\Dq \ll \Gamma\varphi_0^2$. 
Within the free segment,  $H_q(t)$  generates diffusion in phase space along the direction $(\varphi_0\cos(2\pi \flc t),e\sin(2\pi\flc t))$ with normalized  diffusion constant $D= D_q/\varphi_0^2$ in   units where $e=\varphi_0=1$.
As a result, $H_q(t)$ generates correlated flux and charge displacements with variances both given by $\sigma_q^2 = D\TLC/8$ in the free segment. 
We estimate the error rate to be the phase  weight displaced beyond a distance $1/2$ by this difussion along the directions of either of the mode quadratures, $p_{\rm error}\sim e^{-{1}/{8\sigma_q^2}}$. 
Using $C=\frac{e^2}{\pi  h\flc}$, this leads to~\footnote{{The inverse scaling with $\flc$ arises due to the parametrization we use for charge noise:  the  coupling between system and  charge noise  is proportional to $C^{-2}$  while $C\propto\flc^{-1}$, due to  Eq.~\eqref{impedance_cond} fixing the resonator impedance. 
The  charge noise-induced diffusion  rate thus scales with $\flc^{2}$ for fixed $\gamma_q$, while the tolerance for diffusion rate scales with $\flc^{-1}$.
As a result, smaller $\flc$ leads to more stability with our parametrization.}} 
\be 
p_{\rm error}(\gamma_q) \sim e^{ -\frac{e^2}{4\pi^2\gamma_q \flc}}
\ee 
For a tolerated error rate of $\pmax=0.0003$ and $\flc\sim 1\,{\rm GHz}$ resonators, we thus require $\gamma_q \lesssim 3\times 10^{-7} \frac{e^2}{{\rm Hz}}$. }

\addFN{\textbf{Flux noise.}}
We  finally consider flux noise, which we model as a white-noise fluctuating flux $\xi_{\varphi}(t)$ coupled to the system through $H_\varphi = { \xi_\varphi(t)\varphi}/{L}$, where   $\langle \xi_{\varphi}(t)\xi_{\varphi}(t')\rangle = \gamma_{\varphi}\delta(t-t')$. 
On its own, this term generates random diffusion of the charge, and causes a phase space displacement during the free segment, $(\Delta \varphi,\Delta q)$. 
Since no mechanism counteracts charge diffusion during the stabilizer segment, flux noise  also generates a charge displacement in the stabilizer segment with diffusion constant $\Df = \gamma_{\rm \varphi}/L^2$. This displacement is  only corrected  in the following cycle, where it has been mapped to a flux 
displacement. 
Over this duration, the flux-noise induced diffusion kernel has acquired a variance $\sigma_\varphi^2 = \Df \TLC/8+\Df \Ts/2$ along the charge quadrature, and a variance $\sigma_\varphi^2 = \Df \varphi^2_0\TLC/8e^2$  along the flux quadrature. 
Combining the diffusion during the stabilizer and free segment, an analysis similar to the one performed for charge noise results in the error rate estimate 
\be 
p_{\rm error}(\gamma_\varphi) \sim e^{ -\frac{\varphi_0^2}{\pi^2[4+\zs/2\pi] \gamma_\varphi \flc}}
\ee 
Using $\zs \sim 1$, 
$\pmax\sim 0.0003$, and $\flc \sim 1\,{\rm GHz}$, 
we obtain the requirement 
$\gamma_\varphi  \lesssim 3 \times 10^{-7}\frac{\varphi_0^2}{\rm Hz}$.

\subsection{Operation timescales}
We  estimate the readout and  initialization times  using the native  protocol  described in Sec.~\ref{sec:Readout}, assuming the squared supercurrent of the Josephson junction can be detected within $\sim 75\,{\rm ns}$ with a Josephson-device based magnetometer~\cite{danilin_2018,budoyo_2020}. 
The  detection time is estimated 
assuming the  magnetometer is located $5\,\mu {\rm m}$ from the Josephson junction, and has a   sensitivity  of $\sim 10 \,{\rm pT}/\sqrt{{\rm Hz}}$, as has been realized recently~\cite{danilin_2018,budoyo_2020}.
The initialization time is estimated  assuming a $50\%$ probability for successful initialization at each attempt (see Sec.~\ref{sec:Readout}), implying that an average of $4$ attempts are required to initialize the qubit. 

The gate times are estimated using the native gates of the qubit, described in Sec.~\ref{sec:scgates}: the $S$ gate $e^{-i \pi \sigma_z /4}$ is generated by a stabilizer segment with duration  $1/(2\pi \flc)$, while the Hadamard gate is generated by an free segment, with duration $1/(4\flc)$. 
See Sec.~\ref{sec:scgates} for further discussion of the gates. 


\section{Numerical Results}
\label{sec:Numerics}
We now verify our analytic results above with numerical simulations of the qubit. 

To demonstrate the self-correcting properties of  the qubit, we  include charge noise throughout the    simulations, modelling  the system via the master equation 
\be 
\partial_t \rho  =\Big(\mathcal L_{\rm s}(t)+  \mathcal L_{\rm noise}(t)\Big)[\rho], 
\label{eq:numerics_equation}
\ee 
where $\mathcal L_s(t) $ and $\mathcal L_{\rm noise}(t)$ are  the time-evolution generators (Liouvillians) of the stabilization protocol and charge noise, respectively. 
\addFN{We use the Universal Lindblad Equation (ULE) to model the dissipative dynamics in the stabilizer segment~\cite{Kirsanskas_2018,Phd_thesis,Nathan_2019,Davidovic_2021,Nathan_2020b,Nathan_2024a}, via 
\be 
\mathcal L_{\rm s}(t)[\rho ] = -\frac{i}{\hbar}[H_{\rm S}(t),\rho] +\ws(t)\left[\ell\rho \ell -\frac{1}{2}\{\ell^{\dagger} \ell,\rho\}\right].
\ee 
where  $H_{\rm S}(t) =q^2/2C + \varphi^2/2L - \ws(t)E_J\cos(2\pi \varphi/\varphi_0)$ denotes the non-dissipative component of the system Hamiltonian, and 
\be 
\ell\equiv \frac{1}{\hbar C_{\rm R}}\sum_{mn}|\psi_m\rangle \langle \psi_n|\sqrt{2\pi  J([E_n-E_m]/\hbar)}\langle \psi_m|q|\psi_n\rangle\ee 
denotes the ULE jump operator for the system with the switch activated.
Here  $|\psi_n\rangle $ and $E_n$ denote the energies and eigenstates of  $H_{\rm LCJ}\equiv  q^2/2C + \varphi^2/2L - E_J\cos(2\pi \varphi/\varphi_0)$, and $J(\omega)$ the power spectral density of the resistor.
We model the resistor  as an Ohmic bath at temperature $\Tb$, such that 
$J(\omega)= g^2  \omega(1-e^{-\hbar \omega/k_{\rm B}T} )^{-1}$, for some constant $g$~\cite{GardinerZoller,Breuer,Nathan_2020b}, which fixes the loss rate at $\Gamma = 4 (g e /C_{\rm R}\hbar)^2 \frac{E_J}{\hbar} $; see  Eq.~\eqref{eq:gamma}~\footnote{Here we used that $\varepsilon_0 = 4\pi \lamp^2 E_J$.}. 
The ULE is rigorously proven  to be valid in the regime of weak loss rate relative to the intrinsic correlation timescales of the bath (the latter of order $h/k_{\rm B}T$ for Ohmic baths)~\cite{Nathan_2020b,nathan_2022b}. 
Capturing both relaxation and interwell decoherence, we thus expect this model to provide an accurate representation of the dynamics. }



We model charge noise  through the Liouvillian
\be 
\mathcal L_{\rm noise}[\rho] = -i \frac{\xiq(t)}{\hbar C}[q,\rho],
\ee
 where $\xiq(t)$ is a scalar white-noise field satisfying $\langle \xi(t)\xi(t')\rangle = \gamma_q \delta(t-t')$, with $\gamma_q$ defining the charge noise strength. 
  We  set  $\gamma_q = 10^{-12}\, e^2/{\rm Hz}$ throughout the simulations, unless otherwise noted.

We numerically solve Eq.~\eqref{eq:numerics_equation} for various parameter sets, using the Stochastic Schrodinger equation (SSE)~\cite{Dalibard_1992,Carmichael_1993}, and with $\ell$  computed through exact diagonalization. 
Being agnostic to the analysis  in Secs.~\ref{sec:Setup}-\ref{sec:Stabilization},   our simulation thus serves as an independent check of its conclusions.

\subsection{Stabilization of GKP states}
\label{numerics_stabilization}
\begin{figure}
    \centering
\includegraphics[width=.99\columnwidth]{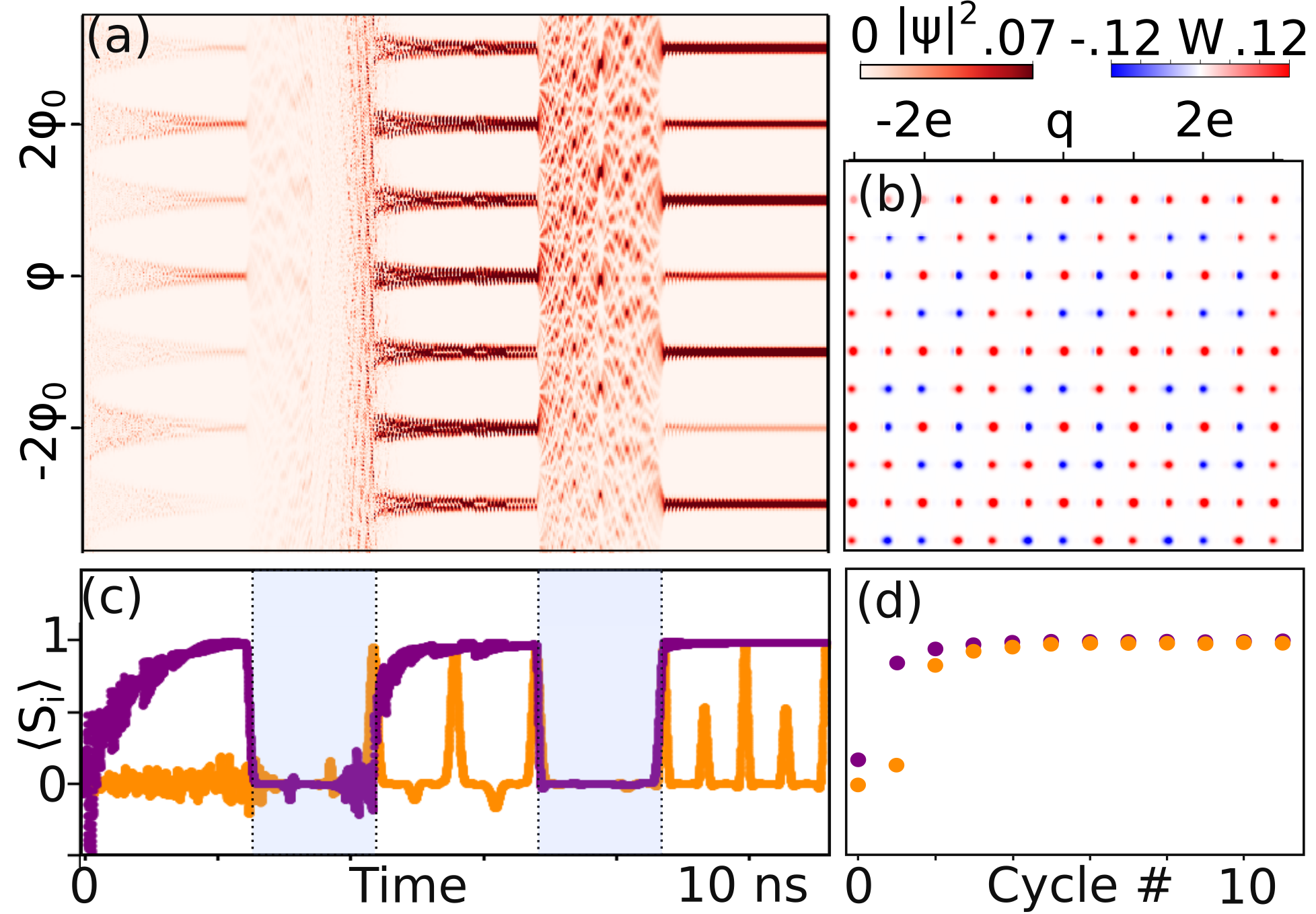}
\caption{\textbf{Dissipation-driven convergence to GKP code subspace}.  See Sec.~\ref{numerics_stabilization} for details.
    (a) Simulated evolution of flux probability distribution over $3$ cycles   obtained for a single SSE trajectory and starting from a random initial state~\cite{initialization}, \addFN{using $E_J/h = 200 \,{\rm GHz}$, $L=2.5\,{\mu}{\rm H}$, $C=15\,{\rm fF}$, $\zs=2$, and $T=200\,{\rm mK}$}.
    (b) Wigner function of the final state and 
    (c) evolution of stabilizers $1$ (purple) and $2$  (orange) for the  trajectory depicted in panel (a).  
    shaded regions indicate stabilizer segments.
    Note that $\langle S_2\rangle$ is only required to take value near $1$  at the end of each stabilizer segment for GKP states to be stabilized.
    (d) Mean value of stabilizers   at the end of each stabilizer segment for the first $12$ cycles, averaged over $100$ trajectories. 
     }
    \label{fig:stab_fig}
\end{figure}

To verify that the protocol stabilizes GKP states, we  initialized the system in a random high-energy state far outside the code subspace~\cite{initialization}, and computed the resulting evolution under the protocol via the SSE, \addFN{for parameters  $E_J/h = 200 \,{\rm GHz}$, $L=2.5\,{\mu}{\rm H}$, $C=15\,{\rm fF}$, $\zs=2$, and $T=200\,{\rm mK}$}. 
 Fig.~\ref{fig:stab_fig}(a)  shows the evolution of the flux probability density for a representative SSE trajectory from this simulation. 
The single trajectory  approaches the center of the wells of the Josephson potential (integer multiples of $\varphi_0$) in each  stabilizer segment, while retaining support in different wells---reflecting maintenance of inter-well coherence, consistent with our discussion  in Sec.~\ref{sec:Stabilization}.
After $3$  cycles, the Wigner function of the trajectory has the characteristic  GKP grid structure  [Fig.~\ref{fig:stab_fig}(b)], indicating successful convergence to the code subspace. 
Indeed,  the expectations of the two stabilizers have relaxed to near-unity after $2$ cycles [Fig.~\ref{fig:stab_fig}(c)]. 
Sampling   over $100$ SSE trajectories, we confirm that stabilization of GKP states is achieved within   $2-3$ cycles of the protocol ($\sim 8\,{\rm ns}$)  [Fig.~\ref{fig:stab_fig}(d)].

\subsection{Dissipative error correction}
\label{sec:dec_numerics}
\begin{figure}
\includegraphics[width=.99\columnwidth]{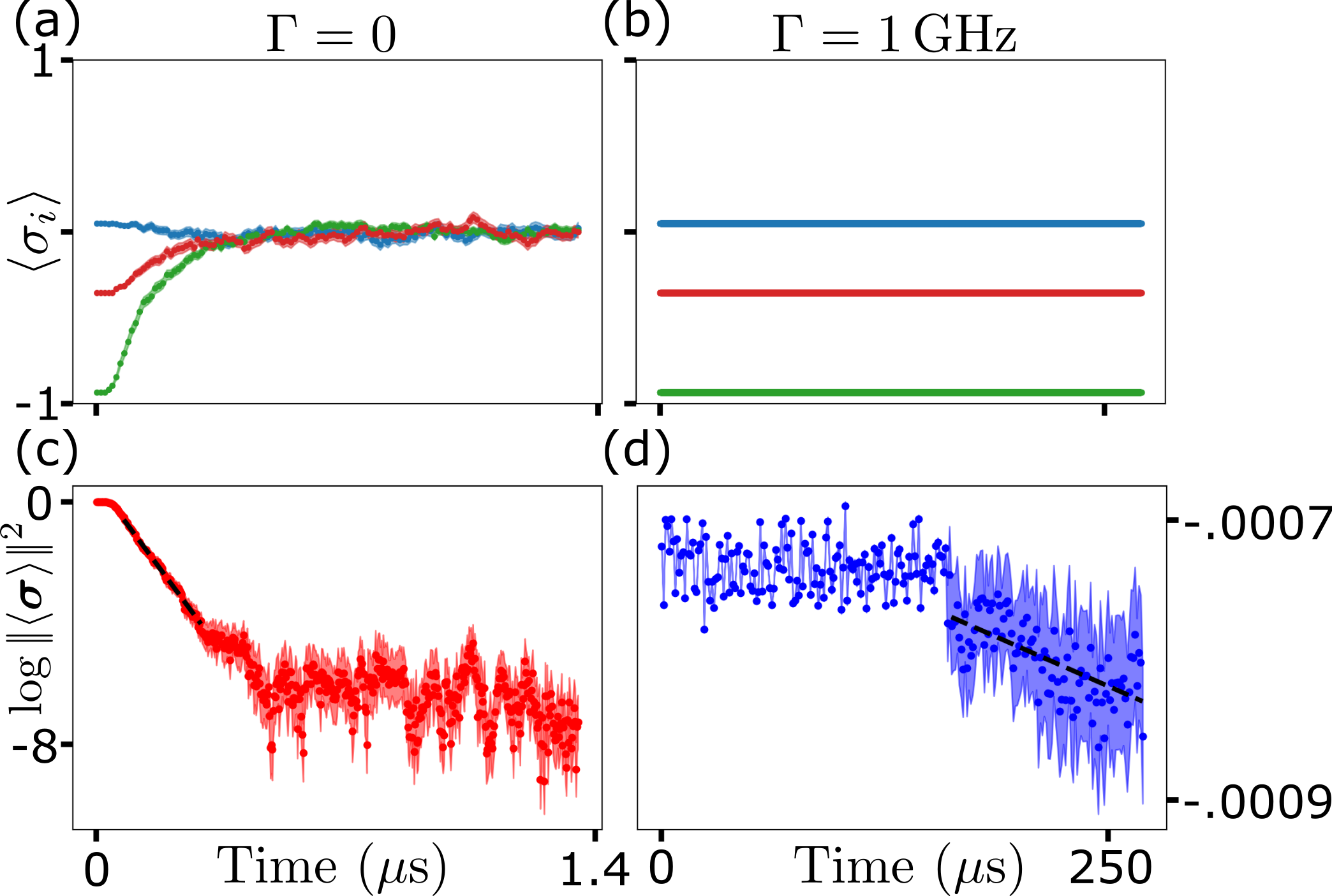}
\caption{   \textbf{Demonstration of dissipative quantum error correction}. \addFN{We use parameters  $E_J/h=200\,{\rm GHz}$, $L=10\,\mu{\rm H}$, $C=60\,{\rm fF}$, $\zs=8$, and $T=40\,{\rm mK}$.}
   (a,b) Simulated evolution of  logical operators in the absence (a) and presence (b) of coupling to the resistor, averaged over 200 SSE trajectories starting from a randomly chosen logical state (see main text).  
   (c,d) Logarithm of logical  {Bloch vector length}, $\log\norm{\bs\sigma}^2$, versus time, in the absence (c) and presence (d) of coupling to the resistor---note the different $x$- and $y$- axis scales in the panels. 
   Dashed lines indicate fits used to estimate qubit lifetimes. 
   Data in panel (d) are obtained by time coarse-graining evolution  of $\norm{\bs\sigma}^2$  over $400$ driving periods before taking the logarithm. 
   Shaded regions, where visible, indicate  standard error of the mean in panels (a,b), and standard deviation from bootstrap resampling of SSE trajectories in panels (c,d). 
    }
    \label{lifetime}
\end{figure}

To confirm that the logical information is dissipatively error corrected by the protocol,  
we compute the evolution of the system for many subsequent cycles, after initializing the system in a  randomly selected computational state~\footnote{Specifically, the system was initialized in a superposition of Gaussian wavefunctions with a Gaussian envelope, with the wavefunction $\Psi(\varphi)= \cos\left(\frac{\theta}{2}\right) \Psi_0(\varphi)   + \sin\left(\frac{\theta}{2}\right)e^{i\varphi}\Psi_1(\varphi)$, where $\Psi_s(\varphi) = \mathcal{N} \sum_{k}
e^{-( 2k+s)^2\pi^2 \lamp^2/4}e^{-(\varphi-[2k+s]\varphi_0)^2/2\lamp^2\varphi_0^2}$, with $\mathcal{N}$ 
a normalization constant and $(\theta,\phi) = (1.93,1.62)$ 
chosen randomly on the unit sphere.
The state above is well confined in the code subspace, and realizes the logical state $\cos\left(\frac{\theta}{2}\right) |0\rangle  + \sin\left(\frac{\theta}{2}\right)e^{i\varphi}|1\rangle$. \addFN{We also note that the results do not rely on the particular details of the initialization, except for the initial expectation values of the logical operators. After a few cycles of the protocol, the resonator has reached a thermal ensemble of generalized GKP state with no memory of the initialization, except the initially encoded logical information.}}.
In Fig.~\ref{lifetime}   we show the  resulting evolution of logical operators  for $\Gamma=0$ and $\Gamma =1\, {\rm GHz}$,  averaged over 200 trajectories, \addFN{and using  parameters  $E_J/h = 200 \,{\rm GHz}$, $L=10\,{\mu}{\rm H}$, $C=60\,{\rm fF}$, $\zs=8$, and $T=40\,{\rm mK}$}.
Whereas the logical operator expectations quickly decay in the absence of the resistor, for   $\Gamma =1\, {\rm GHz}$, the  logical operators remain  stationary over the entire window we simulate. 

To further illustrate the dissipative error correction of the qubit, Fig.~\ref{fig_1}(d) shows the stabilizers and logical operator evolution for representative SSE trajectories at $3$ values of $\Gamma$. 
Evidently, increasing  $\Gamma$ causes the fluctuations of the stabilizers away from  unity to decrease, and the logical operator trajectories to become stationary, implying stabilization of encoded information. 
Note that the logical operators  
for $\Gamma=1\,{\rm GHz}$ 
remains stationary in the presence of significant thermal fluctuations of the  stabilizers, and hence also the state. 
This demonstrates that the encoded information is successfully  decoupled from the thermal  noise from the resistor. 
Interestingly, it is also possible to distinguish individual logical error  events for $\Gamma= 0.25\, {\rm GHz}$  in Fig.~\ref{fig_1}(d): here stabilizers only reach negative values at a few instances, where rare (but significant) noise-induced  fluctuations  takes the system over the energy barrier  that  protects the qubit. Indeed, the logical operator remains near-stationary between these instances, but changes abruptly at instances where the stabilizers obtain negative values.

We estimate the qubit lifetime  via the decay of the logical state {Bloch vector length}, $\norm{\langle {\bs \sigma}\rangle}^2$.
Fig.~\ref{lifetime}(c) shows the stroboscopic evolution of $\log\norm{\langle {\bs \sigma}\rangle}^2$ for $\Gamma = 0$. 
The {Bloch vector length}  remains near-unity for a brief initial period of little decay, which we expect is due to the finite time required for the system's phase space support to reach the domain boundaries of the logical operators. 
Beyond this point,  the data shows a clear  linear trend consistent with exponential decay of the {Bloch vector length}. From a linear fit [dashed line in Fig. \ref{lifetime}(c)], we estimate a lifetime of $63^{+19}_{-9}$ ns, with errors indicating $95\%$ confidence interval from bootstrap resampling of SSE trajectories.
In Fig.~\ref{lifetime}(d), we show the  evolution of $\log\norm{\langle {\bs \sigma}\rangle}^2$ for $\Gamma = 1 {\rm GHz}$ [note the different $x$- and  $y$-scale compared to panel (c)]\footnote{Due to the long qubit lifetime in this regime, the level of decay we can resolve  on simulated timescale is very small, and thus subject to significant statistical fluctuations between SSE trajectories. To overcome this, we show  time coarse-grained  data, obtained by time-averaging the {Bloch vector length} over $400$ driving periods. We also use the time-coarse grained data for our fit. We have confirmed that the  lifetime fit varies little with the chosen window length, while the quality of the linear fit increases with $n$.}. 
The logarithm of the {Bloch vector length}   exhibits a clear linear decrease after an initial period of $\sim 150\, \mu{\rm s}$ where the information is near-stationary~\footnote{We speculate that the near-stationary interval arises from a small effective Lieb-Robinson velocity in phase space, which implies it takes a long time for  exponential tails of the phase-space support of the state to spread  to the domain boundaries of the logical operators.}.  
From a linear fit of the data after the onset of exponential decay [dashed line in Fig. \ref{lifetime}(d)]~\footnote{The values of $\log{|\!|\langle {\bs\sigma} \rangle|\!|^2}$ shown in Fig.~\ref{lifetime}(d) are very small, and hence the logarithm can be well approximated by its first order Taylor series over this range of values. Thus, $\log\overline{|\!|\langle {\bs\sigma} \rangle|\!|^2}$ being linear over this range is not conclusive evidence that the {Bloch vector length} decays exponentially in time; nonetheless, the inverse slope still defines a meaningful lifetime for the qubit.}, we estimate a lifetime of $1.8^{+9.9}_{-1.1}$ s (where the errors are the $95\%$ confidence interval from bootstrap resampling). While there is significant uncertainty in our estimate of the lifetime, it is clear that the presence of the resistor enhances the timescales over which quantum information is preserved up to macroscopic timescales.

To investigate the scaling of qubit lifetime with the dissipation strength, in Fig.~\ref{fig_1}(b), we  show the evolution of the obtained coherence times as a function of $\Gamma$, for the device parameters from column $3$ of Table~\ref{tab:parameters}. 
The data reveal an exponential trend that continues beyond 
the $1\,{\rm ms}$ range for $\Gamma\gtrsim 0.6\, {\rm GHz}$, indicating a potential for significant qubit stability against phase-space local noise. 

\subsection{Robustness of gates against switch mistiming}
\label{sec:gate_numerics}
\begin{figure}
\includegraphics[width=0.99\columnwidth]{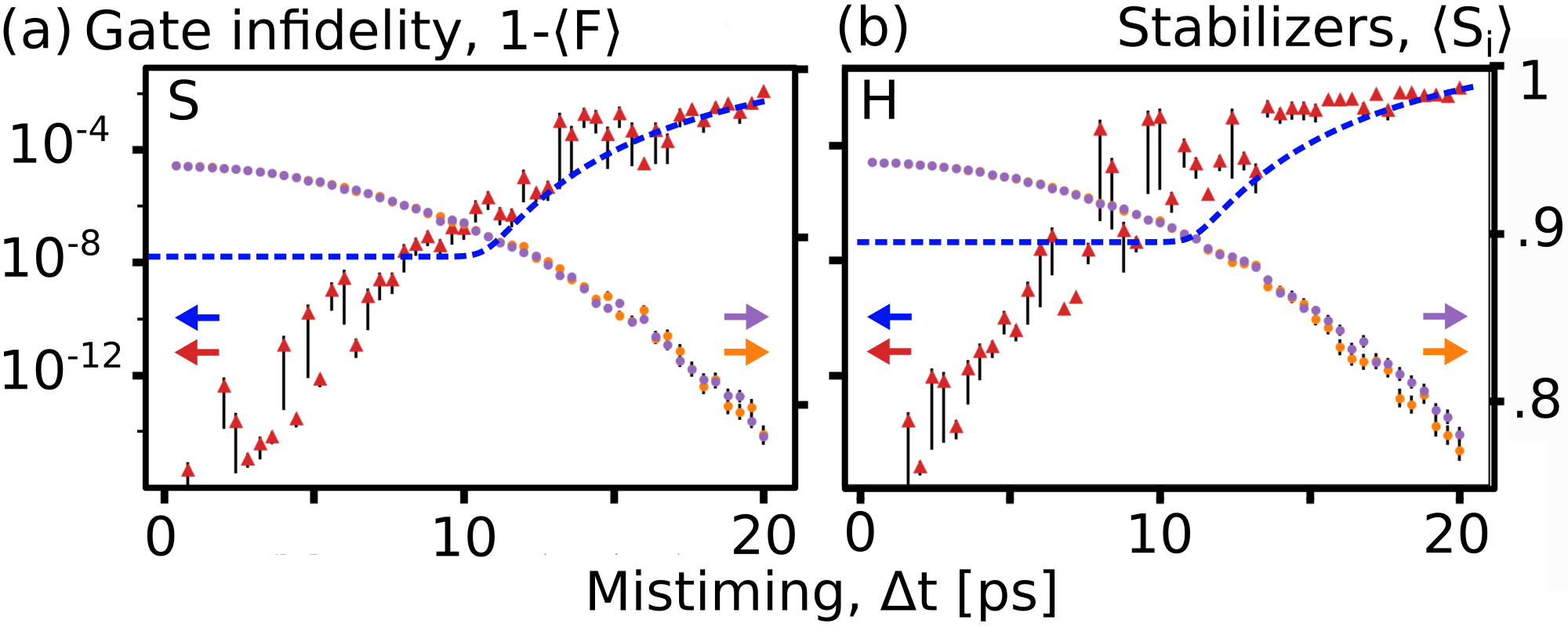}
\caption{\addFN{\textbf{Exponential suppresison of  gate infidelity}.  We use parameters  $E_J/h=200\,{\rm GHz}$,  $L=2.5\,\mu{\rm H}$, $C=15\,{\rm fF}$, $\Gamma =2 \,{\rm GHz}$, and  $T=40\,{\rm mK}$. Gate infidelity (red; left axis) and steady-state values of the GKP stabilizers (purple/orange; right axis) for the native (a) $S$ and (b) $H$ gates of the protocol, as a function of switch mistiming. 
Blue dashed lines indicate the analytically estimated  error rate bound from Eq.~\eqref{eq:error_vs_mistiming}; see main text for more details.  All data points shown are averaged over $10^3$ SSE trajectories. }  }
\label{fig:gates}
\end{figure}
\addFN{We next investigate the qubit's resilience to  switch mistiming, $\Delta t$. Such mistiming qualitatively captures the effects of both imperfect control, as well as the finite rise time of the switched Josephson junction. 
This analysis also serves as a check of the gate infidelity for the native $S$ and $H$ gates of the qubit (recall that these gates---or a power thereof---are generated by each stabilizer and free segment, respectively.}

\addFN{
To analyze the effects of finite $\Delta t$, for each cycle of the protocol we randomly changed  the duration of each stabilizer and free segment by $\delta t$ and $-\delta t$ respectively, preserving the total cycle duration; We made this choice to  reflect that the period of the signal is precisely controllable by available electronics. 
For each cycle, $\delta t$ is drawn uniformly on the interval $[-\Delta t/2,\Delta t/2]$, with $\Delta t$ a parameter we vary. 
We expect this randomly mistimed  protocol   also captures the dynamics of a smooth ramp of a realistic switch at a qualitative level. 
}

\addFN{
In Fig.~\ref{fig:gates}, we plot the error rate per $S$ gate (stabilizer segment) and $H$ gate (free segment), as a function of $\Delta t$, using parameters  $E_J/h=200\,{\rm GHz}$,  $L=2.5\,\mu{\rm H}$, $C=15\,{\rm fF}$, $\Gamma =2 \,{\rm GHz}$, and  $T=40\,{\rm mK}$.
We also include an analtyical  estimate based on Eq.~\eqref{eq:error_vs_mistiming}, combined with the beaseline level from Eq.~\eqref{eq:error_rate_result}, $p(\Delta t)=e^{-\lamp^2/4\pi \flc^2 \Delta t^2}+p_0 $, with $p_0=e^{-1/4\pi \lamp^2}$ the estimated upper bound on the error rate per cycle in the absence of any protocol imperfections. 
Note that $p_0$ dominates for $\Delta t \leq 12 \,{\rm ps}$. Evidently, the error rate exhibits a clear exponential dependence on $\Delta t$, that follows the analytical estimate reasonably well down to $p_0$, beyond which the exponential decrease continues down to a rate consistent with the exponential lifetime enhancement observed in Fig. \ref{fig_1}(b),  significantly undershooting $p_0$. This possibly reflects  that $p_0$ is a loose upper bound on the error rate, as discussed in Sec.~\ref{sec:Stabilization}. We also note that  the error per gate remains around $\sim 10^{-6}$ for mistiming $\Delta t \sim 10\,$ps. Given the availability of pulse train generators with rise times below $10\,$ps~\cite{Afshari_2005}, this suggests our device may be within reach of current experimental capabilities, provided one can control the Josephson Junction on these time scales (which we discuss in the next section). }

\subsection{Readout}
\label{numerics_state_prep}
We finally simulated the readout protocol from Sec.~\ref{sec:Readout}, using parameters $E_J/h = 200\,{\rm GHz}$, $L=10\,\mu{\rm H}$, $C=60\,{\rm fF}$, $T=40\,{\rm mK}$, $\zs=8$.
We first generated characteristic $\ket{0}$ and $\ket{1}$ logical states of the protocol by evolving initial states with support in only even and only odd wells of the cosine potential, respectively, for 18 cycles. 
We then simulated the evolution of the system during the readout protocol from Sec.~\ref{sec:Readout}. 
During  step  $3$ of the protocol, we decreased the effective resistor conductance to $g e /C_{\rm R} \hbar = 0.00016$, \addFN{in order to extend the relaxation time of the  supercurrent signal to the $\sim 100 {\rm ns}$ range, where it can possibly be  detected  (see Sec.~\ref{sec:Readout}).}
In the simulation, we used the charge noise strength $\gamma_q = 0.1 \times 10^{-13}\, e^2/{\rm Hz}$, because  the smaller value of $\kappa e/C_{\rm R} \hbar$ leads to a lower tolerance for charge noise.

\begin{figure}

\centering
\includegraphics[width=.99\columnwidth]{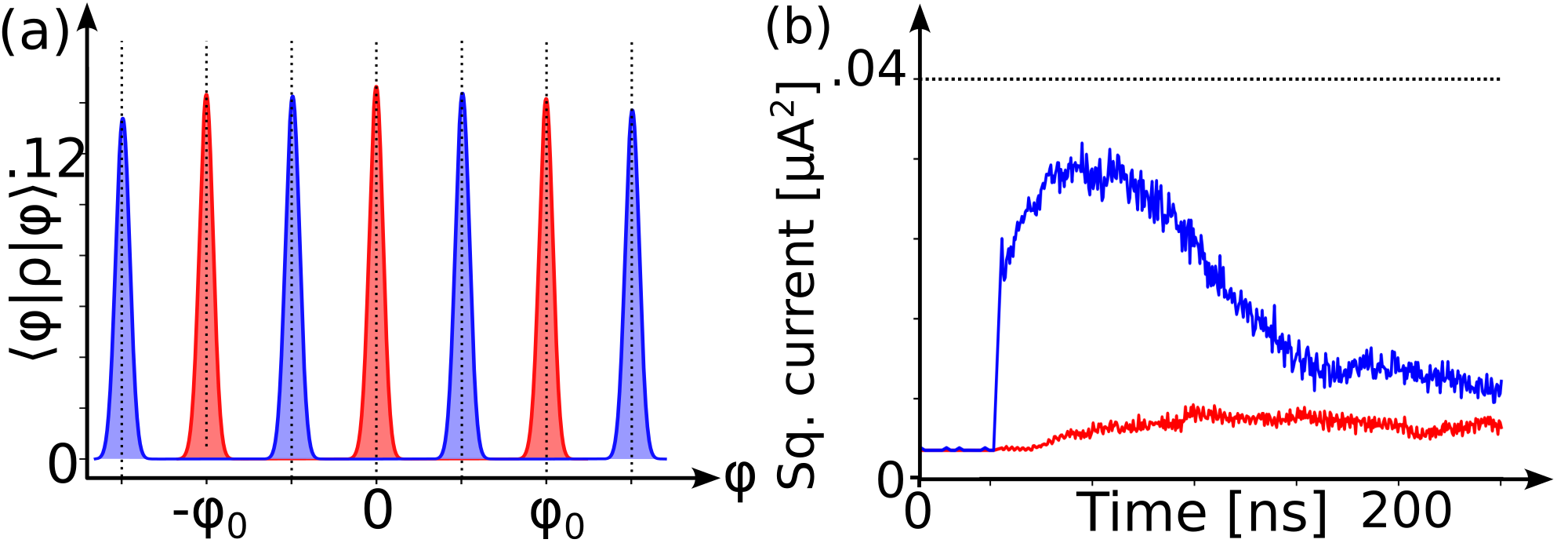}
        \caption{\textbf{Numerical Simulations of  readout protocol}; see  Sec.~\ref{sec:Readout}. \addFN{We use parameters $E_J/h = 200\,{\rm GHz}$, $L=10\,\mu{\rm H}$, $C=60\,{\rm fF}$, $T=40\,{\rm mK}$.} (a) Flux probability density, $\langle \varphi|\rho|\varphi\rangle$,  at the onset of step $3$ of the protocol, immediately before the measurement of squared supercurrent. Here the system is initially in a logical state produced by the protocol with $\sigma_z$ expectation $0$ (red) and $1$ (blue). 
    (b) Evolution of squared supercurrent during step $3$ of the protocol (supercurrent measurement), for the two initializations. }
    \label{measurement_fig}
    
\end{figure}

In Fig.~\ref{measurement_fig}(a) we show the flux probability density, $\bra{\phi}\rho\ket{\phi}$, at the onset of step $3$ of the protocol, for the two different initializations, averaged over $50$ SSE trajectories. 
As described in Sec.~\ref{sec:Readout}, the  distributions resulting from the two logical states are  confined near integer and half-integer multiples of $\varphi_0$, respectively.
In Fig.~\ref{measurement_fig}(b), we show the evolution of the squared supercurrent during step $3$, \addFN{$\langle I^2\rangle  = (2eE_J/h)^2 \sin^2 (2\pi \varphi/\varphi_0) $}, with $t=0$ denoting the onset of the readout protocol. 
Evidently the two different logical states result in very different supercurrent signals, that could be detected by a readout device. 

\label{noise}

\section{Discussion}
\label{sec:Discussion} 
\addFN{In this work, we proposed a circuit-QED architecture for a dissipatively error corrected GKP qubit. 
Our analytical and numerical results indicate that the self-correcting property of the qubit gives rise to an exponentially scaling lifetime, even in the presence of extrinsic noise or device and control imperfections.  %
 dissipative error correction leads to  exponential lifetime increase. 
We moreover demonstrated that the qubit supports a set of native, rapidly-operated, and {self-correcting}  single-qubit Clifford gates, whose infidelity we expect will be exponentially suppressed.
Also enhancing its  appeal, the qubit supports a native readout/initialization protocol via the Josephson junction supercurrent.
\addFN{If similar protocols can be identified for multi-qubit and magic gates, our results raise the possibility for  a {self-correcting quantum information  processor}.}}
\addFN{
\subsection{Experimental considerations}
\label{sec:experimental_considerations}
The key technological challenges we foresee for  our proposal is the realization of a  resonator with impedance $12.91\,{\rm k}\Omega$, along with a  controllable Josephson junction with a rapid rise  time $\Delta t$. 
For a given tolerated error rate per cycle, $p_{\rm error}$, the maximal rise time $\Delta t$ is  
is determined from Eq.~\eqref{eq:rise_time_condition}. 
For instance, with a tolerated error rate of  $p_{\rm error}=0.0003$ (4 standard deviations), which requires setting $E_J$ large enough that $\lamp\lesssim 0.1$ [through Eq.~\eqref{eq:error_rate_result}, we require $\Delta t \lesssim 0.02/\flc$. 
Using $\flc = \frac{h}{4\pi e^2 L}$, the condition on rise time for a 4-standard deviation error tolerance thus becomes 
\be 
\Delta t_{\rm max} \approx   8\,{\rm ps}\times L\,[\mu {\rm H}]. 
\label{eq:rise_time_condition}
\ee 
We expect this to be a reasonable estimate for the practically required rise time of the qubit. Note that  the estimate is in agreement with our simulations (Fig.~\ref{fig:gates}). 
}


\addFN{
The above condition on the rise time is in principle compatible with existing technology. For instance, electromagnetic  resonators with  $L\sim 2.5\,\mu {\rm H}$,  and impedance $>12.91\,{\rm k}\Omega$ were  realized with granular aluminium-based superconductors~\cite{Pechenezhskiy_2020}; see also Refs.~\cite{Kuzmin_2019,Peruzzo_2020,Smith_2022} for other recent realizations of high-impedance resonators. 
Moreover,  pulse train generators exist with  rise times below 10 ps~\cite{Afshari_2005}, well within the window dictated by  Eq.~\eqref{eq:rise_time_condition}.  
The main experimental challenge we anticipate for our proposal, then, is to integrate the  pulse to a fast-rise-time Josephson coupler.
Such an integration could, e.g., be realized with high-mobility semiconductor-based Josephson junctions controlled by gate electrodes~\cite{casparis_superconducting_2018,Samizadeh_2020}, Squid junctions controlled by flux lines~\cite{Kemppinen_2008,kolesnikow_2023_GKP}, or  voltage biasing of one or more  junction terminals (see Ref.~\cite{Geier_2024} for a recent proposal)---or combinations of the above approaches.
We also note that the condition on the rise time can be relaxed if resonators with inductances larger than $2.5\,\mu{\rm H}$ can be realized.}
With 
$2.5\,\mu{\rm H}$ inductor  we estimate the gate times of the qubit can be  of order $0.2-0.3\,{\rm ns}$, while readout and initialization times can be of order $75\,{\rm ns}$ and $300\,{\rm ns}$, respectively (see Sec.~\ref{sec:params}). 

\addFN{
Our results show that the self-correcting GKP qubit is  exponentially robust against phase-space local noise. At the same time, its lifetime   may be limited by phase-space-{\it nonlocal} noise, including 
residual  effective Josephson energy of the junction in the free segment due to imperfect deactivation. Any spurious Cooper pair tunneling during due to imperfect deactivation may directly induce non-correctible logical errors by generating nonlocal displacements in phase space.
We thus anticipate a hierarchy of independent success criteria  for the suppression rate of the switch:} 
\begin{enumerate}[label=(\alph*)]
\addFN{
\item Suppressing $E_J$ well below $h\flc$ will likely lead to dissipative {\it generation  of GKP states}, since the oscillator evolution in this case will be dominated by  free phase space rotation 
in the free segment. 
\item 
Suppressing $E_J$ by many orders of magnitude below $h\flc$ may moreover lead to significant lifetime enhancement from {\it dissipative error correction}, due to  spurious cooper pair tunneling being effectively absent over a large number of cycles.} 
\end{enumerate}

\addFN{
We also expect quasiparticles to be an important phase-space nonlocal noise source that could  limit  the qubit  lifetime. 
Quasiparticle poisoning events may effectively translate $q$ by  up to $e$, constituting a phase-space nonlocal process that can  potentially cause non-correctable logical errors~\cite{glazman_bogoliubov_2021}. 
At the same time, we expect this noise source to only matter when quasiparticles are accumulated in the capacitor, i.e.,  when the charge fluctuation  created by the quasiparticle has significant and persistent mutual capacitance with the resonator mode encoding the qubit. Hence,  quasiparticle trapping~\cite{Riwar_2016} and device engineering may offer routes to preventing  quasiparticle poisioning from adversely affecting the qubit. 
Moreover, experiments report poisoning rates in the ms$^{-1}$ range can be reached~\cite{acharya_quantum_2025,Aghaee_2024}, suggesting that quasiparticle-induced errors  can be rare enough to be  efficiently mitigated by active quantum error correction. 
}

\subsection{Outlook}
\addFN{Our results suggest that  the  architecture proposed here will result in  a  qubit which is autonomously error corrected on the physical qubit level, with  lifetime that cales exponentially, up to limits set by quasiparticle events and inverse residual Josephson energy of the junction in the deactivated mode. 
The qubit can moreover operate at temperatures in the $\sim 0.1-1\,{\rm K}$ range, and does not require pristine resonators} (we estimate  required quality factor down to  the $200-1000$ range; see Table~\ref{tab:parameters}) or perfect control over driving  signal or device parameters. 
Finally, the qubit supports  rapid single-qubit Clifford gates  with exponentially suppressed gate infidelity. 
These feature s could significantly simplify the qubit's  integration into a quantum information processor, and facilitate active error correction schemes based on the device.

{Dissipative error correction on the physical qubit level also raises the important possibility of parallel control  of many qubits. Specifically, the self-correcting properties of the qubit gives it an intrinsic tolerance for deviation of device parameters and control signals. In particular, varying response to a global control signal arising from device-to-device deviations would be corrected by the native stabilization protocol, provided the device parmaeters errors are within the thresholds we identified in Sec.~\ref{sec:params}. The ability to manufacture and control large numbers of qubits in parallel is a key component to scalable quantum computation, and so far has proved a challenge in superconducting circuits. Our device offers the exciting possibility of achieving this lofty goal.}


\addFN{Due to the  advantages above, we expect the qubit proposed here offers  a promising alternative route to scalable quantum computation,  bypassing key scalability challenges for approaches purely based on active error correction. }


\addFN{An interesting future direction is to explore whether the device supports a {universal} set of native, self-correcting gates, by exploring realizations of  self-correcting  multi-qubit and magic gates.
Particularly interesting, our  platform supports a self-correcting native magic ($T$) gate with the quasimodular  encoding described in Sec.~\ref{sec:T_gate}. 
We speculate that this mechanism can be leveraged as a resource for protected magic gate generation---i.e., a magic factory---in a  future quantum information processing architecture~\cite{magic_states}.}


\section*{Acknowledgements}
We gratefully acknowledge useful discussions with Philippe Campagne-Ibarcq, Max Geier, Lev-Arcady Sellem, Jacob Hastrup, Luca Banzerus,
Karsten Flensberg, Jonathan Conrad, and Dolev Bluvstein.
F.N. was supported by the U.S. Department of Energy, Office of Science, Basic Energy Sciences under award DE-SC0019166, the Simons Foundation under award 623768, and the Carlsberg Foundation, grant CF22-0727. G.R. is grateful for support from the Simons Foundation as well as support from the NSF DMR Grant number 1839271, and from the IQIM, an NSF Physics Frontiers Center. 
L.J. acknowledges support from the ARO(W911NF-23-1-0077), ARO MURI (W911NF-21-1-0325), AFOSR MURI (FA9550-19-1-0399, FA9550-21-1-0209, FA9550-23-1-0338), DARPA (HR0011-24-9-0359, HR0011-24-9-0361), NSF (OMA-1936118, ERC-1941583, OMA-2137642, OSI-2326767, CCF-2312755), Packard Foundation (2020-71479).
This work was performed in part at Aspen Center for Physics, which is supported by National Science Foundation grant PHY-1607611. The computations presented here were, in part, conducted in the Resnick High Performance Computing Center, a facility supported by Resnick Sustainability Institute at the California Institute of Technology.





\bibliography{bibliography_gkp_3.bib}

\begin{thebibliography}{97}
\bibitem [{Fou()}]{Fourier_expansion}%
  \BibitemOpen
  \href@noop {} {}\bibinfo {note} {\addFN{The operator $\Xi(\varphi/\varphi_0)$
  may conveniently be computed as $\sum_n c_n e^{-\pi i n \varphi/\varphi_0}$
  with $\{c_n\}$ the Fourier coefficients of $\Xi$. Note that $c_n$ is only
  nonzero for {odd} $n$ due to the antisymmetry
  $\Xi(x)=-\Xi(x+1)$}}\BibitemShut {NoStop}%
\bibitem [{S2c()}]{S2comm}%
  \BibitemOpen
  \href@noop {} {}\bibinfo {note} {This result can, e.g., be seen by noting
  that $\bar \varphi$ is a $\varphi_0$-periodic function of $\varphi$ and thus
  can be written as $\bar \varphi = \sum_n c_n e^{-2\pi i n \varphi/\varphi_0}$
  for some coefficients $\{c_n\}$. Since $e^{-2\pi i n \varphi/\varphi_0}$
  translates $q$ by $2en$, we see that $[e^{2\pi i n
  \varphi/\varphi_0},S_2]=0$, and hence also $[\bar
  \varphi,S_2]=0$.}\BibitemShut {Stop}%
\bibitem [{tru()}]{truncationscheme}%
  \BibitemOpen
  \href@noop {} {}\bibinfo {note} {More precisely, the evolution of $\tilde
  \rho(t)$ is unaffected up to $\erf(c)$ corrections if we replace $\bar
  \varphi$ with the time-dependent operator $\bar \varphi_c(t)$ in
  Eq.~\eqref{eq:heff}, where $\bar \varphi(t) = \bar \varphi$ for $|\bar
  \varphi|\leq \Delta \varphi(t)$, and $\bar \varphi_c(t) = \Delta \varphi(t)$
  for $|\bar \varphi|\geq \Delta \varphi(t)$.}\BibitemShut {Stop}%
\bibitem [{ini()}]{initialization}%
  \BibitemOpen
  \href@noop {} {}\bibinfo {note} {Specifically, we initialize the system in a
  state with \addFN{wavefunction $\Psi_0(\varphi) =
  \sum_{n=-45}^{45}\sum_{k=0}^{35} c_{nk}\psi_{k}(\varphi-n\varphi_0)$, with
  $\psi_k(\varphi)$ the $k$th Harmonic oscillator eigenstate with vacuum
  fluctuation width $\lambda_0\varphi_0$ (see Fig.~\ref{fig:wavefunctions}),
  and $\{c_{nk}\}$ random complex numbers with zero mean and identical
  variances.}}\BibitemShut {Stop}%
\end{thebibliography}

\appendix

\section{Effective Hamiltonian in comoving frame}
\label{seca:heff_derivation}

\label{seca:neglecting_delta_h}
Here we demonstrate that the Hamiltonian $\Hpz$ describes the evolution in the comoving frame introduced in Sec.~\ref{sec:Stabilization}, up to an exponentially suppressed correction to the state of the system. 
To recap, we consider the time-evolution in the comoving frame, reached through  the transformation 
\be 
V(t) = e^{-i\varepsilon_L M^2 t}, \quad{\rm where} \quad M\equiv\nint(\varphi/{\varphi_0}) .
\ee
We refer to $M$ as the {\it well index operator} throughout this and the following appendices. 
Our goal is to  show that, throughout each stabilizer segment, the system's density matrix $\rhop(t)\equiv V^\dagger(t)\rho(t)V(t)$, is exponentially close to the time-evolution  generated by $\Hpz$, [given in  Eq.~\eqref{eq:heff}],
$
\rhot \equiv e^{-i\Hpz t }\rho_0 e^{i\Hpz t}.
$
Specifically, we will show that 
\be 
\trnorm{\rhop-\rhot}\lesssim e^{-\frac{E_J}{k_{\rm B}T}}\,.
\label{eqa:expcloseresult}
\ee 
with $\trnorm{\cdot}$ denoting the trace norm, $\trnorm{A}=\Tr[\sqrt{A^\dagger A}]$. 


To establish the above results, 
we  exploit that $M$ is an integral of motion of the dynamics of $\rho(t)$ in the lab and comoving frames,   up to corrections of order $e^{-2\frac{E_J}{k_{\rm B}T}}$~\cite{Ankerhold_1995}. 
More precisely, we show in Appendix~\ref{app:aux_result} that, when the system is  confined  in a finite $\varphi$ range at the onset of the stabilizer segment, $-\wmax\varphi_0\leq \varphi\leq \wmax\varphi_0$ with $\wmax\ll E_J/h\flc$ then, for any functions  $f$ and $g$,  
\be 
\trnorm{[U, f(M)] g(M)\rho_0} \lesssim  \sqrt{\frac{\varepsilon_0 t }{ \hbar}}e^{-\frac{E_J}{k_{\rm B}T}}\wmax\norm{g}\norm{f} 
\label{eqa:ufm_result}
\ee 
where 
$\norm{g}=\max_{|w|\leq \wmax }|g(w)|$,
 $U \equiv e^{-iH_{\rm s}t} $ denotes the evolution operator of the full system in the lab frame throughout the stabilizer segment. 
Here and below, the notation  $x \lesssim y$  indicates that $x$ is smaller than or equal to $y$, up to an $\mathcal O(1)$ prefactor; in particular, $x\lesssim y$ allows $x$ to be much smaller than $y$. 
{The initial confinement $|\varphi|\leq \varphi_0 \wmax$ is assumed for the initialization (see beginning of Sec.~\ref{sec:Stabilization}), and remains justified during subsequent protocol cycles, where the flux probability distribution  has a  Gaussian envelope of width $\kth \varphi_0$, where $\kth = \sqrt{\coth(2\varepsilon_0/k_{\rm B} T)}/\pi\lambda_0$ (see Appendix~\ref{app:envelope}), implying we can set $\wmax \sim \kth$ after the first protocol cycle~\cite{FluxTruncation}.}
In particular, recall that $\lambda_0^{-1}\sim (E_J/h\flc)^{1/4}$, while we require $E_J \gg h\flc$, such that $(E_J/h\flc)^{1/4}$, and hence also  $\kappa$ is much smaller than $E_J/h\flc$.

\newcommand{\rhopp}{\rho_{\rm a}}

As a key intermedicate step in establishing Eq.~\eqref{eqa:expcloseresult}, we introduce the {\it ancillary} density matrix $\rhopp$, defined by assigning the phase factor $e^{-i \varepsilon_L t/\hbar m^2}$ to states in well $m$ {\it before}  time  evolving:
\be 
 \rhopp \equiv  UV^\dagger \rho_0 VU^\dagger,
\label{eqa:rhoppdef}\ee 
Below, we will show that $\rhopp$ is  
 close to both $\rhop$ and $\rhot$, thereby establishing Eq.~\eqref{eqa:expcloseresult} via the triangle inequality. 

To bound the distance between $\rhop$ and $\rhopp$, we note that, by definition 
\be 
\rhop \equiv V^\dagger U \rho_0 U^\dagger V.
\ee 
Since spillover of probability support between wells during the time-evolution with $U$ is exponentially suppressed, it makes little difference whether we assign the well-dependent phase factor (through the unitary $V$) {\it before} or {\it after} time-evolving, implying $\rhop \approx \rhopp$.
Specifically, note that  $\rhop-\rhopp = [U,V^\dagger]\rho_0 VU^\dagger + V^\dagger U \rho_0 [V,U^\dagger]$, implying 
$ \trnorm{\rhop-\rhopp}\leq2\trnorm{[U,V^\dagger]\rho_0}
$~\footnote{ 
This holds since $\trnorm{X}=\trnorm{X^\dagger}$, and $\trnorm{\cU X}=\trnorm{X}$ for any unitary $\cU$.} 
Note that $\trnorm{[U,V^\dagger]\rho_0} $ is of the form on the left hand side of Eq.~\eqref{eqa:ufm_result} with $f(x) =e^{-ix^2\varepsilon_L t/\hbar}$ and $g(x) =1$. 
Hence, we find 
\be   
\trnorm{\rhopp- \rhop}\lesssim \wmax \sqrt{\frac{ \varepsilon_0 t}{\hbar }} e^{-\frac{E_J}{k_{\rm B}T}}.
\label{eqa:trho_to_rhop_comparison}\ee 

To bound the distance from $\rhopp$  to $\rhot$,  we take  the time-derivative in  Eq.~\eqref{eqa:rhoppdef}. Using $\partial_t U = - iH_{\rm s}U$, we obtain 
\be 
\partial_t  \rhopp = -\frac{i}{\hbar}  [H_{\rm s},\rhopp]+\frac{i}{\hbar}\varepsilon_L  [U M^2 U^\dagger, \rhopp]
\label{eqa:rhoprime_derivative}
\ee 
We next note that 
$
 U  M^2 U^\dagger\rhopp 
= U M^2 V^\dagger\rho_0 VU^\dagger 
$, 
implying 
\be 
 U  M^2 U^\dagger\rhopp - M^2 \rhopp =  [U, M^2]  V^\dagger \rho_0 V U^\dagger. 
 \label{eqa:um2u}
\ee 
We  use Eq.~\eqref{eqa:ufm_result} to bound the trace norm of the second term above, with $f(x) = x^2$, implying $\norm{f}=\wmax^2 $,  and $g(x)=e^{-ix^2 \varepsilon_L t/\hbar}$, implying $\norm{g}=1$. Doing this, we obtain 
\be 
\trnorm{ [U, M^2]  V^\dagger \rho_0 V U^\dagger} \lesssim  \wmax^3 \sqrt{\frac{\varepsilon_0 t}{\hbar}}e^{-\frac{E_J}{k_{\rm B}T}}.
\label{eqa:umu_bound}
\ee 
Combining Eqs.~\eqref{eqa:um2u} and \eqref{eqa:umu_bound}, we find 
\be 
 \trnorm{[U M^2 U^\dagger, \rhopp] -[M^2,\rhopp] } \lesssim  \wmax^3 \sqrt{\frac{\varepsilon_0 t}{\hbar}}e^{-\frac{E_J}{k_{\rm B}T}}.
\label{eqa:umu_result} \ee 
Using this result in Eq.~\eqref{eqa:rhoprime_derivative}, we arrive at 
\be 
\partial_t  \rhopp = -\frac{i}{\hbar}[H_{\rm s}-M^2 \varepsilon_L,\rhopp] +\delta \dot \rhopp 
\label{eqa:dtrhopp1}
\ee 
where 
$
\trnorm{\delta \dot \rhopp} \lesssim\frac{\varepsilon_L}{\hbar}\sqrt{\frac{\kth^3 \varepsilon_0 t}{\hbar}}e^{-\frac{E_J}{k_{\rm B}T}}
$. 
Comparing with Eq.~\eqref{eq:heff} in the main text, we   recognize $H_{\rm s}-M^2 \varepsilon_L = \Hpz$---this follows when using $\varphi = \bar \varphi + M\varphi_0$. Thus,  
\be 
\partial_t  \rhopp = -\frac{i}{\hbar}[\Hpz,\rhopp] +\delta \dot \rhopp 
\label{eqa:dtrhopp}
\ee 

We finally use the above result to bound the distance between $\rhopp$ and $\rhot \equiv e^{-i \Hpz t}\rho_0e ^{i\Hpz t}$. 
Using $\rhopp(0)=\rho_0$, integrating Eq.~\eqref{eqa:dtrhopp} results in \be 
\rhopp(t) = e^{-i \Hpz t}\rho_0e ^{i\Hpz t} + \int_{0}^t ds e^{-i \Hpz(t-s)}\delta \dot \rhopp(s)e^{i \Hpz(t-s)} ,
\ee 
where we restored the explicit time-dependence. 
Recognizing the first term on the right-hand side above as $\rhot(t)$, we thus find 
\be 
\trnorm{\rhopp(t)-\rhot(t)}\leq \int_0^t ds \trnorm{\delta \dot \rhopp(s)}. 
\ee 
Next, we use  $
\trnorm{\delta \dot \rho} \lesssim\frac{\varepsilon_L}{\hbar } \wmax^3 \sqrt{\frac{ \varepsilon_0 t}{\hbar}}e^{-\frac{E_J}{k_{\rm B}T}}
$, along with $\varepsilon_L \sim \hbar/\trev$.
Thus, within the stabilizer segment ($0\leq t \leq \Ts$), 
\be 
\trnorm{ \rhopp -  \rhot}\lesssim \frac{\Ts\wmax^3 }{\trev}\sqrt{\frac{\varepsilon_0 \Ts}{\hbar}}e^{-\frac{E_J}{k_{\rm B}T}}\label{eqa:trho_to_rhopp_bound}
\ee 
Combining this result with Eq.~\eqref{eqa:trho_to_rhop_comparison}
 and the triangle inequality  $\trnorm{\tilde \rho-\rhop}\leq \trnorm{\tilde\rho-\rhopp}+\trnorm{\rhop-\rhopp}$, and using $\Ts \geq \trev$, it follows that 
\be 
\trnorm{ \rhop - \rhot}\lesssim \frac{\Ts}{\trev}\wmax^3\sqrt{\frac{ \varepsilon_0 \Ts}{\hbar}}e^{-\frac{E_J}{k_{\rm B}T}}\label{eqa:trho_to_rhop_bound}
\ee 
This establishes Eq.~\eqref{eqa:expcloseresult}, which was our goal. 

\section{Confinement in code subspace}
\label{app:speed_limit}
Here we derive the bounds on the spread of $S_2$ support quoted in Eqs.~\eqref{eq:quasicarge_confinement} and \eqref{eq:s2fixedpoint} of the main text. 
To recap, our goal is to bound  the probability support of $ \tilde\rho(t)$ in the domain $|S_2| > s$ as a function of $s$ and time $t$,  
\be 
\tilde{\cP}_{2}(s,t)\equiv \Tr[\theta(s-S_2)\rhot(t)].
\label{eqa:p2_def}
 \ee
where $\rhot(t)\equiv e^{-i\Hpz t}\rho(0)e^{i\Hpz t}$ denotes the evolution generated by the effective Hamiltonian $\Hpz$, and coincides with state of the system in the  comoving frame, up to an exponentially small correction (bounded in the previous Appendix). 

\subsection{Derivation of Eq.~\eqref{eq:quasicarge_confinement}}

We first derive the bound in Eq.~\eqref{eq:quasicarge_confinement}. 
To this end, for a given $k>0$, we introduce the  operator 
\be 
w_k \equiv e^{-k S_2}.
\ee 
The positive semidefiniteness of  $w_k$ and $\theta(s-S_2)$ implies that $w_k \geq e^{-ks}\theta(s-S_2)$ for any $s$. 
Thus,  for any $k>0$ and $s\in [-1,1]$, 
\be 
\tilde{\cP}_2(s,t) \leq   e^{ k s }\langle w_k(t)\rangle, 
\label{eqa:pq_1}
\ee 
 where we introduced the shorthand  $\langle w_k(t)\rangle \equiv \Tr[w_k \tilde \rho(t)]$. 
We now  bound  $\langle w_k(t)\rangle$ by considering its  equation of motion,
\be 
\partial_t \langle w_k(t)\rangle = \frac{i}{\hbar}\Tr\left(\rhot(t)[\Hpz,w_k]\right).
\ee 
We recall from Eq.~\eqref{eq:heff} that  $\Hpz = \Hpz_0+M\varphi_0\bar \varphi/L$ with $\Hpz_0$ defined in Eq.~\eqref{eq:hpz0def}, $\bar \varphi \equiv \varphi \mod \varphi_0$ denoting the quasiflux operator, and $M=\nint(\varphi/\varphi_0)$ denoting the integer-valued well index-operator. Using $[\Hpz_0,S_2] =[\bar \varphi,S_2]=0$, and $[M,S_2]=-2i\sin(2\pi q/e)$, we thus obtain 
\be 
[\Hpz,w_k] =  -2ik  \frac{\bar \varphi \varphi_0}{L} \sin\left(\frac{2\pi q }{e}\right)w_k. 
\label{eqa:wk_eom2}
\ee 
Using  that $[\bar \varphi,e^{-i2\pi q/e}]=0$, we thus find 
\be 
\partial_t \langle w_k(t)\rangle = - \frac{2k\varphi_0}{L\hbar } \Tr\left(\bar \varphi \sin\Bigg[\frac{2\pi q}{e}\Bigg] e ^{-\frac{kS_2}{2}} \tilde \rho(t) e^{-\frac{kS_2}{2}}\right)
\label{eqa:wk_expr}
\ee 
Note that $e ^{-kS_2/2} \tilde \rho(t) e^{-kS_2/2}$ is a positive semidefinite matrix with trace $\langle w_k(t)\rangle$; we can thus view it as a rescaling of the density matrix $\rho_k(t)$, defined in Eq.~\eqref{eq:velocity_def} in the maint text, i.e.,: 
\be 
\rho_k(t) \equiv\frac{1}{\langle w_k(t)\rangle} e ^{-\frac{kS_2}{2}} \rhot(t)e ^{-\frac{kS_2}{2}}.
\label{eqa:rho_k_def}
\ee 
%
Combining Eqs.~\eqref{eqa:wk_expr}-\eqref{eqa:rho_k_def} we thus find 
\be 
\partial_t \langle  w_k(t)\rangle =  - k v_k(t) \langle w_k(t)\rangle.
\label{eqa:vk_eom}
\ee 
where 
\be 
v_k(t) \equiv  \frac{2\varphi_0}{\hbar L} \Tr\left[\rho_k(t)\bar \varphi \sin\left(\frac{2\pi q}{e}\right)\right]. 
\label{eqa:vkdef}
\ee 
Formally integrating Eq.~\eqref{eqa:vk_eom} yields 
\be 
\langle w_k(t)\rangle = e^{-k\int_0^t ds v_k(s)}\langle w_k(0)\rangle 
\label{eqa:wk_evolution}
\ee 
implying 
 \be 
\tilde \cP_{2}(s,t) \leq   e^{k\left[s+\Delta s_k(t)\right]}\langle w_k(0)\rangle.
\label{eqa:lieb_robinson_bound_0}
\ee 
with $\Delta s_k (t) = -k\int_0^t dt' v_k(t')$. 

We now recall from Sec.~\ref{sec:Stabilization} that we assumed the $S_2$ support of $\rho(0)$ [and hence also $\rhot(0)$] to be confined to the region $S_2\geq  s_0$. 
In this case,  $\langle w_k(0)\rangle \leq e^{-ks_0}$, implying that, for any $k\geq 0$, 
 \be 
\tilde\cP_{2}(s,t) \leq   e^{-k\left[s_0-s-\Delta s_k(t)\right]}.
\label{eqa:lieb_robinson_bound_1}
\ee 
Combining Eqs.~\eqref{eqa:rho_k_def},~\eqref{eqa:vkdef},~and~\eqref{eqa:lieb_robinson_bound_1}, this establishes  Eq.~\eqref{eq:quasicarge_confinement} of the main text, which is the goal of this subsection.

\subsection{Derivation of Eq.~\eqref{eq:s2fixedpoint}}
Here we derive Eq.~\eqref{eq:s2fixedpoint} of the main text.
We establish Eq.~\eqref{eq:s2fixedpoint}  from Eq.~\eqref{eqa:lieb_robinson_bound_0} by fixing $k=k_0$, where $k_0 = 1/2\pi \lambda^2$~\footnote{The choice $k=k_0$ is chosen because it results in simple expressions for the bounds we obtain. We emphasize that sharper bounds can possibly be obtained for different values of $k$}, and  computing $\langle w_{k_0}(0)\rangle$ for a state stablized by the protocol. 
To this end, we note that 
\be 
\langle w_{k_0}(0)\rangle = e  \int_{-\infty}^\infty \!\! dx\, p_q(xe)e^{-k_0\cos(2\pi x )}
\label{eqa:wk0expr}\ee 
with  $p_q(x) \equiv \Tr[\delta(q-x)\tilde \rho(0)]$ the charge probability distribution of the system at  the onset of the stabilizer segment. 
Since a free segment maps $\varphi/\varphi_0$ to $q/e$, $p_q(x)$ is given by the flux probability distribution at the end of the previous stabilizer segment with this rescaling. 
Using the flux distribution from Eq.~\eqref{eq:flux_distribution_stabseg}, we thus find 
\be 
p_q(x)\approx  \frac{e^{-x^2 /\lth^2 e^2}}{\sqrt{2\pi}e\lth}.
\ee 

Using the result above, we  compute Eq.~\eqref{eqa:wk0expr} via the saddle point approximation, which is valid when  $\lth\ll 1$ (and hence also $k_0\gg 1$). 
For $k=k_0$, the exponent of the integrand in Eq.~\eqref{eqa:wk0expr} attains its maximum around $x=0$, where it scales as $-\frac{1}{2\pi^2 \lambda^2 } (1+\frac{2}{3}\pi^4 x^4)$.
In the regime $\lth\ll 1$, this integral is dominated by the contribution near $x=0$, and thus 
\be 
\langle w_{k_0}(0)\rangle \approx \frac{1}{\sqrt{2\pi }\lambda}\int_{-\infty}^\infty dx e^{-\frac{1}{2\pi^2 \lambda^2}\left(1+\frac{2}{3}\pi^4 x^4\right)}.
\ee 
Evaluating the integral, we find 
\be 
\langle w_{k_0}(0)\rangle \lesssim \frac{0.5}{\sqrt{\lth}}\,e^{-\frac{1}{2\pi^2\lth}}
\ee 
Inserting this in Eq.~\eqref{eqa:lieb_robinson_bound_0}, we conclude  
 \be 
\tilde\cP_{2}(s,t) \lesssim     \frac{0.5}{\sqrt{\lth}}\exp\left[\frac{-1+s+\Delta s_{k_0}(t)}{2\pi^2 \lambda^2 }\right].
\label{eqa:lieb_robinson_bound_11}
\ee 
In the limit $\Delta s_{k_0}(t)\ll 1 $, and suppressing the subdominant power-law prefactors, we recover $\cP_{2}(s,t) \lesssim    \exp\left[\frac{-1+s}{2\pi^2 \lambda^2 }\right]$. This establishes Eq.~\eqref{eq:s2fixedpoint}, which was the goal of this subsection.

\section{Bound on logical error rate}
\label{app:error_rate_bound}
Here we establish the bound for the logical error rate quoted in  Eq.~\eqref{eq:logical_er_result} of the main text. 
\newcommand{\dts}{\delta\tilde{\sigma}}
To recap, we are interested in bounding the change of the logical expectation values in the evolution generated by $\Hpz$, after the system has converged to  the fixed point in the code subspace (See Sec.~\ref{sec:fixed_point}). 
I.e., we seek to bound $|\dts_i|$, where 
\be 
\dts_i  \equiv \Tr[\rhot(\Ts)\sigma_i]-\Tr[\rhot(0)\sigma_i].
\ee
and $\rhot(0)$ is the density matrix at the onset of the stabilizer segment, after the system has converged to the fixed point of the protocol, as described in Sec.~\ref{sec:fixed_point}.

Below, we establish bounds on $\dts_x,\dts_y$, and $\dts_z$ separately. The derivation is quite independent for each logical operator, and, for $\dts_x$ and $\dts_y$, involve multiple triangle inequalities; we therefore devote a separate subsection to each logical operator, in the order $\dts_z$ (Sec.~\ref{seca:sz}), $\dts_x$ (Sec.~\ref{seca:sx}), and $\dts_y$  (Sec.~\ref{seca:sy}). 
From these derivations, we obtain the following inequalities
\begin{eqnarray}
|\dts_x| & \lesssim &   \frac{\Ts}{\lth^{\frac{3}{2}} \trev}e^{-\frac{1}{\pi^2 \lth^2}}, \label{eqa:sigma_x_bound}
\\ 
|\dts_y|  & \lesssim & \sqrt{ \frac{\varepsilon_0 \Ts}{\hbar}}\frac{\kth^3\Ts ^2}{\sqrt{\lambda}\trev^2} e^{-\frac{E_J}{k_{\rm B}T}}+\frac{\Ts }{\lambda^{\frac{3}{2}}\trev}e^{-\frac{1}{\pi^2\lambda^2}}, 
\label{eqa:sigma_y_bound}
\\ 
|\dts_z| &\lesssim &\sqrt{\frac{ \varepsilon_0 \Ts}{\hbar}}\frac{\kappa^3\Ts}{\trev}e^{-\frac{E_J}{k_{\rm B}T}}
\label{eqa:sigma_z_bound}
\end{eqnarray}
where $\kappa = \sqrt{\coth(2\varepsilon_0/k_{\rm B}T)}/\pi\lambda_0$ denotes the dimensionless GKP envelope width, and $\varepsilon_0 = \sqrt{4\pi h\flc  E_J}$ denotes the characteristic excitation energies in the wells of the cosine potential from the Josephson junction.
As in the main text $x\lesssim y$ indicates that $x$ is bounded from above by  $y$, up to multiplication by some $\mathcal O(1)$ constant.
In particular, we emphasize that $x\lesssim y$ allows $x$ to be arbitrarily smaller than $y$.

Using that we work in the regime $\lambda \ll 1$, $\Ts \geq \trev$, and $ \lambda^2 \leq  \lambda_0^2 + k_{\rm B }T/\pi^2 E_J $ (see Appendix~\ref{app:lambdaineq}), the results above thus imply 
\be 
|\dts_i |\lesssim  \frac{\Ts^2\kappa^3}{\trev^2\lth}\sqrt{\frac{   \varepsilon_0 \Ts}{\hbar}}  \exp\left[-\frac{1}{k_{\rm B}T/E_J+\pi^2 \lambda_0^2}\right]\label{eqa:logical_error_rate_bound}
\ee 
Suppressing the power-law prefactors, this establishes Eq.~\eqref{eq:error_rate_result} in the main text.

\subsection{Bound on   $\dts_z$} 
\label{seca:sz}
We first establish the bound  for $\dts_z$ in Eq.~\eqref{eqa:sigma_z_bound}.
The derivation is structured as follows: first, we show that $\dts_z$ is exponentially close to the change of the expectation of  $\sigma_z$ in the {\it lab frame} over the stabilizer segment. Secondly, we show that the latter---giving the net flux of probability support between the wells of the Josephson potential---is bounded via an Arrhenius law. 

To relate $\dts_z$ to the change of $\sigma_z$ in the lab frame, we recall that $\trnorm{\rhop(t)-\rhot(t)}\lesssim \frac{\kappa^3\Ts}{\trev}\sqrt{{ \varepsilon_0 \Ts}/{\hbar}}e^{-\frac{E_J}{k_{\rm B}T}}$ [Eq.~\eqref{eqa:trho_to_rhop_bound}], with $\rhop(t)\equiv V^\dagger(t) \rho(t)V(t)$ and $V(t)$ denoting the transformation to the comoving frame introduced in Eq.~\eqref{eq:cmftransf}. 
Next, we note that $V$ commutes with $\sigma_z$, implying that  $\Tr[\rhop(t)\sigma_z]= \Tr[\rho(t)\sigma_z]$. 
We now 
\be 
|\Tr[AB]|\leq \trnorm{A}\norm{B},\label{eqa:trnorm_ineq}
\ee 
where, for any operator $A$,  $\trnorm{A}$ and $\norm{A}$ denote its trace and singular value norms, respectively, $\trnorm{A} \equiv \Tr[\sqrt{A^\dagger A}]$ and $\norm{A}\equiv \sup_{\psi,\phi}{|\langle \psi|A|\phi\rangle|}/{\sqrt{|\langle \psi|\psi\rangle\langle \phi|\phi\rangle|}}$. 
Using $\norm{\sigma_z}=1$, we thus find
\be 
|\Tr[\rho(t)\sigma_z]-\Tr[\rhot(t)\sigma_z]|\lesssim  \frac{\kappa^3\Ts}{\trev}\sqrt{\frac{ \varepsilon_0 \Ts}{\hbar}}e^{-\frac{E_J}{k_{\rm B}T}}
\label{eqa:szl-szt}
\ee

We next note that $\Tr[\rho(t)\sigma_z]$ gives the net imbalance of probability weight between  the  even and odd wells of the cosine potential from the Josephson junction in the state $\rho$. 
The flow of probability support between the wells is suppressed via the   Arrhenius law, i.e, exponentially small  in $2\frac{E_J}{k_{\rm B}T}$.
For example, Ref.~\cite{Ankerhold_1995} showed that  the flow rate of probability support between even and odd wells is  of order $\sim \frac{\varepsilon_0}{\hbar} e^{-2\frac{E_J}{k_{\rm B}T}}$, leading to 
\be 
\partial_t \Tr[\rho\sigma_z] \sim \frac{\varepsilon_0}{\hbar } e^{-\frac{2E_J}{k_{\rm B}T}}.
\ee 
To relate e.g., using the triangle inquality, $|f(t)-f(0)|\leq \int_0^t dt' |\partial_t f(t')|$, it follows that 
\be 
|\Tr[\rho(t)\sigma_z]-\Tr[\rho(0)\sigma_z]|\lesssim\frac{\varepsilon_0 t}{\hbar} e^{-\frac{2E_J}{k_{\rm B}T}}.
\label{eqa:zlabframechange}
\ee 
By combining the above result with Eq.~\eqref{eqa:szl-szt} and using the triangle inequality, we thus obtain 
\be 
    |\dts_z|\lesssim \frac{\kappa^3\Ts}{\trev}\sqrt{\frac{ \varepsilon_0 \Ts}{\hbar}}e^{-\frac{E_J}{k_{\rm B}T}}
\label{eqa:szl-szt_2}
\ee 
which was what we wanted to show 

\subsection{Bound on $\dts_x$}
\label{seca:sx}
\begin{figure}
    \centering
    \includegraphics[width=0.99\columnwidth]{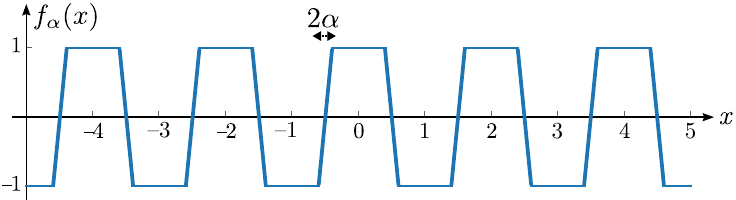}
    \caption{Plot of the function $\fr(x)$}
    \label{figa:falpha}
\end{figure}
We next establish the bound on $|\dts_x|$ in Eq.~\eqref{eqa:sigma_x_bound}.
The derivation has  2 steps:
Firstly, we show that 
$\Tr[\rhot(t)\sigma_x]\approx \Tr[\rhot(t)\tau_x(t)]$, for a {\it regularized} logical operator 
\be 
    \tau_x \equiv \fr(q/e),
\label{eqa:sigmap_def}\ee 
where $\fr(x)$ (plotted in Fig.~\ref{figa:falpha}) is given by 
\be
\fr (x) = \left\{\begin{array}{ll}
    1, \quad & |x|<1/2-\alpha\\ 
    (1/2-|x|)/\alpha, \quad  &|x|>1/2-\alpha 
\end{array}\right.,
\label{eqa:falpha_def}
\ee 
where $\alpha=\lambda/\sqrt{2}$ and $\fr(x)=-\fr(x+1)$ defines  $\fr(x)$ for all other values of $x$. 
The function $f_{\rm r}(x)$  can be viewed as a  continuous generalization of the crennelation function $\Xi(x)$. 
Specifically, in Sec.~\ref{seca:sigmaxp_def}, we show that, for states stabilized by the protocol,
\be 
|\Tr[\rhot(t) {\sigma}_x] -\Tr[\rhot(t) {\tau}_x] | \lesssim  \frac{1}{\sqrt{\lambda}} e^{-\frac{1}{\pi^2 \lth^2}}.
\label{eqa:stospbound}\ee 
Secondly, in Sec.~\ref{seca:sigmaxp_bound}, we  show that $\Tr[\rhot(t) {\tau}_x] $ is  near-constant throughout the stabilizer segment:
\be 
|  \Tr[\rhot(t) \tau_x]- \Tr[\rhot(0)\tau_x] |\lesssim    \frac{\Ts}{\lth^{\frac{3}{2}} \trev}e^{-\frac{1}{\pi^2  \lth^2}}. 
\label{eqa:sigmap_bound}\ee 
Eq.~\eqref{eqa:sigma_x_bound} follows when combining these two results along with $\lambda \ll 1$, $\Ts\geq \trev$ and  using the triangle inequality.

We now proceed to derive Eqs.~\eqref{eqa:stospbound} and \eqref{eqa:sigmap_bound}.
Throughout the remainder of this section will suppress time-dependence of all quantities, unless otherwise noted. 

\subsubsection{Derivation of Eq.~\eqref{eqa:stospbound}} 

\label{seca:sigmaxp_def}
To establish Eq.~\eqref{eqa:stospbound} we first note that $\fr(x) = \Xi(x)$ for $|x|<1/2 - \alpha$. 
As a result,  $\sigma_x-\tau_x$ only has support in the region 
$S_2 \leq s_\alpha$, with $s_\alpha \equiv -\cos(2\pi \alpha)$.
Hence, $\sigma_x-\tau_x =  P_2(s_\alpha)(\sigma_x-\tau_x)  P_2(s_\alpha)$, where $ P_2(s)\equiv \theta(s-S_2)$, implying  
\be 
|\Tr[\rhot  \tau_x] - \Tr[\rhot  \sigma_x]| = |\Tr[ P_2(s_\alpha)\rhot P_2(s_\alpha)(\sigma_x-\tau_x)]|.
\label{eq:d8}\ee 
Using $|\Tr[AB]|\leq \trnorm{A}\norm{B}$, we find 
\be 
|\Tr[\rhot\tau_x] - \Tr[\rhot  \sigma_x]|\leq \trnorm{ P_2(s_\alpha)\rhot P_2(s_\alpha)}\norm{\sigma_x-\tau_x}.
\label{eq:d81}\ee 
Since $\sigma_x-\tau_x = \Xi(q/e)-\fr(q/e)$, the eigenvalues of $\sigma_x-\tau_x$ are bounded by $1$, implying  $\norm{\sigma_x-\tau_x} \leq  1$.
Moreover, since $ P_2(s_\alpha)\rhot P_2(s_\alpha)$ is  positive semidefinite,  $\trnorm{A}=\Tr[A]$ for positive semidefinite $A$, and $P^2_2(s)=P_2(s)$, we find $\trnorm{ P_2(s_\alpha)\tilde \rho P_2(s_\alpha)}=\Tr[\tilde \rho  P_2(s_\alpha)]$. 
Recognizing $\Tr[\tilde \rho  P_2(s_\alpha)]={\cP}_{2}(s_\alpha)$, with ${\cP}_{2}(s_\alpha)$ defined above Eq.~\eqref{eq:quasicarge_confinement} of the main text,  we conclude that 
\be 
|\Tr[\rhot\tau_x] - \Tr[\rhot  \sigma_x]| \leq  {\cP}_{2}(s_\alpha )\,. 
\label{eqa:sigmap_relation}
\ee

Next, we recall from Eq.~\eqref{eqa:lieb_robinson_bound_11}, that, for states stabilized by the protocol, and with $\Delta s_{k_0}(t)\ll 1$
\be 
\cP_2(s_\alpha) \lesssim \frac{1}{\sqrt{\lambda}} e^{-\frac{1-s_\alpha }{2\pi^2  \lth^2}}.
\label{eqa:cp2_bound}\ee 
with $s_\alpha \equiv -\cos(2\pi \alpha)$, and with $\mathcal O(1)$ prefactors suppressed.
Since 
$ 
\alpha = \lth/\sqrt{2\pi}
$  
 and we work in the regime $\lth \ll 1$,  we find  $s_\alpha \approx -1 + \pi  \lth^2$, implying 
\be 
{\cP}_2(s_\alpha) \lesssim \frac{1}{\sqrt{\lambda}} e^{-\frac{1}{\pi^2 \lth^2}}. 
\label{eqa:alpha_choice_result}\ee 
where we again suppressed $\mathcal O(1)$ prefactors.
Using this in Eq.~\eqref{eqa:sigmap_relation}, we recover Eq.~\eqref{eqa:stospbound}, which was our goal.

\subsubsection{Derivation of Eq.~\eqref{eqa:sigmap_bound}} 
\label{seca:sigmaxp_bound}
To establish Eq.~\eqref{eqa:sigmap_bound}, we consider the equation of motion for $\Tr[\tau_x\rhot]$, which can be written 
\be 
\partial_t \Tr[\rhot\tau_x] 
= -\frac{i}{\hbar}\Tr \left([\tau_x,\Hpz]\tilde \rho\right).
\label{eqa:sigmap_eom}\ee 
where $\Hpz$ is given in Eq.~\eqref{eq:heff} of the main text. 
To evaluate the commutator , we recall $\Hpz=\Hpz_0 +M\varphi_0 \bar \varphi/L$, where $M=\nint(\varphi/\varphi_0)$, $\bar \varphi=\varphi-M\varphi_0$, while $\bar H$---defined in Eq.~\eqref{eq:hpz0def} of the main text---is exclusively a function of $\bar \varphi$, $q$, $H_{\rm R}$, and $Q_{\rm R}$.
Since $\tau_x$ is a $2e$-periodic function of $q$, it commutes with any $\varphi_0$-periodic function of $\varphi$, i.e.,  operators that can be written as functions of $\bar \varphi$. 
As a result $[\tau_x,\bar \varphi]=[\tau_x,\Hpz_0]=0$, implying 
$ [\tau_x,\Hpz]=[\tau_x,M] \varphi_0\bar \varphi/L.
$
We next note that $[\tau_x,M]=\frac{1}{\varphi_0}[\tau_x,\varphi-\bar \varphi]$, implying $[\tau_x,M]=\frac{1}{\varphi_0}[\tau_x,\varphi]$. 
Since $[q,\varphi]=-i\hbar $, we thus  conclude 
\be 
[\tau_x,M] = -i \frac{\hbar}{e} \fr'(q/e)
\label{eqa:mcomm}
\ee 
Thus we find~\footnote{To see that the right-hand side  is Hermitian, note that the $2e$  periodicity of $f'_\alpha(q/e)$ implies  $[\bar \varphi , f'_\alpha(q/e)]=0$} 
\be 
[\tau_x,\Hpz]  = i\frac{\bar \varphi \hbar }{ Le}  f'_\alpha(q/e).
 \label{eqa:hsigmaprime_expr}
\ee

We now use that 
\be 
f'_\alpha(x) = \frac{X}{\alpha} 
\label{eqa:fprime_expression}
,\quad X \equiv \sum_z \theta(\alpha-|q/e-z-1/2|)(-1)^z.
\ee 
Using this together with Eq.~\eqref{eqa:sigmap_eom}, we find 
\be 
\partial_t \Tr[\rhot \tau_x]  = \frac{1}{ eL\alpha }\Tr[\tilde \rho  \bar \varphi X ] 
\ee
Note that $X$ only has support in the subspace with $S_2 <s_\alpha$. Thus $X= P_2(s_\alpha) X P_2(s_\alpha)$. We also recall $[\bar \varphi,S_2]=0$~\cite{S2comm}, implying  $[\bar \varphi,P_2(s)]=0$. As a result,  $\Tr[\tilde \rho  \bar \varphi X ] = \Tr[P_2(s_\alpha)\tilde \rho P_2(s_\alpha)\bar \varphi X]$. 
Using  $|\Tr[AB]|\leq \trnorm{A}\norm{B}$  along with  $\trnorm{P_2(s_\alpha)\tilde \rho P_2(s_\alpha)}=\tilde \cP_2(s_\alpha)$ and $\norm{\bar \varphi X}\leq \norm{\bar\varphi}\norm{X}=\varphi_0/2$, we conclude 
\be 
|\partial_t \Tr[\rhot \tau_x] |  \leq    \frac{\varphi_0}{2 e L \alpha}\tilde\cP_2(s_\alpha) . 
\label{eqa:sigmaxp123}\ee

To simplify the prefactor, we finally use that $\varphi_0/e = h/2e^2 $, which coincides with the resonator impedance $\sqrt{L/C}$. 
Also using $\sqrt{LC}=\trev$, we find that $L = \trev \varphi_0/e $.
Thus, we find 
\be 
|\partial_t \Tr[\rhot \tau_x] |  \leq    \frac{ \cP_{2}(s_\alpha)}{\alpha\trev } .
\label{eqa:sigmap_bound1}
\ee
Recalling that $\alpha= \lambda/\sqrt{2\pi}$, and using Eq.~\eqref{eqa:alpha_choice_result}, we thus find 
\be 
|\partial_t \Tr[\rhot \tau_x] |\lesssim      \frac{1}{\lambda^{\frac{3}{2}} \trev } e^{-\frac{1}{ \pi^2  \lth^2}}.  
\ee
where we also used $\sqrt{2\pi }\sim 1$. 
We obtain Eq.~\eqref{eqa:sigmap_bound} by integrating over time,  using the triangle inequality. This  concludes this subsection. 
\subsection{Bound on $\dts_y$}
\label{seca:sy}
\newcommand{\cY}{\mathcal{Y}}
We finally establish the bound on $\dts_y(t)$ in Eq.~\eqref{eqa:sigma_y_bound}. 
The derivation proceeds in 2 steps: 
First, we introduce a scalar quantity $\cY(t)$, which we show is approximately identical to $\Tr[\tilde\rho(t)\sigma_y]$ (Sec.~\ref{seca:y_dist}).
Specifically, 
\be 
\cY(t) \equiv    i\Tr[\tau_x U(t) V^\dagger(t) \sigma_z \rho(0) V(t) U^\dagger(t)] 
\label{eqa:ydef}
\ee 
where $U(t)=e^{-iH_{\rm s}t}$ is the evolution operator in the lab frame, and $V(t)$ is  generates the transformation to the comoving frame [Eq.~\eqref{eq:cmftransf}]. 
In Sec.~\ref{seca:y_dist}, we show that 
\begin{equation}
|\cY(t)-\Tr[\rhot(t) \sigma_y]|\lesssim \sqrt{\frac{\varepsilon_0 t}{\hbar}} \frac{\kth^3\Ts e^{-\frac{E_J}{k_{\rm B}T}}}{\trev} +\frac{e^{-\frac{1}{\pi^2\lambda^2}}}{\sqrt{\lambda}}.
\label{eqa:ybound0}
\end{equation} 

Secondly, in Sec.~\ref{seca:ychange}, we show that $\cY(t)$ is near-constant over the stabilizer segment: 
\begin{equation}
|\cY(\Ts)-\cY(0)| \lesssim\sqrt{ \frac{\varepsilon_0 \Ts}{\hbar}}\left[\frac{\kth^3\Ts ^2 e^{-\frac{E_J}{k_{\rm B}T}}}{\sqrt{\lambda}\trev^2}+\frac{\Ts e^{-\frac{1}{\pi^2\lambda^2}}}{\lambda^{\frac{3}{2}}\trev}\right].
\label{eqa:ychangebound}
\end{equation}
Combining these two results via the triangle inequality, and using $ \Ts\geq \trev,t$  and $\lambda \ll 1$, we establish the bound on $\dts_y$ quoted in Eq.~\eqref{eqa:sigma_y_bound}. 

The remainder of this subsection is devoted to deriving Eqs.~\eqref{eqa:ybound0} (Sec.~\ref{seca:y_dist}) and Eq.~\eqref{eqa:ychangebound} (\ref{seca:ychange}). In this discussion, we suppress time dependence of all quantities for brevity. 
\subsubsection{Derivation of Eq.~\eqref{eqa:ybound0}} 
\label{seca:y_dist}
To establish Eq.~\eqref{eqa:ybound0}, we first use the triangle inequality along with $\sigma_y = i\sigma_x\sigma_z$ to write 
\be 
|\cY-\Tr[\tilde \rho \sigma_y]|\leq |\cY-i\Tr[\tilde \rho \tau_x\sigma_z]|+|i\Tr[\tilde \rho (\tau_x\sigma_z- \sigma_x\sigma_z)|
\label{eqa:ytriangle}\ee 
We now bound the two terms on the right hand side separately. 

To bound $|\cY-i\Tr[\tilde \rho \tau_x\sigma_z]|$, we  note that $Y-i\Tr[\tilde \rho \tau_x\sigma_z] =i\Tr[\tau_x (Z-\sigma_z \rhot)]$, where 
\be 
Z\equiv U V^\dagger  \sigma_z\rho_0 V U^\dagger.
\label{eqa:y_in_terms_of_s}
\ee 
This follows from Eq.~\eqref{eqa:ydef}.
Using $|\Tr[AB]|\leq \trnorm{A}\norm{B}$ along with $\norm{\tau_x}=1$, 
we find 
\be 
|\cY-\Tr[\tilde \rho \sigma_y]|\leq \trnorm{Z-\sigma_z  \rhot}.
\label{eqa:abcd}\ee 
To bound the right-hand side, we introduce 
 the ancillary density matrix $\rhopp \equiv  UV^\dagger \rho_0 V U^\dagger$ that was also considered   in  Appendix~\ref{seca:heff_derivation} [see Eq.~\eqref{eqa:rhoppdef}].
Using the triangle inequality, we find 
\be 
\trnorm{Z-\sigma_z  \rhot} \leq \trnorm{Z-\sigma_z\rhopp}+\trnorm{\sigma_z\rhopp-\sigma_z \rhot}
\label{eqa:trianc}\ee 

We bound the second term on the right hand side above using $\trnorm{AB}\leq\norm{A}\trnorm{B}$~\footnote{To see this, note that $\trnorm{AB}=\Tr[\cU AB]$ for some unitary $\cU$ that generates the polar decomposition of $AB$. The result follows uing $|\Tr(AB)|\leq \norm{A}\trnorm{B}$} and $\norm{\sigma_z}=1$, implying 
$\trnorm{\sigma_z\rhopp-\sigma_z \rhot}\leq\trnorm{\rhopp-\rhot}$.
We now recall from Eq.~\eqref{eqa:trho_to_rhopp_bound} that 
\be 
\trnorm{ \rhopp -  \rhot}\lesssim \frac{\Ts}{\trev}\wmax^3 \sqrt{\frac{\varepsilon_0 \Ts}{\hbar}}e^{-\frac{E_J}{k_{\rm B}T}}\label{eqa:trho_to_rhopp_bound2}
\ee 
where $\wmax$ denotes the $\varphi$ support range of $\rho_0$  in units of $\varphi_0$, such that  $\rho_0$ has $\varphi$ support  confined within the interval $[-\wmax\varphi_0,\wmax \varphi_0]$. Since we consider the dynamics of stabilized states, the flux distribution of $\rho_0$ has a Gaussian envelope of characteristic width $\kappa$ (see Sec.~\ref{sec:fixed_point} of the main text), implying we may set $\wmax \sim \kappa$~\cite{FluxTruncation}.
Using again that  $\trnorm{AB}\leq \norm{A}\trnorm{B}$ and $\norm{\sigma_z}=1$, we find 
\be 
\trnorm{ \sigma_z \rhot  -  \sigma_z \rhopp }\lesssim \frac{\Ts\kth^3 }{\trev}\sqrt{\frac{\varepsilon_0 \Ts}{\hbar}}e^{-\frac{E_J}{k_{\rm B}T}}\label{eqa:trho_to_rhopp_bound3}
\ee 

To bound the first term on the right hand side of Eq.~\eqref{eqa:trianc}, we use the definition of $Z$ in   Eq.~\eqref{eqa:y_in_terms_of_s} to write 
\be 
\trnorm{ Z- \sigma_z \rhopp} = \trnorm{[\sigma_z,U]V^\dagger \rho_0 }
\label{eqa:s_rhop_bound}
\ee 
Here we used that $\trnorm{\cU X}=\trnorm{X}$ for any unitary operator  $\cU$, and $[V^\dagger,\sigma_z]=0$. 
We next use the inequality in Eq.~\eqref{eqa:ufm_result}, noting that  the right-hand side above is of the same form as the left hand side  in Eq.~\eqref{eqa:ufm_result}, with   $g(x)=e^{-ix^2\varepsilon_L t/\hbar} $ and $f(x) = (-1)^x$.  
Using that $\norm{f}=\norm{g}=1$,  we find 
\be 
\trnorm{ S- \sigma_z \rhopp}\lesssim 
\kth \sqrt{\frac{\varepsilon_0 t}{\hbar}}e^{-\frac{E_J}{k_{\rm B}T}
}.
\label{eqa:s_bound1}
\ee 
Combining this with Eq.~\eqref{eqa:abcd}~and~\eqref{eqa:trho_to_rhopp_bound3}, we obtain 
\be |\cY-  i\Tr[\tilde \rho \tau_x\sigma_z]|\lesssim \frac{\Ts\kth^3}{\trev} \sqrt{\frac{\varepsilon_0 \Ts}{\hbar}}e^{-\frac{E_J}{k_{\rm B}T}}
\label{eqa:firsttermbound}
\ee
This bounds the first term on the right hand side of Eq.~\eqref{eqa:ytriangle}. 

We next  bound the second term on the right hand side of Eq.~\eqref{eqa:ytriangle}, $|\Tr[\tilde \rho (\tau_x\sigma_z - \sigma_x\sigma_z)|$. We first  recall that  $\sigma_x-\tau_x = P_2(s_\alpha)(\sigma_x-\tau_x)  P_2(s_\alpha)$, where $ P_2(s)\equiv \theta(s-S_2)$  [see text above  Eq.~\eqref{eq:d8}].
Thus 
\be 
|Tr[(\tau_x-\sigma_x)\sigma_z \tilde \rho]| =  |\Tr[\tilde \rho P_2(s_\alpha) (\sigma_x-\tau_x)P_2(s_\alpha)\sigma_z]|. 
\ee 
Note  that $P_2(s)$ is a function of $S_2$, and hence, $[P_2(s),\sigma_z]=0$.
Using this fact, along with $|\Tr(AB)|\leq \trnorm{A}\norm{B}$,  $\norm{\sigma_z (\sigma_x-\tau_x)}\leq 1$, and $\trnorm{ P_2(s_\alpha)\tilde \rho  P_2(s_\alpha)}=\cP_2(s_\alpha)$, we find 
\be 
|\Tr[(\tau_x-\sigma_x)\sigma_z \tilde \rho]|\leq  \cP_2(s_\alpha). 
\ee 
Since $\alpha = \lth/\sqrt{2\pi}$ and  $\cP_2(s_\alpha)\lesssim\frac{1}{\sqrt{\lth}} e^{-\frac{1}{\pi^2\lth^2}}$ [Eq.~\eqref{eqa:alpha_choice_result}], we find  
\be 
|\Tr[(\tau_x-\sigma_x)\sigma_z \tilde \rho]|\lesssim\frac{1}{\sqrt{\lth}} e^{-\frac{1}{\pi^2\lth^2}}. 
\label{eqa:ybound3}
\ee 
Combining Eqs.~\eqref{eqa:ybound3},~\eqref{eqa:firsttermbound}, and 
\eqref{eqa:ytriangle}, we obtain 
\be 
|Y-\Tr[\rhot \sigma_y]|\lesssim  \frac{\kth^3\Ts}{\trev} \sqrt{\frac{\varepsilon_0 t}{\hbar}}e^{-\frac{E_J}{k_{\rm B}T}}+\frac{1}{\sqrt{\lambda}}e^{-\frac{1}{\pi^2\lambda^2}}.
\label{eqa:ybound01}
\ee 
This  establishes Eq.~\eqref{eqa:ybound0}, which was the goal of this subsection. 
\subsubsection{Derivation of Eq.~\eqref{eqa:ychangebound}}
\label{seca:ychange}
We next derive Eq.~\eqref{eqa:ychangebound}, which  bounds the change of $\cY$ over the stabilizer segment. 
To this end, we note from Eq.~\eqref{eqa:y_in_terms_of_s} that 
\be 
\partial_t \cY = \Tr[\tau_x \partial_t Z],
\ee 
with $  Z\equiv U V^\dagger  \sigma_z\rho_0 V U^\dagger.
$ defined  in Eq.~\eqref{eqa:y_in_terms_of_s}. 
Using $V=e^{-i\varepsilon_L t M^2/\hbar}$, $U=e^{-it H_{\rm s}}$ and $\Hpz =H_{\rm s}-\varepsilon_L M^2$, where  $M\equiv\nint(\varphi/\varphi_0)$, explicit computation yields 
\begin{align}
\partial_t Z &= -\frac{i}{\hbar} [\Hpz,Z] + \delta \dot Z, \quad {\rm where}\notag\\ 
 \delta \dot Z &\equiv \frac{i\varepsilon_L}{\hbar} [(U M^2U^\dagger  -M^2),Z]. 
\end{align}
Thus
\be 
\partial_t Y =  \frac{i}{\hbar} \Tr[S  [\Hpz,\tau_x]]+\Tr[\tau_x \delta \dot Z]. 
\label{eqa:dty_expr}
\ee 
We now bound  the two terms on the right hand side above separately.

We first bound   $\Tr[\tau_x \delta \dot Z]$. 
We first note   that, since $|\Tr[AB]|\leq \trnorm{A}\norm{B}$ and $\norm{\tau_x}=1$ , $|\Tr[\tau_x \delta \dot Z]| \leq \trnorm{\delta \dot Z}$.
To bound  $\trnorm{\delta \dot Z}$ we note from the definition of $Z$ in Eq.~\eqref{eqa:y_in_terms_of_s} that
\be 
\delta \dot Z=  \frac{i\varepsilon_L}{\hbar} \left( [U,M^2]V^\dagger \sigma_z\rho_0  V U^\dagger+ U V^\dagger \sigma_z\rho_0  V [U,M^2] \right).
\ee 
Using $|\Tr[\tau_x \delta \dot Z]| \leq \trnorm{\delta \dot Z}$ along with the triangle inequality, we thus find 
\begin{equation} 
|\Tr[\tau_x \delta \dot Z]| \leq\frac{\varepsilon_L}{\hbar} \trnorm{ [U,M^2]V^\dagger \sigma_z \rho_0 }+\frac{\varepsilon_L}{\hbar} \trnorm{[U^\dagger,M^2]V\rho_0}\,.
\end{equation}
We can use Eq.~\eqref{eqa:ufm_result} to bound the two terms on the right, with $f(x)=x^2$ and $g(x)= e^{-i\varepsilon_L tx^2/\hbar -i \pi x} $  and $g(x)= e^{-i\varepsilon_L tx^2/\hbar }$  for the first and second term, respectively. Using $\norm{f}=\wmax^2 \sim \kth^2$ (see Footnote~\cite{FluxTruncation}), this analysis yields 
\be 
|\Tr[\tau_x \delta \dot Z]|\leq \frac{ \kth ^3}{\trev}\sqrt{ \frac{\varepsilon_0 t}{\hbar}}e^{-\frac{E_J}{k_{\rm B}T}}.
\label{eqa:deltasdotbound}
\ee 

We next bound the first term on the right hand side of Eq.~\eqref{eqa:dty_expr}, $\frac{i}{\hbar}\Tr([\tilde H_{\rm s},\tau_x]S)$. 
To this end, we first recall  
$
 [\Hpz,\tau_x]  = \frac{\bar \varphi \hbar }{ Le\alpha} X ,
$
 where $X  \equiv \sum_z \theta(\alpha-|q/e-z- 1/2|)(-1)^z$ 
 [See Eqs.~\eqref{eqa:hsigmaprime_expr}~and~\eqref{eqa:fprime_expression}]. 
Thus
\be 
 \Tr[Z [\Hpz,\tau_x]]=\frac{\hbar }{ Le \alpha}\Tr[Z\bar \varphi X].
\ee 
Next, we insert $Z= \sigma_z \tilde \rho + (Z-\sigma_z \tilde \rho)$, resulting in 
\begin{align}
 \Tr[Z [\Hpz,\tau_x]]&=\hbar \frac{\Tr[\sigma_z \tilde \rho \bar \varphi X ]}{Le\alpha}+\hbar\frac{\Tr[ (Z-\sigma_z \tilde \rho)\bar\varphi X]}{Le\alpha}. 
 \label{eqa:scommbound}
\end{align}

We first bound the second term in the numerator above. To this end, we use $|\Tr[AB]|\leq \trnorm{A}\norm{B}$ and $\norm{\bar \varphi X}\leq\norm{X}\norm{\bar \varphi}= \varphi_0/2$,  finding 
\be 
|\Tr[\bar\varphi X ( Z-\sigma_z  \rhot)]|\lesssim \frac{\varphi_0}{2}\trnorm{( Z-\sigma_z  \rhot)}
\ee 
Next, we  use the triangle inequality  \be 
\trnorm{( Z-\sigma_z  \rhot)} \leq \trnorm{(Z-\sigma_z  \rhopp)}+\trnorm{\sigma_z(\rhot-\rhopp)}.
\ee
Recalling from  Eqs.~\eqref{eqa:s_bound1}~and~\eqref{eqa:trho_to_rhopp_bound3} that 
\begin{align}
    \trnorm{( Z-\sigma_z  \rhopp)}&\lesssim  \kth \sqrt{ \frac{\varepsilon_0 t}{\hbar}}e^{-\frac{E_J}{k_{\rm B}T}}, \\ \trnorm{\sigma_z(\rhot-\rhopp)}&\lesssim  \frac{t\kth^3}{\trev} \sqrt{\frac{\varepsilon_0 t}{\hbar}}e^{-\frac{E_J}{k_{\rm B}T}},
\end{align}
 we thus find 
\be 
|\Tr[ ( Z-\sigma_z \tilde \rho)\bar\varphi X]|\lesssim \frac{ \varphi_0t\kth^3}{2\trev}\sqrt{\frac{ \varepsilon_0 t}{\hbar}}e^{-\frac{E_J}{k_{\rm B}T}}.
\label{eqa:scommbound1}
\ee 

We now bound the first term  on the right-hand side of Eq.~\eqref{eqa:scommbound}. To this end, we exploit that $X$ only has support for $S_2 \leq s_\alpha$, implying 
$ 
X=  P_2(s_\alpha)X  P_2(s_\alpha).
$
Also using $|\Tr[AB]|\leq \norm{A}\trnorm{B}$ 
along with  $[ P_2(s_\alpha), \varphi]=[ P_2(s_\alpha),\sigma_z]=0$ and $\trnorm{ P_2(s)\tilde \rho P_2(s)}= \cP_2(s)$, we find
\be 
|\Tr[\sigma_z \tilde \rho \bar \varphi X]|\leq \cP_2(s_\alpha) \frac{\varphi_0}{2}
.\ee 
Recalling that  $\cP_2(s_\alpha) \lesssim \frac{1}{\sqrt{\lambda}} e^{-\frac{1}{\pi^2 \lth^2}}$ [Eq.~\eqref{eqa:alpha_choice_result}], we obtaiun 
\be 
|\Tr[\sigma_z \tilde \rho \bar \varphi X]|\lesssim \frac{\varphi_0}{2\sqrt{\lambda}} e^{-\frac{1}{\pi^2\lth^2}}.
\label{eqa:scommbound2}
\ee

Combining Eqs.~\eqref{eqa:scommbound1} and \eqref{eqa:scommbound2}  with Eq.~\eqref{eqa:scommbound}, and recalling  $\alpha=\lambda /\sqrt{2\pi}$, we conclude 
\begin{equation}
\frac{1}{\hbar}|\Tr[Z [\tilde H_{\rm s},\tau_x]]|\lesssim \frac{ \varphi_0}{Le}\left[\sqrt{\frac{ \varepsilon_0 t}{\hbar}}\frac{\kth^3 te^{-\frac{E_J}{k_{\rm B}T}}}{\sqrt{\lambda}\trev}+\frac{e^{-\frac{1}{\pi^2\lambda^2}}}{\lambda^{\frac{3}{2}}}\right]
\label{eqa:comms_bound}
\end{equation}
Using $\frac{\varphi_0}{Le}  = \frac{1}{L}\frac{\sqrt{L}}{\sqrt{C}}=\frac{1}{\trev}$, and combining the above result with Eqs.~\eqref{eqa:deltasdotbound}~and~\eqref{eqa:dty_expr}, we find 
\begin{equation}
    |\partial_t \cY| \lesssim \sqrt{ \frac{\varepsilon_0 t}{\hbar}}\frac{ \kth ^3(1+{t}/{\sqrt{\lambda}\trev})}{\trev}e^{-\frac{E_J}{k_{\rm B}T}}
    +\frac{e^{-\frac{1}{\pi^2\lambda^2}}}{\lambda^{\frac{3}{2}}\trev}. 
\end{equation}
Integrating the right-hand side from $t=0$ to $t=\Ts$, and using $\kth \gg 1$, $\lth \ll 1$, and $\Ts \geq \trev$, we  obtain Eq.~\eqref{eqa:ychangebound}, which was the goal of this subsection.

\section{Derivation of Eq.~\eqref{eqa:ufm_result}}
\label{app:aux_result}

Here we derive  Eq.~\eqref{eqa:ufm_result}, which is used in Appendices~\ref{seca:heff_derivation} and \ref{app:error_rate_bound}. Specifically, we consider an initial state $\rho_0$ which is confined inside the wells of the cosine potential from the Josephson junctions (see Sec.~\ref{sec:Stabilization}), and has its $\varphi$ support confined to a finite interval, $|\varphi|\leq \wmax \varphi_0$, where $\wmax\ll E_J/h\flc$. 
Below we show that, for this state, and for any two functions $f$ and $g$ of $M = \nint(\varphi/\varphi_0)$, 
\be 
\norm{[U (t),f(M)]g(M)\rho_0}_{\rm tr}  \lesssim \norm{f}\norm{g}\wmax \sqrt{\frac{\varepsilon_0 t}{ \hbar}}e^{-\frac{E_J}{k_{\rm B}T}}
\label{eqa:auxresult}\ee 
where   we suppressed $\mathcal O(1)$ prefactors, $\norm{g}=\max_{|x|\leq \wmax}|g(x)|$, and
$U(t)=^{-iH_{\rm s}t}$ is the evolution operator of the combined circuit-resistor system during the stabilizer segment. 

\newcommand{\cO}{\mathcal{O}}
To establish Eq.~\eqref{eqa:auxresult}, we first introduce the shorthand for the argument on the left hand side above, 
\be 
A \equiv [U(t), f(M)] g(M)\rho_0.
\ee 
Our goal thus is to bound $\trnorm{A}$. 
To this end, we first 
insert $1 = \sum_z P_z$ where $P_z = \theta([z+1/2]-\varphi)\theta(\varphi-[z-1/2])$, to obtain
\be 
A=  \sum_ w \sum_{z \neq w}  P_z U(t)P_w \rho_0 [f(z)-f(w)]g(w)
\ee 
where $ \sum_{z\neq w} $ denotes the sum over all integers $z$  between $-\wmax$ and $\wmax$ \textit{distinct} from  $w$, while $\sum_w$ implicitly sums between $-\wmax$ and $\wmax$.  
Next, we use that, for any operator $\cO$, $\trnorm{\cO \rho }\leq \sqrt{\Tr[\cO^\dagger \cO \rho]}$~\footnote{Specifically, $\trnorm{\cO\rho}=\Tr[\cU \cO\rho]$ for some unitary $\cU$ that generates the polar decompositon of $\cO\rho$. We identify $\Tr[\cU \cO\rho]= (\sqrt{\rho},\sqrt{\rho}\cU \cO)_{\rm HS}$, with $(x,y)_{\rm HS}=\Tr[x^\dagger y] $ the Hilbert Schmidt inner product. The result follows when using the Cauchy-Schwartz inequality, $(x,y)_{\rm HS}\leq \sqrt{(x,x)_{\rm HS}(y,y)_{\rm HS}}$.}, leading to 
\be 
\trnorm{A}\leq  \sum_ w  \sum_{z\neq w} |g(w)|\sqrt{J_{wz}[f(z)-f(w)]^2 }\,, 
\ee 
Where  $J_{wz}\equiv \Tr(U P_w \rho_0P_wU^\dagger P_z)$.
Next, we use that  $|f(z)-f(w)|\leq 2\norm{f}$ and $|g(w)|\leq \norm{g}$ to write 
\be 
\trnorm{A}\leq  2\norm{f}\norm{g}\sum_ w  \sum_{z\neq w} \sqrt{J_{wz}}\,.
\label{eqa:trnormaeq}
\ee 
We  recognize $J_{wz}$ as the total probability flow from well $w$ to well $z$  over the time interval from $0$ to $t$.
 Due to the energy barrier of order $2E_J$ between the wells, the rate of probability support escaping well $w$ is of order $ p_we^{-2\frac{E_J}{k_{\rm B}T}}\frac{\varepsilon_0}{\hbar} $, with $p_w \equiv \Tr[P_w \rho_0]$ the initial probability weight in well $w$~\cite{Ankerhold_1995}. 
As a result, 
\be 
\sum_{z\neq w}|J_{wz}|\lesssim  e^{-2\frac{E_J}{k_{\rm B}T}} \varepsilon_0 t p_w /\hbar  .
\ee 
Using that the sum runs over $2\wmax + 1$ terms, and that, for any vector $(v_1,\ldots v_N)$, $\sum_{i=1}^N \sqrt{|v_i|} \leq \sqrt{N}\sqrt{\sum_i|v_i|} $ (a consequence of the Cauchy-Schwartz inequality), we thus find 
\be 
\sum_{z\neq w} \sqrt{J_{wz}} \lesssim e^{-\frac{E_J}{k_{\rm B}T}}\sqrt{ \frac{ \varepsilon_0 t }{\hbar}}\sqrt{p_w} \sqrt{\wmax}.
\ee 
Using  $\sum_{i=1}^N \sqrt{|v_i|} \leq \sqrt{N}\sqrt{\sum_i|v_i|}$ again, along with $\sum_{w}p_w = 1$, we find $\sum_w \sqrt{p_w}\lesssim\sqrt{\wmax}$. Thus, 
\be 
\sum_w \sum_{z\neq w} \sqrt{J_{wz}} \lesssim e^{-\frac{E_J}{k_{\rm B}T}}\sqrt{ \frac{ \varepsilon_0 t }{\hbar}} \wmax.
\ee 
Combining this with Eq.~\eqref{eqa:trnormaeq}, we obtain 
\be 
\trnorm{A}\lesssim   \norm{f}\norm{g} e^{-\frac{E_J}{k_{\rm B}T}}\sqrt{ \frac{ \varepsilon_0 t }{\hbar}} \wmax
\ee 
This establishes Eq.~\eqref{eqa:auxresult}, which was the goal of this appendix.

\section{Inequality for squeezing parameter}
\label{app:lambdaineq}
Here we show that the thermally-renormalized GKP squeezing parameter, $\lambda \equiv \sqrt{\coth(2 \varepsilon_0/k_{\rm B}T)}\lambda_0$, satisfies 
\be 
\lambda^2 \leq  \frac{k_{\rm B}T}{\pi^2 E_J}+\lambda_0^2
\label{eqa:lambda_result}
\ee 
where $\varepsilon_0 = \sqrt{4\pi h\flc  E_J}$ denotes the characteristic excitation energies in the wells of the Josephson  potential, and $\lambda_0 \equiv   \left({ h\flc  }/{4\pi^3 E_J}\right)^{1/4}$ denotes the zero-point fluctuation of  flux  in these wells, in units of $\varphi_0$. 
This result is used in Appendix~\ref{app:error_rate_bound}. 
To this end, we use $\coth(x)\leq 1+x^{-1}$, implying $ 
\lambda^2  \leq  \lambda_0^2 + {\lambda_0^2}k_{\rm B}T/{ \varepsilon_0}.
$ 
Eq.~\eqref{eqa:lambda_result} follows by inserting $\varepsilon_0 = 4\pi^2 \lambda_0^2 E_J$ and using $4\pi^
2 \geq \pi^2$. 
Note also that a tighter bound can be established using $\coth(x)\leq \sqrt{1+x^{-2}}$, leading to $\lambda^2  \leq  \sqrt{\lambda_0^4 + (4\pi^2  E_J/k_{\rm B}T)^{-2}}$.

\section{Envelope of charge and flux probability distributions}
\label{app:envelope}
In this Appendix, we show that the envelope of the charge and flux probability distributions for stabilized states in the system are approximately given by Gaussians of standard deviations $\kth e $ and $\kth \varphi_0$, respectively, where 
\be 
\kth \equiv \frac{\sqrt{\coth(2\varepsilon_0/k_{\rm B}T)}}{\pi \lambda_0 }, \quad\lambda_0 \equiv  \left(\frac{ h\flc  }{4\pi^3 E_J}\right)^{1/4}.
\ee

To obtain this result, it is convenient to  first consider the charge probability density of a stabilized state, convolved with a Gaussian smoothening kernel of some given width $\sigma$,
\be 
\hat W_\sigma (q_0) =\frac{1}{\sqrt{2\pi}\sigma}\int_{-\infty}^\infty dq_1  e^{-(q_1-q_0)^2/2\sigma^2} \delta(\hat q-q_1).
\label{eqa:wsig_def}
\ee 
with $\hat q$ the charge operator; here we introduced the $\hat {\cdot}$ accent to avoid confusion between operators and scalars. 
The above definition implies  that $\langle \hat W_\sigma (q_0)\rangle =  \frac{1}{\sqrt{2\pi}\sigma}\int dq_1  e^{-(q_1-q_0)^2/2\sigma^2}p_q(q_1)$ with $p_q(x)= \langle \delta(\hat q-x)\rangle$ the charge probability distribution of the system. 
In this sense, $\Tr[\hat \rho \hat W_\sigma (q_0)]$ gives the  charge probability distribution in the state $\hat \rho$, when smoothed by a Gaussian kernel of width $\sigma$. 

We now compute $\langle \hat W_\sigma(q)\rangle $ for a state stabilized by the protocol in the main text, $\hat \rho$. 
Below, we will  exploit that stabilized states have all their $\varphi$-support confined near integer multiples of $\varphi_0$. 
In order to do this, we first compute the matrix elements of $\hat W_\sigma(q)$ in the $\varphi$-basis, $\{|\varphi\rangle\}$ with $|\varphi\rangle$ the  eigenstate of $\hat \varphi$ with eigenvalue $\varphi$ and normalization $\langle \varphi|\varphi'\rangle = \delta(\varphi-\varphi')$. 
Using   that $\langle \varphi|\delta(\hat q-q_1)|\varphi'\rangle  = \frac{1}{2\pi \hbar } e^{-iq_1 (\varphi-\varphi')/\hbar}$ in Eq.~\eqref{eqa:wsig_def} and completing the square leads  us to  
\be 
\langle \varphi|\hat W_\varphi(q_0)|\varphi'\rangle = \frac{1}{2\pi\hbar } e^{-i\frac{q_0}{\hbar} (\varphi-\varphi')  -\frac{(\varphi-\varphi')^2}{2\Delta \varphi_\sigma}} .
\label{eqa:wsig_elements}
\ee 
with $\Delta \varphi_\sigma \equiv \hbar/\sigma$. 
When $\sigma \gg e$, $\Delta \varphi_\sigma$ is much smaller than $\hbar/e = \varphi_0/\pi$. 
As a result, $\langle \varphi|\hat W_\varphi(q_0)|\varphi'\rangle$ is only nonzero when $|\varphi-\varphi_0|\ll \varphi_0$. 
Since $\langle \varphi|\hat \rho|\varphi'\rangle$ is only nonzero if $\varphi$ and $\varphi'$ are within a distance $\sim \lth\varphi_0$ from multiples of $\varphi_0$, we thus find, 
\be 
\langle\hat W_{\sigma}(q)\rangle   = \sum_z \int_{-[z+\frac{1}{2}]\varphi_0}^{[z+\frac{1}{2}]\varphi_0} \!\! d^2 \varphi  \langle \varphi|\hat \rho|\varphi'\rangle e^{-\frac{(\varphi-\varphi')^2}{2\Delta \varphi_\sigma^2}-i\frac{q_0}{\hbar} (\varphi-\varphi')}. 
\ee 
with the shorthand $\int_{a}^{b}d^2\varphi = \int_{a}^b d\varphi\int_{a}^{b} d\varphi'$. 
This result holds for for $\sigma \gg e$ and $\lth\ll 1$. 
We recognize the integral above as $\int_{-\infty}^{\infty} d^2 \varphi  \langle \varphi|\hat \rho_z|\varphi'\rangle e^{-\frac{(\varphi-\varphi')^2}{2\Delta \varphi_\sigma^2}-i\frac{q_0}{\hbar} (\varphi-\varphi')}$, with $\hat \rho_z$ the projection of $\hat \rho$ into well $z$, defined as the $\varphi$ interval $[z-1/2]\varphi_0 < \hat \varphi < [z+1/2\varphi_0]$. 
Using Eq.~\eqref{eqa:wsig_elements}, we can rewrite the integral as $\frac{1}{2\pi\hbar } \sum_z \Tr[W_\sigma(q) \rho_z]$, leading us to 
\be 
\langle \hat W_\sigma(q_0)\rangle = \frac{1}{2\pi\hbar } \sum_z \Tr[W_\sigma(q) \rho_z] .
\label{eqa:wsig_expval_zsum}
\ee 

We next recall that $\rho_z$ describes the thermal steady state of  a Harmonic oscillator of vacuum fluctuation length $\lambda_0$ and excitation energy $\varepsilon_0$ (weighted by the total probability of finding the system in well $z$, $p_z \equiv \Tr[\rho_z]$).
The charge probability distribution for this state is given by 
\be 
\Tr[\hat \rho_z\delta(\hat q-q_1)] \approx  \frac{p_z}{\sqrt{\pi} e \kth }e^{-\frac{q_1^2}{\kth ^2e ^2}}
\label{eqa:charge_probablity_of_well_matrix}
\ee 

Combining  Eqs.~\eqref{eqa:charge_probablity_of_well_matrix} and ~\eqref{eqa:wsig_def}, and using the convolution rule for  Gaussian distributions, we find 
\be 
\Tr[\hat W_\sigma(q) \hat \rho_z] = \frac{p_z}{\sqrt{\pi[2\sigma^2+\kth ^2e^2]}} e^{-\frac{q^2}{\kappa _{\rm th}^2 e^ + 2\sigma^2}}.
\ee 
Inserting this into Eq.~\eqref{eqa:wsig_expval_zsum}, and using $\sum_z p_z=1$, we finally obtain 
\be 
\langle \hat W_\sigma(q)\rangle \approx \frac{1}{\sqrt{2\pi[\kth^2e^2/2 + \sigma^2]}} e^{-\frac{q^2}{\kth ^2 e^2  + 2\sigma^2}}
\ee 
Recalling the convolution rules for Gaussian distribution, this result is consistent with the envelope function for the charge distribution being given by a Gaussian with width $\kth  e$.

To infer the envelope of the flux probability distribution we note that this envelope is given by that of the charge probability distribution, after the rescaling $q/e\to \varphi/\varphi_0$; thus, the envelope of the flux proability distribution is a Gaussian of width $\kth \varphi_0$.

\section{Emergence of peak structure in the charge probability distribution}
\label{Appendix_Measurement}
Here, we demonstrate how the characteristic fractal peak structure of the charge probability distribution in Fig.~\ref{fig:measurement} emerges---leveraged in the readout protocol discussed in Sec.~\ref{sec:Readout}.


When dissipatively stabilized, the system will initially be in a thermal mixture of coherent superpositions of {low-energy} states of the wells of the cosine potential from the Josephson junction. 
We consider one of such superpositions, writing it as 
$|\psi(0)\rangle = \sum_{m,\mu} c_{m\mu} |m,\mu\rangle$, where $|m,\mu\rangle$  is the state with wavefunction $\langle \varphi |m,\mu\rangle = \psi_\mu(\varphi-m\varphi_0)$, with $\psi_\mu(\varphi) \equiv e^{-\frac{\varphi^2}{2\lambda_0^2\varphi_0^2}} H_\mu\left(\varphi/\lambda_0\varphi_0\right){(2^\mu \mu!2\pi\lambda_0\sqrt{\pi})}^{-1/2}$, where  $H_\mu(x)$ denotes the $\mu$th  Hermite polynomial; i.e., the $\mu$th excited state of the Harmonic oscillator corresponding to well $m$. 
For superpositions of low-energy-well states, $c_{m,\mu}$ is only nonzero for  $\mu \ll \sqrt{1/\lambda_0}$.  Also note  that $c_{m\mu}$ is nonzero only for even or odd $m$ when the system is in an $+1$ or $-1$ eigenstate of $\sigma_z$, respectively. 
Since the system is dissipatively stabilized, $\langle S_2\rangle \approx 1$, and the superposition is phase-coherent: $c_{m\mu}\approx c_{m+2,\mu}$. 

We now consider the evolution of $|\psi(0)\rangle$ during the stabilizer segment (neglecting the effects of dissipation, which can be analyzed via the approach from Sec.~\ref{sec:Stabilization}).  
After evolution with $H_{\rm s}$ for a time $t$, the state of the system is hence given by
$
e^{-iH_{\rm s}t}|\psi(0)\rangle \approx \sum_m c_{m\mu} e^{-i[m^2\varepsilon_L+\mu\varepsilon_0]t/\hbar} |m,\mu\rangle
$. 

To obtain the charge probability distribution, we consider the evolution of the Wigner function of the system, $W(\varphi,q,t)\equiv \frac{1}{\pi\hbar}\int_{-\infty}^\infty d\varphi'\, \langle \varphi +\varphi'|\rho(t)|\varphi-\varphi'\rangle e^{2iq\varphi/\hbar}$, from which we may obtain the charge probability distribution through $p(q,t)=\int d\varphi W(\varphi,q,t)$. 
A straightforward derivation 
shows that 
\begin{widetext}
\be
W(\varphi,q,t) = \! \sum_{m,n,\mu,\eta}
c_{m\mu}^* c_{n\eta} e^{\pi i \frac{  q(m-n)}{e}+i[\varepsilon_L  (m^2-n^2)+\varepsilon_0(\mu-\eta)]t/\hbar}w_{\mu\eta}\left(\frac{\varphi-\varphi_0(k+l)}{2} ,q\right)\,, 
\label{eq:wigner1}
\ee
where $w_{\mu\eta}(\phi,q) $ is the cross-term Wigner function of eigenstates $\mu$ and $\eta$ of the Harmonic oscillator corresponding to the central well:
\be 
w_{\mu\eta}(\varphi,q) = \frac{1}{\pi\hbar}\int_{-\infty}^{\infty}d\varphi\, \psi^*_\mu(\varphi+\varphi')\psi_\eta(\varphi-\varphi')e^{2iq\varphi'/\hbar }.
\ee 
Introducing $l=m+n$, such that $(m^2-n^2)=l(l-2n)$, along with $\varepsilon_L = \pi\hbar2\pi \flc/2$,  we find  
\be
W(\varphi,q,t) 
     = \sum_{l \in \mathbb Z}\sum_{\mu,\eta}e^{-i\varepsilon_0(\mu-\eta)t/\hbar}w_{\mu\eta}(\varphi- l\varphi_0/2,q)f_l^{\mu\eta}\left(q-le  \frac{2\pi \flc  t}{2}\right), \quad f_l^{\mu\eta}(q)\equiv \sum_{n} c_{l-n,\mu}^* c_{n,\eta} \, e^{-\pi i q(l-2n)/e}.
   \label{eq:wigner}
 \ee 
 \end{widetext}
Since $c_{n,\mu}\approx c_{n+2,\mu}$, each $f^{\mu\eta}_n(q)$ is sharply peaked around $q\approx ze$ for $z\in \mathbb Z$; i.e.,  each $f_n(q)$ is a nascent Dirac comb with periodicity $e$. 
Since $c_l$ is only nonzero when $l=s\mod 2$, the sign of the peaks alternate based on the parity of $n$. 
Since  $w_{\mu\eta}(\phi,q)$ is sharply peaked around $\phi=0$,  $W(\phi,q,0)$ has its support confined as peaks near $(\varphi,q)=( n_1\varphi_0/2 ,n_2 e/2)$ for integer $n_1$, $n_2$, 
 with corresponding sign $(-1)^{(s+n_1)n_2}$. 
This grid structure is clearly visible in Fig.~\ref{fig:measurement}(b) of the main text, where we plot the Wigner function numerically obtained for the system in panel (a) at $t=0$. 
As $t$ increases, Eq.~\eqref{eq:wigner} the column of  peaks where $\varphi = \varphi_0 n_1$ shifts in the positive $q$-direction with  velocity  $\pi n_1  e \flc $.

When $ t = \frac{a}{b}\frac{\TLC}{2\pi}$ for integers $a,b$, peaks from  columns where $a n_1 \sim   c \, (\mod b)$ align  at values $q/e = c/2b$. 
Thus, the charge probability distribution, $p(q,t)=\int d\varphi W(\varphi,q,t)$, has its support confined around values  $q=ec/2b$ for  each integer $c$. 
However, due to the alternating sign  of peaks in columns where $n_1$ and $s$ have opposite parities, the peaks of the Wigner function will only interfere constructively at  values of $c$ with parity $s $, provided $b$ is even. 
This interference is in indicated with the dashed arrows  in Fig.~\ref{fig:measurement}(a) at $ t =\TLC/8\pi$.
Consequently, $p(q,t)$ consists of peaks centered at values $q/e = (2n + s)/b$ where $n \in \mathbb Z$.

\end{document}